\documentclass[twocolumn]{aastex631} 
\usepackage[utf8]{inputenc}
\usepackage{longtable}

\AtBeginEnvironment{longtable}{\tiny}

\usepackage{amsmath}

\usepackage[bottom]{footmisc}

\usepackage{xcolor}

\usepackage[figuresright]{rotating}
\usepackage{graphicx}
\shortauthors{R. Huang et al.}
\shorttitle{XMM-Newton Legacy Survey of M31}

\begin{document}

\title{An XMM-{\color{red}New}ton View of the {\color{red}AN}dromeda {\color{red}G}alaxy as {\color{red}E}xplored in a {\color{red}L}egacy {\color{red}S}urvey (New-ANGELS) I: the X-ray Source Catalogue}

\author[0000-0001-7900-4204]{Rui Huang}
\affiliation{Department of Astronomy, Tsinghua University, Beijing 100084, China}

\author[0000-0001-6239-3821]{Jiang-Tao Li}
\affiliation{Department of Astronomy, University of Michigan, 311 West Hall, 1085 S. University Ave, Ann Arbor, MI, 48109-1107, U.S.A.}

\author[0000-0002-6324-5772]{Wei Cui}
\affiliation{Department of Astronomy, Tsinghua University, Beijing 100084, China}

\author[0000-0001-6276-9526]{Joel N. Bregman}
\affiliation{Department of Astronomy, University of Michigan, 311 West Hall, 1085 S. University Ave, Ann Arbor, MI, 48109-1107, U.S.A.}

\author[0000-0002-0584-8145]{Xiang-Dong Li}
\affiliation{School of Astronomy and Space Science, Nanjing University, Nanjing 210023, People’s Republic of China}
\affiliation{Key Laboratory of Modern Astronomy and Astrophysics, Nanjing University, Ministry of Education, Nanjing 210023, People’s Republic of China}

\author[0000-0003-0293-3608]{Gabriele Ponti}
\affiliation{INAF-Osservatorio Astronomico di Brera, Via E. Bianchi 46, 23807 Merate (LC), Italy}
\affiliation{Max-Planck-Institut für Extraterrestrische Physik, Giessenbachstrasse, 85748 Garching, Germany}

\author[0000-0002-2941-646X]{Zhijie Qu}
\affiliation{Department of Astronomy \& Astrophysics, The University of Chicago, Chicago, IL 60637, U.S.A}

\author[0000-0002-9279-4041]{Q. Daniel Wang}
\affiliation{Department of Astronomy, University of Massachusetts, Amherst, MA 01003, U.S.A.}

\author[0000-0001-8632-904X]{Yi Zhang}
\affiliation{INAF-Osservatorio Astronomico di Brera, Via E. Bianchi 46, 23807 Merate (LC), Italy}

\correspondingauthor{Jiang-Tao Li}
\email{pandataotao@gmail.com}

\begin{abstract}
We introduce the New-ANGELS program, an \emph{XMM-Newton} survey of $\sim7.2\rm~deg^2$ area around M\,31, which aims to study the X-ray populations in M\,31 disk and the X-ray emitting hot gas in the inner halo of M\,31 up to 30 kpc. In this first paper, we report the catalogue of 4506 detected X-ray sources, and attempt to cross-identify or roughly classify them. 
We identify 352 single stars in the foreground, 35 globular clusters and 27 supernova remnants associated with M\,31, as well as 62 AGNs, 59 galaxies, and 1 galaxy clusters in the background. We uniquely classify 236 foreground stars and 17 supersoft sources based on their X-ray colors. X-ray binaries (83 LMXBs, 1 HMXBs) are classified based on their X-ray colors and X-ray variabilities.
The remaining X-ray sources either have too low S/N to calculate their X-ray colors or do not have a unique classification, so are regarded as unclassified. The X-ray source catalogue is published online. Study of the X-ray source populations and the contribution of X-ray sources in the unresolved X-ray emissions based on this catalogue will be published in companion papers.
\end{abstract}


\section{Introduction} \label{sec:Intro}

X-ray studies of stellar sources provide a direct probe of the formation and evolution of different stellar populations (e.g., \citealt{Fabbiano06}).
Compared to X-ray sources in the MW (e.g., \citealt{Liu07,Revnivtsev07,Revnivtsev11,Zhu18}) with large distance uncertainties (e.g., \citealt{Jonker04}), sources in a nearby galaxy are essentially at the same distance and are thus optimized to study the science based on accurate luminosity measurements [e.g., the X-ray source luminosity functions (\citealt{Gilfanov04}).]. 
For a local galaxy with a distance of a few Mpc and a typical \emph{Chandra} or \emph{XMM-Newton} exposure time of a few hundred ks, a point source detection limit of $\sim10^{36-37}\rm~ergs~s^{-1}$ in the 0.5-7 keV band, for example, is typically reached 
(e.g., \citealt{Kim04,Kim09,Fabbiano06,Liu07,Revnivtsev08,Lehmer10,Mineo12}).
This detection limit is insufficient to characterize the properties of individual X-ray binaries down to the LF break at $\lesssim10^{37}\rm~ergs~s^{-1}$ (e.g., \citealt{Gilfanov04}). 
Most nearby galaxies with moderate X-ray observations only have about $\sim10^2$ detected X-ray sources (e.g., \citealt{Irwin02,LiJ08,LiJ13a}), which is too few for many follow-up analyses, such as estimating the X-ray-stellar mass ratio ($N_{\rm X}/M_*$ or $L_{\rm X}/M_*$; the latter could be highly biased by one or two luminous X-ray sources; e.g., \citealt{Revnivtsev08,Lehmer10}). 
This highlights the importance of an X-ray survey of a nearby galaxy to a detection limit as low as $\lesssim10^{36}\rm~ergs~s^{-1}$.

The Andromeda galaxy M\,31 is the closest massive spiral galaxy to the MW (distance $d=761\rm~kpc$, $\mathrm{5^{\prime\prime} \approx 18\rm~pc}$; \citealt{liSub2DistanceM312021}). It is the most massive galaxy in the Local Group and one of the most massive spiral galaxy in the local Universe, with a stellar mass of $M_*=(1-1.5)\times10^{11}\rm~M_\odot$ and a dark matter halo mass of $M_{\rm 200}=(8-11)\times10^{11}\rm~M_\odot$ \citep{Tamm12}. M\,31 is very quiescent in star formation, with a star formation rate of only $\rm SFR\sim0.4\rm~M_\odot~yr^{-1}$ \citep{Barmby06}. The Galactic foreground absorption toward M\,31 ($N_{\mathrm{H}} \sim 5\times 10^{21}~\mathrm{cm^{-2}}$) is not too high \citep{HI4PI2016}. The small distance, large stellar and gravitational mass, low SFR, and large galactic disk inclination make M\,31 ideal to study the population, luminosity function, and vertical distribution of old stellar X-ray sources in a large luminosity range.

X-ray sources toward the direction of M\,31 has been extensively studied with various X-ray telescopes (e.g., \citealt{vanspeybroeckObservationsXraySources1979,1991ApJ...382...82T,1997A&A...317..328S,2001A&A...373...63S}), especially the bulge (e.g.,
\citealt{vossStudyPopulationLMXBs2007}, \citealt{stieleTimeVariabilityXray2008}) and disk area (e.g., 
\citealt{kongChandraStudiesRay2003},  \citealt{sasakiDeepXMMNewtonObservations2018}).
The first catalogue of the \emph{XMM-Newton} survey of M\,31 was published by \cite{Pietsch05}, which consisted of 856 sources. \cite{stieleDeepXMMNewtonSurvey2011a} then updated the catalogue which contains 1948 sources. However, these observations toward the disk and bulge are insufficient to characterize the spatial variation of X-ray sources over large physical scales, which could be affected by the formation, evolution, and migration of different source populations (e.g., \citealt{Zuo08}).
 
\begin{figure*}
\centering
\includegraphics[width=0.97\textwidth]{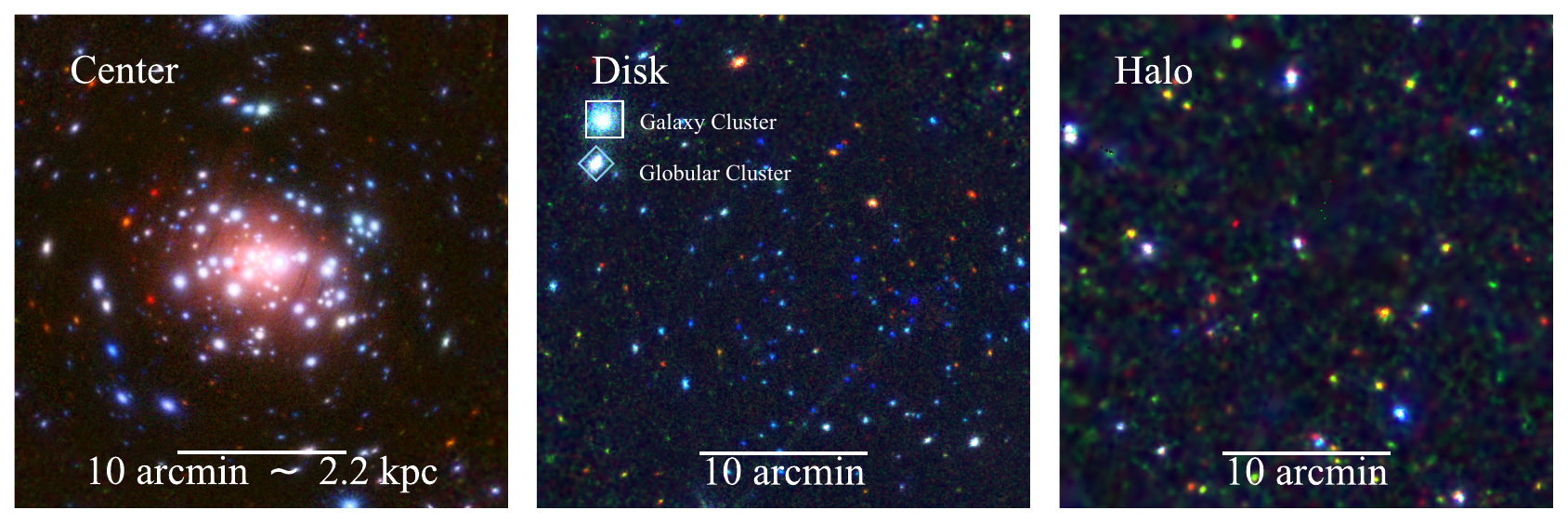}
\includegraphics[width=0.97 \textwidth]{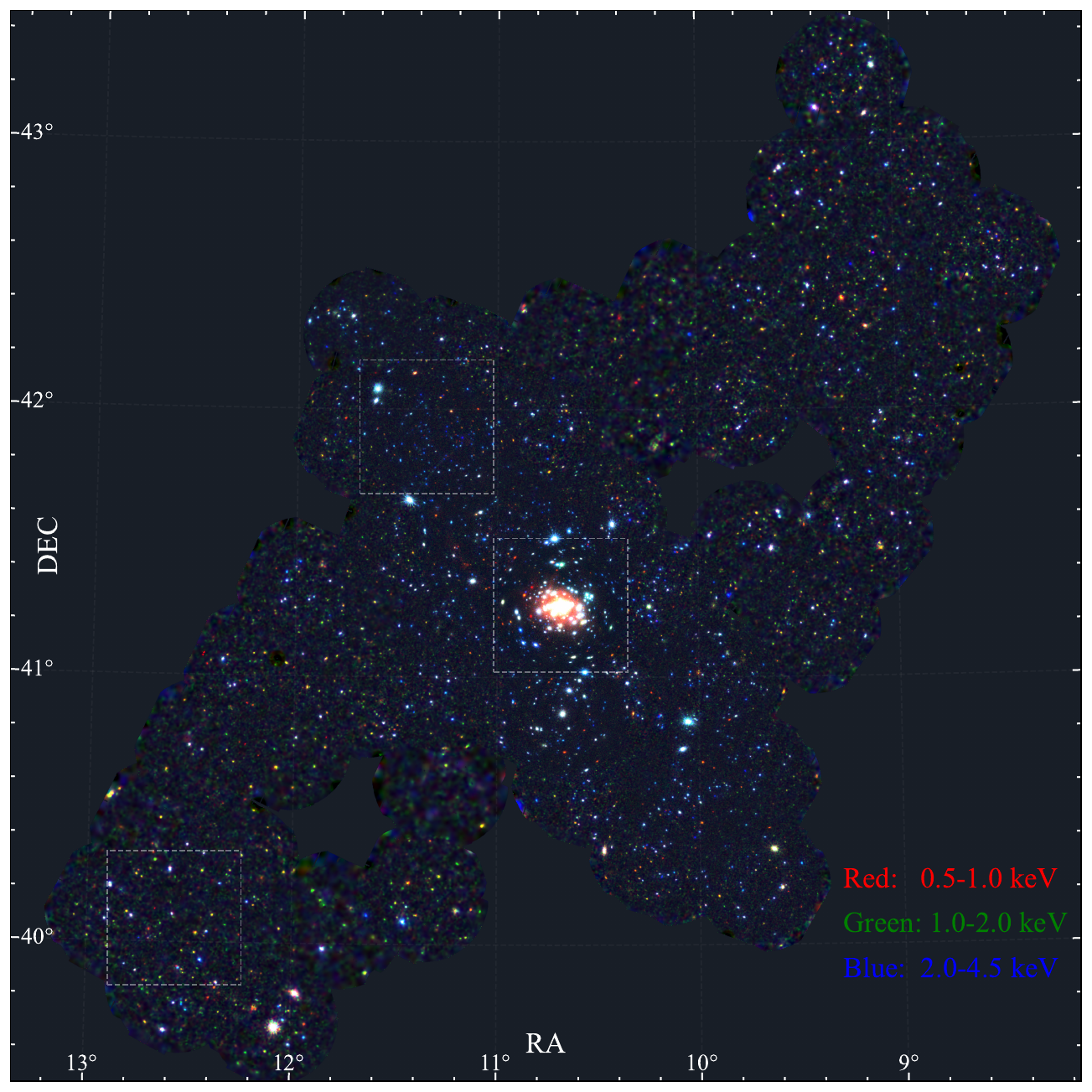}
\caption{Mosaicked false-color image of the \emph{XMM-Newton} observations of M\,31. We also show three zoom-in $30^{\prime} \times30^{\prime}$ regions in the center, disk, and halo. The count rate images are background subtracted, exposure corrected, adaptively smoothed, but not vignetting corrected. The scale bars of $10^{\prime}=2.2~\mathrm{kpc}$ in the inserted panels are calculated at the distance of M\,31 (761 kpc). Because the center region is much brighter, we adjust the color scale of the upper left panel to capture more details.}
\label{fig:image}
\end{figure*}

An XMM-{\color{red}New}ton View of the {\color{red}AN}dromeda {\color{red}G}alaxy as {\color{red}E}xplored in a {\color{red}L}egacy {\color{red}S}urvey (New-ANGELS) is a legacy X-ray survey of the M\,31 disk and bulge, as well as its inner halo around the disk. The program is based on an \emph{XMM-Newton} AO-16 large program surveying the M\,31 halo at $r\lesssim30\rm~kpc$ (PI: Li, Jiang-Tao), plus many archival data mainly covering its disk and bulge (Figs.~\ref{fig:ExpMap}, \ref{fig:image}; Table~\ref{table:XMMDatalist}). The major scientific goals include not only the population, spatial distribution, and X-ray/IR ratio of stellar sources, but also the survey of foreground/background sources, as well as the large-scale diffuse X-ray emission from the hot gas (e.g.,  \citealt{liChandraDetectionDiffuse2007}, \citealt{bogdanUnresolvedEmissionIonized2008},\citealt{bogdanUnresolvedXrayEmission2010}, \citealt{2020A&A...637A..12K}), the hot CGM potentially produced by accretion instead of galactic feedback in a representative massive quiescent spiral galaxy (e.g., \citealt{LiJ16,LiJ17,LiJ18}).

As the first paper of the New-ANGELS project, we herein present by far the most comprehensive X-ray source catalog extracted from a $\approx\mathrm{7.2~deg^2}$ field around M\,31, down to a flux limit of $F_{\rm 1.0-2.0 keV}=3.6\times10^{-15}\rm~ergs~s^{-1}~cm^{-2}$ (reached in 90\% of the sky area covered in the survey) and the luminosity limit of $L_{\rm 1.0-2.0 keV}=\mathrm{2.5 \times 10^{35}~ergs~s^{-1}}$ for M\,31 sources. Further statistical analysis and scientific discussions based on this X-ray source catalog (X-ray source LFs, spatial distribution, and the X-ray/IR ratio of different X-ray source populations), as well as analysis of the diffuse X-ray emission from the hot gas, will be presented in follow-up papers. The present paper is organized as follows: 
In \S\ref{sec:Data}, we introduce the \emph{XMM-Newton} observations used in the New-ANGELS project, as well as the data reduction and source detection procedures. We then discuss the source identification and classification based on either the multi-wavelength cross-identification or the multi-band color and/or extent in \S\ref{sec:identification}. The complete source catalog is described in detail in \S\ref{sec:Catalog}.
All errors are quoted at 1~$\sigma$ confidence level throughout the paper unless specially noted.

\section{Observations and Data Reduction}\label{sec:Data}

\begin{figure*}
\centering
\includegraphics[width=1.\textwidth]{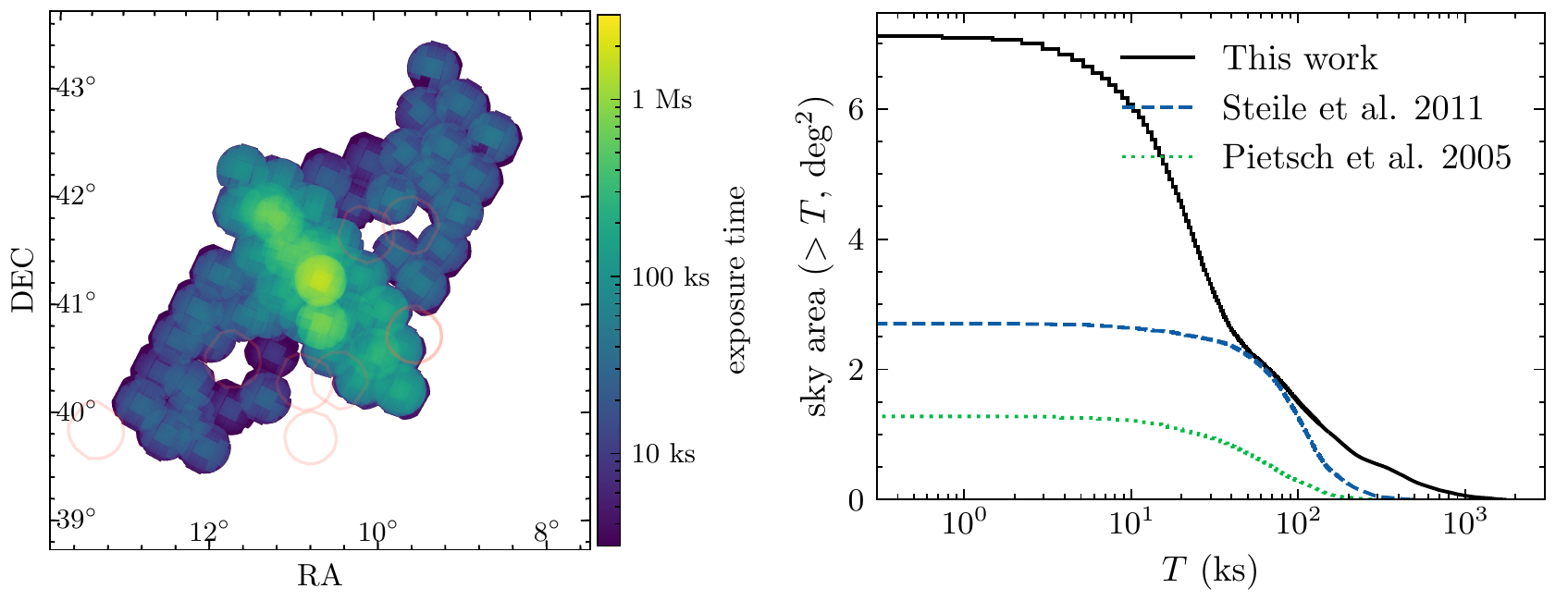}
\caption{Left: All-EPIC exposure map of the \emph{XMM-Newton} observations around M\,31. The color denotes the effective exposure time accounting for the vignetting. The pink circles represent observations that are excluded from data reduction and analysis due to background flaring.
Right: The sky coverage of the observations above the given all-EPIC exposure time. The sky coverage of this work, \citet{stieleDeepXMMNewtonSurvey2011a}, and \citet{Pietsch05} are shown in black solid line, blue dashed line, and green dotted line, respectively. 
\label{fig:ExpMap}}
\end{figure*}

In order to have a complete survey of the X-ray sources in and around M\,31, we make use of all the available \emph{XMM-Newton} observations which have a pointing position within the projected radius $r\leq30\rm~kpc$ ($\approx2.2^\circ $, adopting a distance of $d=761\rm~kpc$; \citealt{liSub2DistanceM312021}) around the center of M\,31.  
In summary, there are 137 observations matching our selection criteria, but 28 of them are significantly impacted by strong background flares. Eleven observations pointing to the center of M\,31, with less than 20 ks of exposure time, are excluded as the central region already had sufficient exposure time, resulting in a final selection of 98 observations for subsequent analysis.
Table~\ref{table:XMMDatalist} in Appendix~\ref{appendsec:XMMDatalist} provides a summary of them. The total/net (after removing strong background flares) exposure time is $\approx 3.1/2.5$, $3.1/2.6$, and $2.7/1.9\rm~Ms$ for MOS-1, MOS-2, and PN, respectively.
The mosaicked \emph{XMM-Newton} exposure map of all the observations is presented in Fig.~\ref{fig:ExpMap}. 

We reprocess the \emph{XMM-Newton} data using the \emph{XMM-Newton} Science Analysis Software (SAS) v19.1.0. SAS does not have a fixed version of calibration files. Instead, these files are updated continuously. We use the latest calibration files until May 2022. 
We first reproduce the event lists of the three EPIC cameras (MOS-1, MOS-2, and PN) using the SAS tools \texttt{emchain} and \texttt{pnchain} for MOS and PN, respectively. We then identify and tag the MOS CCDs which are in anomalous states with \texttt{emtagnoise}. The registered events from these tagged CCDs are removed in the following analysis. We identify good time intervals (GTIs) by removing strong background flares mainly caused by elevated soft proton contamination using \texttt{mos-filter} and \texttt{pn-filter}. 
We further filter the events with the \#XMMEA\_EM filter in \texttt{evselect} for MOS and ``FLAG == 0'' for PN which provide a canned screening set of FLAG values for the event. These filters exclude `wrong' or `suspicious' events, e.g., near hot pixels or outside the field of view (FOV) of the telescope. 
The GTIs of different observations (identified with the observation ID: ObsID) after these cleanings are summarized in Table~\ref{table:XMMDatalist}.

\begin{figure}
\includegraphics[width=1.0 \columnwidth]{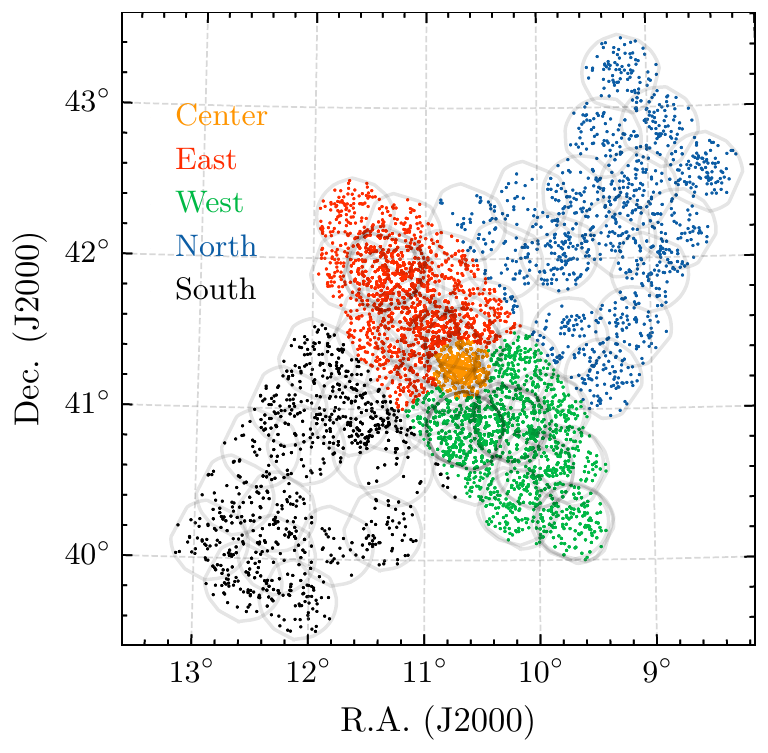}
\caption{Location of detected X-ray sources. Sources detected in different areas as defined in Table~\ref{table:region} are plotted in different colors.
}
\label{fig:source map}
\end{figure}

After reprocessing the event list, we perform standardized EPIC source detection with the task \texttt{edetect\_stack}\footnote{\url{http://xmm-tools.cosmos.esa.int/external/sas/current/doc/eboxdetect/}} (\citealt{traulsenXMMNewtonSerendipitousSurvey2019,Traulsen20}) in five bands: 0.2-0.5~keV, 0.5-1.0~keV, 1.0-2.0~keV, 2.0-4.5~keV, and 4.5-12.0~keV, which are the standard bands used in many \emph{XMM-Newton} catalogues (e.g., the 4XMM-DR12 catalog; \citealt{Webb20}). The count rates measured in these bands are also used to define the hardness ratios of the source. The hardness ratio is defined as ${HR}_\mathrm{i}=(R_{\mathrm{i+1}}-R_{\mathrm{i}})/(R_{\mathrm{i+1}}+R_{\mathrm{i}})$, where the $R_\mathrm{1}, R_\mathrm{2}, R_\mathrm{3}, R_\mathrm{4}$, and $R_\mathrm{5}$ are the count rates of all-EPIC in 0.2-0.5 keV, 0.5-1.0 keV, 1.0-2.0 keV, 2.0-4.5 keV, and 4.5-12.0 keV, respectively.  \texttt{edetect\_stack} is a tool chain to perform source detection on multiple mosaicked or stacked observations. It greatly improves the sensitivity of the source detection in the overlapping areas of different observations. 
Considering the efficiency, the source detection edetect\_stack is not applied to the entire New-ANGLES footprint. Instead, we split the \emph{XMM-Newton} survey area around M\,31 into five smaller areas (North, South, East, West, and Center, as summarized in Table~\ref{table:region} and plotted in different colors in Fig.~\ref{fig:source map}) and perform source detection separately in each of them. For each of these five areas, we only include observations whose pointing direction is within or $\leq10^\prime$ from the boundary of different areas in the source detection procedure.

\begin{table}
\begin{center}
\caption{Definition of sky areas used for source detection}
\begin{tabular}{lcrr}
\hline\noalign{\smallskip}
\hline\noalign{\smallskip}
\multicolumn{1}{c}{Region} &
\multicolumn{1}{c}{Definition}  \\
\noalign{\smallskip}\hline\noalign{\smallskip}

North &  $\rm d1\geq30^\prime$ \\
South &  $\rm d1\leq-30^\prime$ \\
East &  $\rm -30^\prime<d1<30^\prime$, $\rm d2 < 0^\prime$ and $\rm R>12^\prime $ \\
West &  $\rm -30^\prime<d1<30^\prime$, $\rm d2\geq 0^\prime$ and $\rm R>12^\prime $  \\
Center &  $\rm R\leq12^\prime$ \\

\noalign{\smallskip}
\hline
\noalign{\smallskip}
\end{tabular}
\label{table:region}
\end{center}
\tablecomments{d1, d2, and R are the angular distance to the major axis, minor axis, and center of M\,31, assuming the position angle of M\,31 is $37.3^{\circ}$. Source detected in these different areas are plotted in different colors in Fig.~\ref{fig:source map}.}
\end{table}

We use \texttt{edetect\_stack} to detect sources with a two-stage procedure. We first run \texttt{edetect\_stack} to the original cleaned data, which calls \texttt{evselect}, \texttt{eexpmap}, and \texttt{emask} to create the counts, exposure, and mask maps, respectively. 
We further modify the mask maps to limit the source detection area within the boundary as defined in Table~\ref{table:region} and Fig.~\ref{fig:source map} to speed up the source detection. 
In the second stage, we run \texttt{edetect\_stack} again to perform the source detection within the five sub-areas, separately. 
It first calls \texttt{eboxdetect} to search sources in two steps: in the first step we use a local background to detect and remove the sources, then create global background maps; while in the second step, we use these global background files to perform sliding box search again.
After this initial source searching, \texttt{edetect\_stack} then calls \texttt{emldetect} to acquire the parameters of the sources by performing joint maximum likelihood point spread function (PSF) fitting on all the involved images of the given source simultaneously (see more detail in \citealt{traulsenXMMNewtonSerendipitousSurvey2019}, \citealt{Traulsen20}). 
\texttt{edetect\_stack} then produce the final EPIC summary source list by calling \texttt{srcmatch}.

Sometimes a faint source could fall into the gaps or bad pixels of one of the three EPIC cameras. 
In this case, \texttt{edetect\_stack} still performs maximum likelihood multi-source PSF fitting (by calling \texttt{emldetect}) at the same position based on the detection on the other one or two cameras. 
It may obtain a zero flux of the source on a certain camera.
This will significantly bias the final parameters of the source in the summary list. Therefore, we mark the `zero flux' of detected sources as non-detection due to their falling into the CCD gaps or bad pixels in one or two cameras. We then recalculate the combined parameters of the sources based only on the detection on other cameras (flux, detection likelihood with an updated degree of freedom, etc.). 

\section{X-ray Source Identification and Classification}\label{sec:identification}

M\,31 is unique for systematic studies of different populations of X-ray sources either associated with the galaxy or in the foreground/background (e.g., \citealt{Fabbiano06}), including low- and high-mass X-ray binaries (LMXBs and HMXBs), ultra-luminous X-ray sources (ULXs), and other accreting or non-accreting stellar binary X-ray sources (e.g., supersoft sources, SSSs), open or globular clusters, isolated neutron stars and magnetars, supernova remnants (SNRs) and pulsar wind nebulae (PWNe), background galaxies/groups/clusters, AGN, and transient X-ray sources (e.g., SNe, GRB), etc. In this section, we will identify or classify the detected X-ray sources around M\,31 in two ways: either by identification of the sources based on cross-matching them with other firmly identified multi-wavelength counterparts (\S\ref{sec:correlation}), or by rough classification of the sources based on their multi-band (including X-ray) colors, morphology, and/or variability (\S\ref{sec:diagram}). The former way is often a firm unique identification of the nature of the source, while the latter one is just a rough classification of the source, and a single source may be classified as candidates of different types of objects.

\subsection{Multi-wavelength catalogues and cross-match criteria}\label{subsec:CatalogCriteria} 

We compare our \emph{XMM-Newton} source catalogue in and around M\,31 to other multi-wavelength catalogues to cross-identify the sources. 
The multi-wavelength catalogues used in the cross-identification include (as summarized in Table~\ref{Tab:classification}): the SIMBAD \citep{Wenger2000}, Pan-STARRS (PS1) DR1 \citep{Chambers2016}, and AllWISE \citep{2014yCat.2328....0C} databases for various types of objects; the Local Group Galaxies Survey (LGGS; \citealt{Massey2006}) for optical sources on the M\,31 disk; GAIA DR3 for stars \citep{Gaia2016,Gaia2021}; the Revised Bologna Catalog (\citealt{Galleti2004}) for globular clusters (GlCs); \citet{Jennings2014} and \citet{Lee2014} for SNRs; the Million Quasars (Milliquas) catalog v7.2 (an update of \citealt{fleschHalfMillionQuasars2015}) for AGN.

We adopt different cross-match criteria for different types of sources based on their different spatial extension and multi-wavelength properties. Details of the cross-identification procedure for different types of sources will be described in \S\ref{subsubsec:StellarSrcsForeground}, \ref{subsubsec:XraySourcesM31}, and \ref{subsubsec:BackgroundSources}, while a brief summary of the criteria is listed in Table~\ref{Tab:classification}. In most of the cases, we define the cross match criteria based on the X-ray position uncertainty $\sigma_{\mathrm{pos}}$ of the \emph{XMM-Newton} sources (joint fit in five bands), which is determined via PSF fitting.
To include the additional position uncertainty, we use the following  empirical formula: \begin{equation}
\sigma_{\mathrm{combined}}=\sqrt{0.89\times \sigma_{\mathrm{\mathrm{pos}}}^2+0.42^2} 
\label{eq:position_uncerainty}
\end{equation}
More information can be found in 
Appendix~\ref{appendsec:astrometry}.
For most of the point-like sources, we cross match the multi-wavelength counterparts with $\Delta {\mathrm{pos}}<3.44\sigma_{\mathrm{combined}}$ from the detected X-ray sources. For extended sources such as galaxies, galaxy clusters, or SNRs, however, in addition to an initial cross-match in a larger radial range, we manually check each cross-matched sources on the multi-wavelength images to avoid mis-match. 

A more complicated case is the foreground (fg) star, which we use the Gaia DR3 catalogue to identify \citep{Gaia2016, Gaia2021}. As the number density of foreground star is very high, we often identify multiple stars within $3.44~\sigma_{\mathrm{combined}}$ from the X-ray sources. We therefore adopt a different procedure to cross match the stars to our X-ray sources. We generate a foreground star catalogue from Gaia DR3 with PSS $>$ 90\% (PSS is the column in GAIA DR3 referring to the probability as a single star).
We then use the \texttt{NWAY}\footnote{https://github.com/JohannesBuchner/nway}(a Bayesian algorithm for cross-matching multiple catalogs, \citealt{salvatoFindingCounterpartsAllsky2018,nway2021}) algorithm to cross-match the foreground stars to our X-ray catalogue by considering their X-ray/optical locations, the X-ray-to-optical flux ratio, and the X-ray hardness ratio HR2. Details of the cross-matching procedure is presented in the appendix (\S\ref{appendsec:CrossMatchGAIA}). We use the \texttt{NWAY} parameter p\_any which is the probability that the X-ray source has at least a reliable counterpart, instead of just the separation between the X-ray and optical sources, to describe the likelihood of a GAIA source being the counterpart of an X-ray source. Typically, p\_any$>$0.9 corresponds to a false detection rate of $<5\%$. 

\begin{table*}
\begin{center}
\caption{Summary of the criteria and results of source identification and classification in and around M\,31}
\label{Tab:classification}
\begin{tabular} { |p{2cm}|p{3.5cm}|p{3.6cm}|p{3.6cm}|p{1.4cm}|p{1.5cm}| }
\hline \hline
Type & Cross-Matched Catalogs & Identification & Classification & Confirmed & Candidates \\
\hline \hline
fg star & GAIA DR3 &  1. NWAY p\_any$^a$$>0.9$; 2. PSS$^b$$>90\%$; 3. point source & 1. soft$^c$ in X-ray; 2. not AGN or confirmed fg star; 3. point source &  352 & 236 
 \tabularnewline
\hline
AGN  &  The Million Quasars (Milliquas) catalog, v7.2 & $\Delta \mathrm{pos}<3.44 \sigma_{\mathrm{combined}}$$^d$ from a confirmed AGN & $\Delta \mathrm{pos}<3.44\sigma_{\mathrm{combined}}$ from an AGN candidate & 62 & 505 \tabularnewline
\hline
Globular cluster (GlC) in M\,31 &  Revised Bologna Catalog, v5; SIMABD & $\Delta \mathrm{pos}<3.44\sigma_{\mathrm{combined}}$ from a confirmed GlC & $\Delta \mathrm{pos}<3.44\sigma_{\mathrm{combined}}$ from a GlC candidate & 35 & 0 \tabularnewline
\hline
Supernova remnant (SNR) in M\,31  & \cite{Jennings2014,Lee2014} & 1. no AGN or confirmed fg star at $\Delta \mathrm{pos}<10^{\prime\prime}$; 2. $\Delta \mathrm{pos}<3.44\sigma_{\mathrm{combined}}$ from a SNR candidate; 3. manual check & 1. no AGN or confirmed fg star at $\Delta \mathrm{pos}<10^{\prime\prime}$; 2. $3.44\sigma_{\mathrm{combined}}\leq \Delta \mathrm{pos} <10^{\prime\prime}$ from a SNR candidate; 3. manual check & 27 & 4 \tabularnewline
\hline
Galaxy (Gal) & SIMABD & 1. no AGN or confirmed fg star at $\Delta \mathrm{pos}<6^{\prime\prime}$; 2. $\Delta \mathrm{pos}<3.44\sigma_{\mathrm{combined}}$ from a galaxy; 3. manual check & 1. no AGN or confirmed fg star at $\Delta \mathrm{pos}<6^{\prime\prime}$; 2. $3.44\sigma_{\mathrm{combined}}\leq \Delta \mathrm{pos}<6^{\prime\prime}$ from a galaxy; 3. manual check  & 59 & 3 \tabularnewline
\hline
Galaxy cluster (GCl) &  SIMBAD & 1. extended$^e$ source in X-ray; 2. close to a confirmed galaxy cluster based on manual check & 1. extended source in X-ray; 2. close to a galaxy cluster candidate based on manual check & 1 & 5 \tabularnewline
\hline
Super soft source (SSS) & N/A & 1. HR1$<$0, HR2$<$-0.9; 2. point source &  1. only detected in  0.2-0.5 keV; 2. point source & 11 & 6 \tabularnewline %
\hline
High mass X-ray binary (HMXB) & 1. Pan-STARRS DR1; 2. LGGS  & 1. hard$^c$ in X-ray; 2. blue$^f$ in optical but not AGN or confirmed fg star; 3. coincided with OB association or strongly variable$^g$; 4. point source  & see unclassified sources & 1  &  0 \tabularnewline
\hline
Low mass X-ray binary (LMXB) & 1. Pan-STARRS DR1; 2. LGGS & 1. hard in X-ray; 2. red or no confirmed optical counterpart, but not AGN or confirmed fg star;
3. strongly variable; 4. point source & see unclassified sources &  83 & 0  \tabularnewline
\hline
total classified (with unique classification) & & &  & 631 & 759 \tabularnewline
\hline
\hline
unclassified (with no or more than one classification) & & 1. not uniquely classified; 2. low S/N$^h$ or extended & 1. not uniquely classified; 2. hard in X-ray; 3a. blue in optical: AGN or HMXB; 3b. red or no confirmed optical counterpart: AGN or LMXB; 4. Additional labels added to distinguish AGNs and XRBs based on HR2 and HR3$^i$. & 2327 & 789 \tabularnewline
\hline
\end{tabular} 
\end{center}
The table summarizes the multi-wavelength catalogues used for cross-identification of different types of X-ray sources, as well as the criteria of identifying confirmed sources, or classifying source candidates. The number of different types of firmly identified or roughly classified sources are listed in the last two columns.
\end{table*}
\begin{table*}[t!]
Table~\ref{Tab:classification} ---continued.\\
a. We use the NWAY algorithm to match the GAIA sources to our X-ray sources, considering both the X-ray to optical flux ratio and the X-ray hardness (see text for details). p\_any is the matching probability defined by NWAY. The false matching rate is $<5\%$ when p\_any$>$0.9. \\
b. PSS (between 0 and 100\%) is the parameter in GAIA defining the probability of a source being classified as a star. \\
c. We define X-ray sources with HR2-HR1+0.53$>$0 ($\leq$0) as `hard' (`soft'). HR1 (HR2) is the X-ray hardness ratio between the 0.2-0.5~keV and 0.5-1.0~keV (0.5-1.0~keV and 1.0-2.0~keV) bands, as detailed in the \S\ref{sec:Data}. \\
d. $\sigma_{\mathrm{combined}}$ is the modified position uncertainty defined in Equ.~\ref{eq:position_uncerainty}\\
e. We define sources as ``extended'', when the detected extent radius (\texttt{EXTENT}, see Appendix~\ref{appendsec:CatalogColumn}) of sources exceed $6^{\prime\prime}$. \\
f. We define Pan-STARRS counterpart with $g-r>0$ ($<0$) as `red' (`blue'). \\ 
g. We define sources with FLUXVAR$>$10 as `strongly variable'. \\
h. We only use HR1 and HR2 in high S/N detections to classify sources, when $\sigma_{\mathrm{HR1}}<0.2$ and $\sigma_{\mathrm{HR2}}<0.2$, where $\sigma_{\mathrm{HR1}}$ and $\sigma_{\mathrm{HR2}}$ are the uncertainties in the measured HR1 and HR2. \\
i. We prefer unidentified hard X-ray source with $HR3+1.1\times HR2>0$ as XRB.
\end{table*}

\subsection{Multi-Wavelength Cross Identification} \label{sec:correlation}

There are basically three types of X-ray sources detected in the field: foreground sources which are mainly stellar sources (single or binary stars) in the MW, various types of X-ray sources associated with M\,31 (SNRs, globular clusters, and many stellar sources), and distant sources in the background (AGN, galaxies, and galaxy clusters). The most reliable way to firmly identify these X-ray sources is often to search for their identified multi-wavelength counterparts, using the catalogues and cross-match algorithm described in the above section (\S\ref{subsec:CatalogCriteria}). This is often more reliable than a simple classification of the sources based on their X-ray or multi-band colors or other multi-wavelength parameters (\S\ref{sec:diagram}).

\subsubsection{Stellar X-ray sources in the foreground}\label{subsubsec:StellarSrcsForeground} 

Since M\,31 is located in the direction off the Galactic plane, most of the foreground X-ray sources should be relatively old stars, including both single and binary stars in the MW.
Although in principle we can also detect other types of sources such as SNRs or globular clusters, they could in principle be easily identified and unlikely to be the dominant population in the direction of M\,31 (in fact, we do not find any SNRs or globular clusters in the MW toward the direction of M\,31 in our catalogue). 

We will discuss the classification of binary X-ray sources in \S\ref{sec:diagram}. Here we only focus on single star X-ray sources. Single stars tend to be individually faint in X-ray. Typically, K and M stars have $L_{X}\lesssim10^{29}\rm~ergs~s^{-1}$ (e.g., \citealt{Schmitt95}). O, B stars can be as luminous as $L_{X}\sim10^{33}\rm~ergs~s^{-1}$ (e.g., \citealt{Sana06}), which is marginally detectable at the distance of M\,31 with very deep X-ray observations (the lowest detection limit toward the nuclear region of M\,31 in our NEW-ANGELS project is $L_{\rm 0.2-12~keV}\sim10^{34}\rm~ergs~s^{-1}$; see Paper~II for details). However, these young stellar populations are not expected to be located far away from the disk of the MW. Therefore, most of the X-ray emitting single stars cross-matched with the GAIA catalogue should be within the solar neighbourhood. We have firmly identified 352 X-ray sources as single stars in the GAIA catalogue, of which, 234 have a reliable distance measurement, and the median value is $d=434_{-207}^{+555}\rm~pc$, confirming their nature as foreground sources.

Cautions should be made here that the GAIA catalogue does not distinguish single star and close binary stars, but simply call them ``single stars'' due to the dominance of single stars in the catalogue \citep{Delchambre22}. We find some sources classified as ``fg stars'' in Table~\ref{Tab:classification} have hard X-ray spectra (Fig.~\ref{fig:color_color_diagram}a), which may indicate some of them are actually accreting XRBs. Nevertheless, we still call fg stars identified in GAIA as ``single stars'' throughout the paper.

\subsubsection{GlCs and SNRs associated with M\,31}\label{subsubsec:XraySourcesM31} 

Most of the discrete X-ray sources associated with M\,31 are neutron star or black hole accreting binaries. Most such binary systems are faint in optical and IR, so are identified mainly based on their X-ray properties (e.g., \citealt{stieleDeepXMMNewtonSurvey2011a}). Classification of these sources will be further discussed in \S\ref{sec:diagram}. We herein only cross-identify two types of X-ray sources associated with M\,31, GlC and SNRs, with the corresponding multi-wavelength catalogues.

X-ray emission from GlCs is often contributed by various types of low-luminosity X-ray sources, [$L_{\rm X}<10^{35}\rm~ergs~s^{-1}$, including active binaries (ABs), cataclysmic variables (CVs), and millisecond pulsars (MSPs), and quiescent LMXBs, etc. (e.g., \citealt{Heinke05,Bahramian20}] or dominated by accreting binaries. Such sources also tend to show significant X-ray variabilities (e.g., \citealt{Heinke05,Bahramian20}). 
The integrated X-ray spectra of GlCs often comprise both thermal and non-thermal components, with the former mainly from ABs and CVs while the latter from MSPs and LMXBs (e.g., \citealt{Heinke05,Zhao22}). Since GlCs are bright and often well identified in optical, we cross match our X-ray catalogue with the Revised Bologna Catalog of M\,31 clusters \citep{Galleti2014yCat}, with the criteria described in \S\ref{subsec:CatalogCriteria} and summarized in Table~\ref{Tab:classification}. We in total identified 35 X-ray bright GlCs in M\,31. 

SNRs are often characterized with a strong thermal plasma X-ray spectrum, although some young ones also have a non-thermal hard X-ray tail. There are $\sim300$ SNRs identified in the MW \citep{Green19}. A large fraction of them reside in the Galactic plane, so suffer from strong foreground extinction. Compared to the SNRs in the MW, studying SNRs in an external galaxy has the advantage of having a typically lower foreground extinction and a more accurate distance. Until now, there are at least a few hundred SNRs detected in external galaxies with multi-wavelength observations (e.g., \citealt{Maggi16,Maggi19,Bozzetto17}). M\,31 is one of the galaxies with the largest number of SNRs identified (e.g., \citealt{Sasaki2012,Lee2014}), only after the face-on and star forming galaxy M\,33 (e.g., \citealt{Long10,White19}) and M\,83 (e.g., \citealt{Blair12,Blair15}).
We merge the optical SNR catalogue of M\,31 from \citet{Jennings2014} and \citet{Lee2014}, resulting in 177 SNR candidates.
Since SNRs are extended sources and the optical/X-ray morphologies are not necessarily well correlated with each other, in additional to an automatic cross match, we also manually check our \emph{XMM-Newton} X-ray and LGGS \citep{Massey2006} H$\alpha$ image to make sure there is an X-ray source associated with the SNR. We regard X-ray sources with a separation of $\Delta \mathrm{pos}<3.44 \sigma_{\mathrm{combined}}$ from an identified SNR as a firmly identified X-ray emitting SNR, while those with $3.44 \sigma_{\mathrm{combined}}\leq \Delta \mathrm{pos} <10^{\prime\prime}$ as an SNR candidate.
We in total identify 27 X-ray emitting SNRs and 4 more candidates.

\subsubsection{Background sources}\label{subsubsec:BackgroundSources} 

Background X-ray sources are anything in the distant universe, including AGNs, normal galaxies, and groups/clusters of galaxies, etc. In principle, their number densities should have a uniform spatial distribution across a small sky area such as that covered by our New-ANGLES survey. The X-ray emission from a normal galaxy without an AGN is comprised of both a thermal hot gas and a non-thermal stellar source component (e.g., \citealt{Mineo12,LiJ13a}). It is in principle distinguishable from an AGN based on their X-ray hardness ratios, especially if the hot gas contribution is significant (see \S\ref{sec:diagram} for details). Groups of galaxies also have softer spectra compared to AGNs, and their X-ray emissions often appear as extended sources with the centroid offset from a galaxy (e.g., \citealt{Sun09}). On the other hand, a galaxy cluster often has stronger X-ray emission from the hot gas than from the AGN, but the hot gas emission is often significantly harder than that of a galaxy (due to the higher temperature, e.g., \citealt{LiJ13b}), so not significantly distinguishable in X-ray hardness from an AGN. However, the X-ray emission of a galaxy cluster is usually extended, and the nearby cluster (e.g., $z<0.5$) is always well identified in optical images. Therefore, it is often not difficult to distinguish these different types of background sources. There may be some difficulties in distinguishing AGNs and XBs just based on their X-ray properties measured with poor counting statistics, but it is quite straightforward to identify an X-ray AGN if it is already identified in other ways (e.g., with an optical spectrum; \citealt{SDSSQSOdr16}).

By cross-matching our X-ray catalog with the Milliquas catalogue \citep{fleschHalfMillionQuasars2015}, we identify 62 confirmed X-ray emitting AGNs and 504 candidates. Since many of the AGN candidates are identified based on their multi-band colors instead of the spectral properties, the identification of an X-ray counterpart makes them a higher chance to be a real AGN. Since our identification of the X-ray emitting AGN is mostly based on the existing AGN catalogues, many of the unclassified hard X-ray sources (especially in the North and South region) could in fact be unidentified AGNs in these catalogues. 

By manually checking their multi-wavelength images (e.g. mostly optical and IR from SDSS, Pan-STARRS, WISE, and HST), we identify 59 X-ray emitting galaxies without an identified AGN, and three more candidates of galaxies. Some of these X-ray sources could be associated with multiple galaxies, so may actually be galaxy groups. Furthermore, we also identify five extended X-ray sources as GCl candidates, while only one is firmly identified as a galaxy cluster at $z\sim0.29$ (2E 171, \citealt{Kotov2006}). This firmly identified GCl is projected in the direction of the disk of M\,31, and appears as an extended hard X-ray source.

\subsection{Classification of X-ray Sources Based on Their Multi-band Properties} \label{sec:diagram}

In the above section (\S\ref{sec:correlation}), we have firmly identified hundreds of single stars in the foreground, GlCs and SNRs associated with M\,31, as well as AGNs, galaxies, and GCls in the background in our \emph{XMM-Newton} X-ray source catalogue. We have also less firmly identified hundreds of candidates of all of these types of sources, except for fg stars (will be discussed below). The cross-identification is mostly based on multi-wavelength cross-match of different identified source catalogues and is in general reliable and unique. However, the total number of sources identified this way is only $\lesssim10^3$, or $\sim20\%$ or the total number of X-ray detected sources. A rough classification of a larger fraction of the X-ray sources, even if not unique and less reliable, is critical in the follow-up statistical analysis of different source populations. In this section, we discuss how to classify the remaining X-ray sources based on their X-ray and/or multi-band colors, or the X-ray variability. 

Among all 62 AGNs and 504 candidates, 160 have an X-ray flux $F_{\rm 2-4.5~keV}>10^{-14}\rm~ergs~s^{-1}$ in the entire $7.2\rm~deg^2$ area covered by our NEW-ANGELS survey, leading to a surface density of $\mathrm{18.3~deg^{-2}}$.
This is only $\sim20\%$ of the prediction based on some cosmological surveys where most of the detected X-ray sources should be AGN (e.g., \citealt{Chen18}).
However, when we zoom in to the inner halo regions (North and South), the number density of all detected sources is comparable to that of those surveys,
the huge difference between the number of classified and expected AGNs may be caused by the large number of unidentified AGNs, which may have poorer optical data and not included in the Milliquas catalogue. Furthermore, most of the stellar X-ray sources, either single or binary stars, are also not unambiguously identified, unless they are foreground sources included in the GAIA catalogue. In many cases, the classifications of these AGNs and stellar X-ray sources may not be unique.

\subsubsection{Types of stellar X-ray sources} \label{subsubsec:TypesStellarXraySources}

Both single stars and binaries can be X-ray sources. Single stars are often too faint in X-ray to be individually detected at the distance of M\,31, so have to be foreground sources (\S\ref{subsubsec:StellarSrcsForeground}). On the other hand, binary X-ray sources have a broad X-ray luminosity range (e.g., \citealt{Gilfanov04}), so could be either foreground sources or associated with M\,31.
There are basically two types of stellar binary X-ray sources: accreting binaries whose X-ray emission is mainly powered by the accretion of materials from the companion star to the compact star, and non-accreting binary sources whose X-ray emission is mainly produced via other mechanisms such as the stellar corona activity. 

The X-ray emission from a non-accreting binary could be produced by the magnetically driven coronal activities (e.g., \citealt{Gudel04}). These sources are often identified as chromospherically active binaries (ABs; e.g., \citealt{Sazonov06}), whose soft X-ray emission shows strong variability in a timescale of a few hours, and the spectrum is characterized by a thermal plasma at a typical temperature of $kT\sim0.5\rm~keV$ (sometimes with more complex temperature distributions; e.g., \citealt{Revnivtsev08, Singh22}). There are also many individually faint stars (either binary stars or single stars) with a thermal plasma X-ray spectrum but only individually detectable with deep X-ray observations within the MW (e.g., \citealt{Rosner85,Schmitt95,Sana06,Revnivtsev09}). Most of these non-accreting binary X-ray sources cannot be individually detected at the distance of M\,31, but could still collectively contribute to the unresolved X-ray emissions.

Accreting X-ray binaries include LMXBs, HMXBs, cataclysmic variables (CVs), and supersoft sources (SSSs), etc., depending on the mass of their donor stars and the nature of the accreting compact star (black hole, neutron star, and white dwarf; e.g., \citealt{Remillard06,Mukai17}). Some special types of X-ray sources, such as ULXs and probably the major X-ray emitting sources in GlCs, are also accreting X-ray binaries (e.g., \citealt{vossStudyPopulationLMXBs2007}). The X-ray emission from an accreting binary is mainly produced in its accretion disk and corona, and appears to be a power law with a photon index typically in the range of $\Gamma\sim1.5-2.0$ \citep{Irwin03,Fabbiano06}. On the other hand, because the temperature of accretion disk around a black hole is expected to fall with the increasing mass of the accreting compact object (e.g., \citealt{Miller04}), the intrinsic X-ray emission tends to be harder than an accreting SMBH (the photon index of an AGN is typically $\Gamma\sim2$; e.g., \citealt{Brightman13,LiJ21}).
However, it is often not straightforward to distinguish an individual X-ray binary either in the foreground or associated with M\,31 from a background AGN only based on their X-ray spectral slopes, due to the limited number of photons and the complex intrinsic absorption. Furthermore, there also exist some special types of accreting X-ray binaries whose X-ray emission is supersoft. These SSSs are accreting white dwarfs, but different from the CVs, their X-ray emission is mostly produced by steady nuclear fusion on the white dwarf's surface, instead of from the accretion disk (e.g., \citealt{Greiner00}). 

Accreting X-ray binaries span a large range of X-ray luminosity, with some most luminous ones could be detected individually in nearby galaxies (e.g., \citealt{Gilfanov04}). Therefore, it is often not easy to distinguish if a source is foreground or associated with M\,31 only based on its spectral shape and X-ray brightness, especially if the source is located outside the disk of M\,31. Nevertheless, based on multi-wavelength cross-identifications, there are various types of accreting X-ray binaries or candidates identified associated with M\,31 in the literature, such as LMXBs (e.g., \citealt{stieleDeepXMMNewtonSurvey2011a,sasakiDeepXMMNewtonObservations2018}), HMXBs (e.g., \citealt{greeningXraySpectralSurvey2009,williamsComparingChandraHubble2018,dalcantonPANCHROMATICHUBBLEANDROMEDA2012,williamsPANCHROMATICHUBBLEANDROMEDA2014,Lazzarini2021}), and SSSs (e.g., \citealt{Pietsch05,PietschOpticalnovae2005,stieleDeepXMMNewtonSurvey2011a,sasakiDeepXMMNewtonObservations2018}). In particular, there are also two confirmed ULXs in M\,31: CXOM31~J004253.1+411422 ($L_{\rm 0.2-10~keV}\approx 3.8\times10^{39}\rm~ergs~s^{-1}$; \citealt{Henze2009ATel,Kaur2012}) and XMMU~J004243.6+412519 ($L_{\rm 0.5-10~keV} \approx 2.3\times10^{39}\rm~ergs~s^{-1}$; \citealt{Henze2012ATel,Esposito2013}). In the next subsection (\S\ref{subsubsec:XraySourcesClassCriteria}), we will introduce a systematic way to classify these stellar X-ray sources and AGNs.

\begin{figure*}
\includegraphics[width=1 \textwidth]{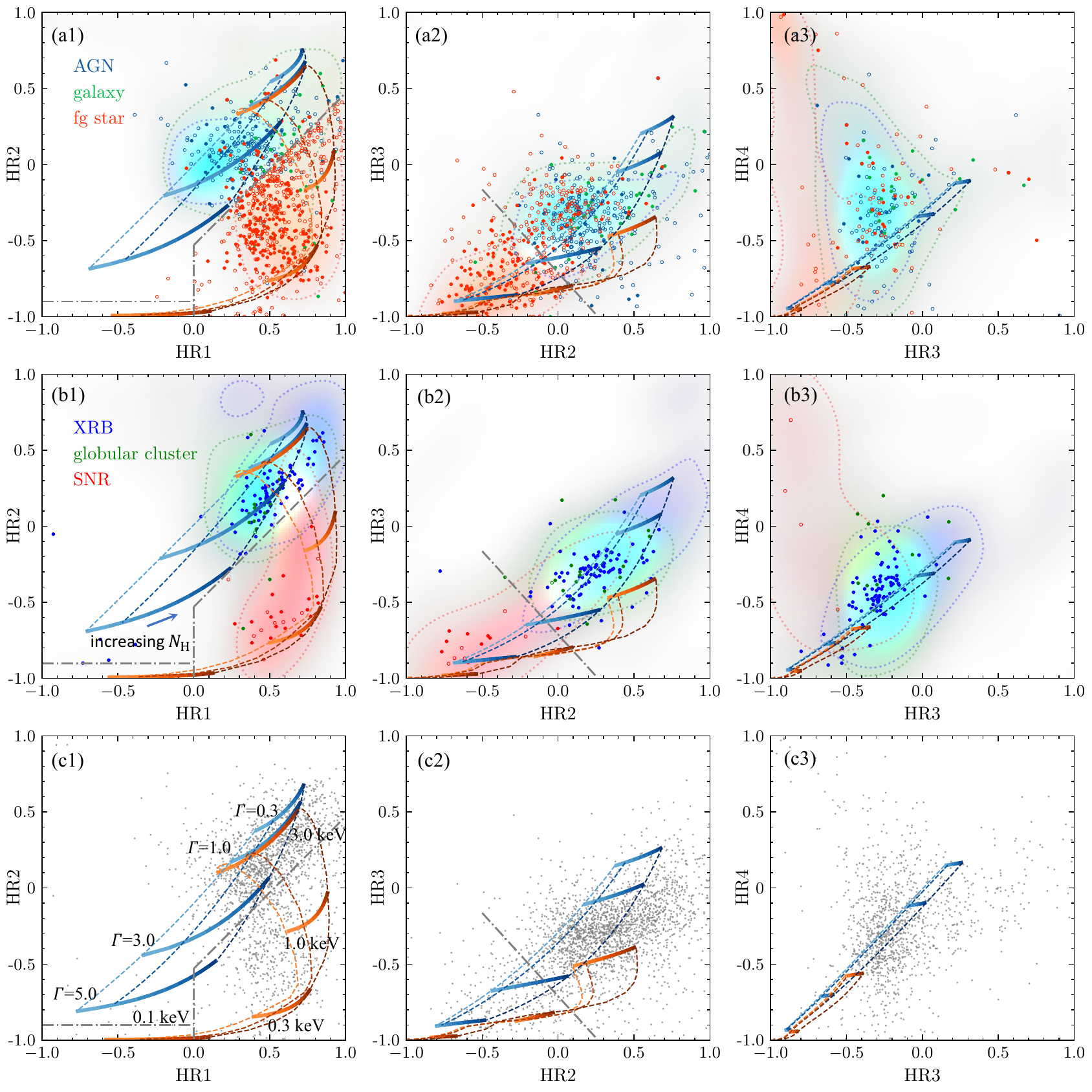}
\caption{Color-color diagram of the detected X-ray sources with HR errors $\sigma_{\mathrm{HR}}<0.2$. Identified fg stars, AGNs and galaxies are plotted in the top panels (a1, a2 and a3), with the firmly identified ones plotted in filled circle, while candidates in open circles. Identified globular clusters, XRBs and SNRs are plotted in the middle panels (b1, b2 and b3). All the sources detected in the present work are plotted in the bottom panels (c1, c2 and c3). The underlying false-color images are the probability distribution of the cross-matched sources between 4XMM-DR11s \citep{Traulsen20,Traulsen22} and SIMBAD (top row: AGN in blue, galaxy in green, fg star in red; middle row: XRB in blue, globular cluster in green, SNR in red). The dotted contours enclose 68\% of the cross-matched sources. We also overlay different power-law (in blue) and APEC (in red) models in the mesh grids. The color of the thick solid curves denote $N_\mathrm{H}$, which changes from 0 to $\rm 3\times10^{21}~\mathrm{cm}^{-2}$, with the three dashed curves mark $N_\mathrm{H}=0$, $5\times10^{20}$, and $3\times10^{21}~\mathrm{cm}^{-2}$, respectively. The upper limit of $N_\mathrm{H}$ is roughly consistent with the peak value in the M\,31 disk. When computing the models, the response files of PN, MOS-1 and MOS-2 are used for the top, middle, and bottom rows, respectively. The dash-dotted lines show our criteria to classify different types of objects (SSSs, ``hard'' and ``soft'' sources; see Table~\ref{Tab:classification} for details)}
\label{fig:color_color_diagram}
\end{figure*}

\begin{figure}
\begin{center}
\includegraphics[width=1.0 \columnwidth]{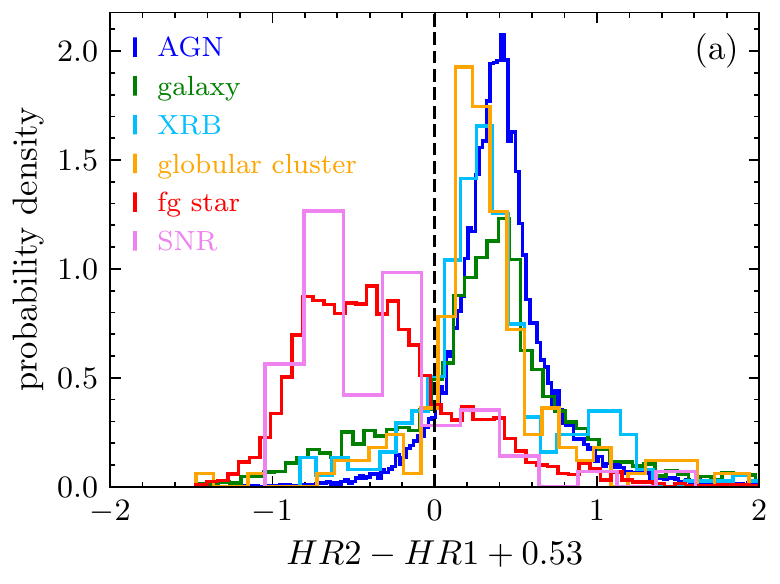}
\includegraphics[width=1.0 \columnwidth]{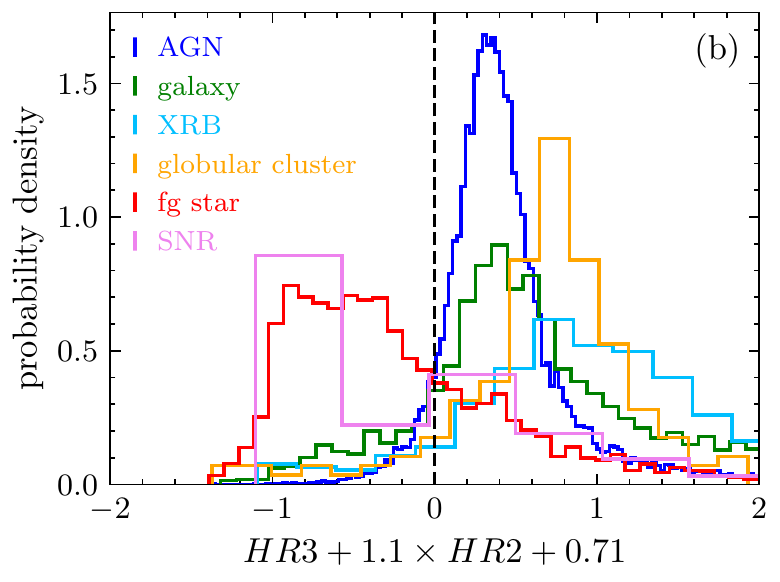}
\end{center}
\caption{The probability density distribution of different types sources in Fig~\ref{fig:color_color_diagram} at different hardness ratios. The vertical dashed lines mark our criteria to separate fg star v.s. AGN and XRB. The HR1-HR2 criterion in the upper panel is used in the source classification as summarized in Table~\ref{Tab:classification}.}\label{fig:HRcriterion}
\end{figure}

\subsubsection{X-ray source classification criteria} \label{subsubsec:XraySourcesClassCriteria}

We first classify the X-ray sources based solely on their X-ray colors, which are defined with a few inter-band hardness ratios as defined in \S\ref{sec:Data}.
Compared to spectral analysis which is often adopted for bright sources, these hardness ratios are especially helpful to roughly characterize the spectral shape of faint sources which have too few counts to extract a spectrum. A few X-ray color-color diagrams based on different hardness ratios are presented in Fig.~\ref{fig:color_color_diagram}, where we also overlay the expected locations of a few simple spectral models (APEC and power law) with different parameters (foreground absorption column density $N_{\rm H}$, thermal plasma temperature $kT$ of the APEC model, or photon index $\Gamma$ of the power law model). 
Three different column densities in the figure represent typical $N_{\rm H}$ values of no absorption with $N_{\rm H}=0~\mathrm{cm}^{-2}$, Galactic foreground absorption to the direction of M\,31 with $N_{\rm H}\approx5\times 10^{20}~\mathrm{cm}^{-2}$, and the absorption of M\,31 disk with $N_{\rm H}\approx3\times 10^{21}~\mathrm{cm}^{-2}$ which are estimated from HI4PI HI map \citep{HI4PI2016}, respectively.
Assuming either a power-law or APEC model, the absorption would cause X-ray sources to shift along the solid line shown in the color-color diagram (Fig.~\ref{fig:color_color_diagram}). Despite M\,31's proximity to the Galactic plane (with a Galactic latitude of $b=-21.4^\circ$), the amount of Galactic foreground absorption is not substantial enough to cause a significant shift in its position on the color-color diagram according to Fig.~\ref{fig:color_color_diagram}.
All the data points plotted in Fig.~\ref{fig:color_color_diagram} are the detected sources in M\,31 with the detection likelihood \texttt{EP\_DET\_ML} $\geq6$, and the uncertainty on the measured hardness ratios $\sigma_{\rm HR}<0.2$. There are in total $\sim2000$ sources plotted in Fig.~\ref{fig:color_color_diagram}. We will only use these sources with a relatively high detection S/N to discuss the source classification criteria below. The remaining ones, if not already uniquely classified as fg~stars, AGNs, GlCs, SNRs, galaxies, or GCls, will then be treated as unclassified sources (Table~\ref{Tab:classification}).

We first characterize the location of the identified X-ray sources in 4XMM-DR11s on the color-color diagram, and compare them to our firmly identified and classified sources around M\,31, in order to determine the source classification criteria.
We cross-match the 4XMM-DR11s catalogue (e.g., \citealt{Traulsen20,Traulsen22}) with the SIMBAD catalogue to identify the X-ray sources, with a strict criteria of the separation between the sources in the two catalogues $\Delta \mathrm{pos}<1^{\prime\prime}$. 
As there are too many identified sources matched to 4XMM-DR11s (18605 AGNs, 3446 galaxies, 7037 stars, 51 SNRs, 379 XRBs, 154 GlCs), we plot the probability density distribution, instead of individual data points, of different types of sources in Fig.~\ref{fig:color_color_diagram}.

It is clear that different types of sources distribute in distinct areas in the X-ray color-color diagram (Fig.~\ref{fig:color_color_diagram}), although some populations are heavily overlapped. In general, AGNs, XRBs and GlCs are hard in X-ray, while SNRs and fg stars are soft. These ``hard'' and ``soft'' sources are easy to distinguish. Galaxies have a wide distribution on the color-color diagram, and includes the area occupied by AGNs, indicating that some of the X-ray sources identified as galaxies may actually host an unidentified X-ray bright AGN. Most of the fg stars appear soft in X-ray. These sources are likely single stars or non-accreting binaries as introduced in \S\ref{subsubsec:StellarSrcsForeground} and \ref{subsubsec:TypesStellarXraySources}. However, a significant fraction of fg stars have a hard X-ray spectrum consistent with that of XRBs. These sources may thus be unidentified accreting X-ray binaries instead of single stars (\S\ref{subsubsec:StellarSrcsForeground}; \citealt{Delchambre22}).

The contamination of AGNs and XRBs in the identification of galaxies and fg stars is more clearly shown in Fig.~\ref{fig:HRcriterion}, where we plot the probability distribution of different types of sources in two X-ray color combinations obtained from the HR1-HR2 and HR2-HR3 diagrams. The dashed lines in the panels a and b of Fig.~\ref{fig:HRcriterion} ($HR2-HR1+0.53=0$, $HR3+1.1\times HR2+0.71=0$) are the same as those in Fig.~\ref{fig:color_color_diagram}a,b, separating the ``hard'' and ``soft'' sources.
Notably, the criterion used in the HR1-HR2 diagram (Fig.~\ref{fig:color_color_diagram}a) is in line with the effect of absorption. Even when there is a significant amount of absorption, modifying the value of $N_{\mathrm{H}}$ would not substantially impact the classification of X-rays as either ``hard' or ``soft''. 
It is clear that galaxies have a similar peak position in X-ray colors as AGNs, but a significantly broader distribution.
This similar peak position indicates many galaxies should indeed host an unidentified AGN. Despite XRBs would dominate the emission of galaxies, the peak of galaxies shown in Fig.~\ref{fig:color_color_diagram}b is still in line with that of AGNs.
On the other hand, fg stars, although the peak in the soft X-ray color domain, have a very asymmetric probability distribution, which is significantly biased to a harder tail. This hard tail should reflect the contribution from unidentified binary stars. Meanwhile, as indicated by Fig.~\ref{fig:HRcriterion}b, XRBs tend to be harder than AGNs.

In addition to the ``hard'' and ``soft'' sources, we also identify some supersoft sources in Fig.~\ref{fig:color_color_diagram}. These three category of sources are defined as:

$\bullet$ Hard: $HR1-HR2<0.53$ and $HR2>-0.5$ in the HR1-HR2 diagram; $HR3+1.1\times HR2>-0.71$ on the HR2-HR3 diagram.

$\bullet$ Soft: $HR1-HR2>0.53$ and $HR1>0$ in the HR1-HR2 diagram; $HR3+1.1\times HR2<-0.71$ on the HR2-HR3 diagram.

$\bullet$ Supersoft: $HR2<-0.9$ and $HR1<0$ in the HR1-HR2 diagram. If a source is only detected in the 0.2-0.5~keV band and we cannot calculate HR2, it is also regarded as a candidate of SSSs (Table~\ref{Tab:classification}). Since SSSs are often not detected in the hard X-ray band, we do not define a criterion to identify them on the HR2-HR3 diagram.

Most ($\approx70\%$) fg stars and SNRs are ``soft'' sources, while most ($\approx90\%$) AGNs, galaxies, XRBs and GlCs are ``hard''. SSSs by definition are in the supersoft domain. They are often optical novae, or accreting white dwarfs, which often have a thermal spectrum with a temperature of $kT\lesssim100\rm~eV$ (e.g., \citealt{DiStefano04,PietschOpticalnovae2005}).

\begin{figure}
\begin{center}
\includegraphics[width=1.0 \columnwidth]{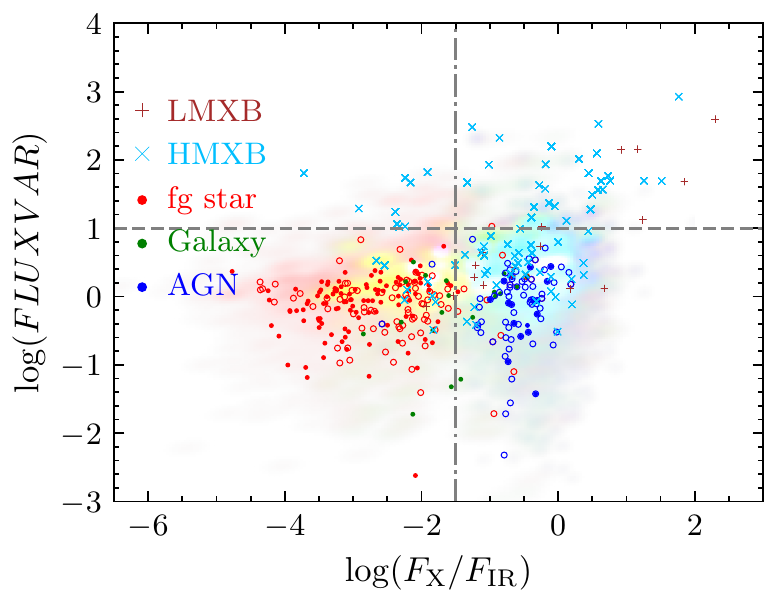}
\end{center}
\caption{The X-ray to infrared flux ratio (X-ray flux calculated in 0.2-12~keV, while IR flux is the WISE 3.4 $\mathrm{\mu m}$ flux) and X-ray variability of different X-ray source populations. Symbols are the same as in Fig.~\ref{fig:color_color_diagram}, except that the firmly identified LMXBs and HMXBs in the literatures (not limited to M\,31; e.g., \citealt{Binder2015AJ,Zhang2011}) are plotted in plus and cross symbols.
The vertical dash-dotted line roughly separates fg stars and AGN, while the horizontal dashed line defines ``highly variable'' sources, which is a key criterion to firmly identify XRBs (Table~\ref{Tab:classification}).}
\label{fig:variability_xoratio}
\end{figure}

In addition to the X-ray colors, we also adopt the X-ray variability and the X-ray to IR flux ratio as two additional criteria to classify different types of sources. Here we characterize the variability of an X-ray source with the parameter \texttt{FLUXVAR}, which is the largest flux difference in terms of the standard deviation $\sigma$ between multiple \emph{XMM-Newton} observations (Table~\ref{appendsec:CatalogColumn}). The $3.4\rm~\mu m$ IR flux of an X-ray source is taken from the AllWISE catalogue \citep{2014yCat.2328....0C}. As shown in Fig.~\ref{fig:variability_xoratio}, the most variable X-ray sources are always XRBs, so we take \texttt{FLUXVAR}$>10$ as an additional criterion for firmly identified XRBs (Table~\ref{Tab:classification}), although many XRBs are indistinguishable from fg stars and AGNs in X-ray variability. AGNs and fg stars are also significantly distinguishable in the X-ray-IR color, we then use $\log(F_{\rm X}/F_{\rm IR})=-1.5$ (the vertical line in Fig.~\ref{fig:variability_xoratio}) as an auxiliary criterion to separate fg stars and AGNs. 
Although this criterion is not directly used in the classification of the sources, because fg stars and AGNs are often well separated in X-ray colors, our results are still reaffirmed, since our identified and classified fg stars and AGNs are also well separated by the $\log(F_{\rm X}/F_{\rm IR})=-1.5$. 
So it is only used to place a caveat if the classification based on the X-ray color is thought to be less reliable, e.g., in case the source has a low S/N, or the two criteria based on the X-ray color and the X-ray-to-IR ratio are clearly conflict with each other.

\subsubsection{Procedure to classify unidentified sources} \label{subsubsec:ClassProcedure}

Based on the above discussions, our final procedure to classify the \emph{unidentified} X-ray sources (stellar sources and AGNs) in our NEW-ANGELS survey is summarized below (also see a summary in Table~\ref{Tab:classification}):

$\bullet$ We first cross-identify confirmed fg stars, confirmed or candidate AGNs, GlCs, SNRs, galaxies, GCls, based on their multi-wavelength counterparts (\S\ref{sec:correlation}). The \emph{XMM-Newton} data of some of these sources may not have high enough S/N to compute their X-ray colors with small uncertainties.

$\bullet$ For sources not identified in the above way, we calculate their X-ray colors HR1 and HR2. Sources with the uncertainties on HR1 and HR2 $\sigma_{\rm HR1}\geq0.2$ or $\sigma_{\rm HR2}\geq0.2$ typically have low S/N, so are often regarded as unclassified, although we sometimes give them a preferred classification. Sources only detected in 0.2-0.5~keV are taken as candidates of SSSs.

$\bullet$ For relatively bright unclassified sources with $\sigma_{\rm HR1}<0.2$ and $\sigma_{\rm HR2}<0.2$, we first classify them into ``hard'', ``soft'', and ``supersoft'' sources, based on their X-ray colors defined in \S\ref{subsubsec:XraySourcesClassCriteria}. ``Supersoft'' sources are directly regarded as confirmed SSSs. ``Soft'' sources are directly taken as candidates of fg stars. The remaining ``hard'' sources are regarded as unclassified.

$\bullet$ For the ``hard'' unclassified sources, we further give them some preferred but non-unique classifications based on some additional multi-wavelength criteria. If the optical counterpart source is blue, they can be either HMXBs or AGNs, with their X-ray color defined with HR2 and HR3 being an additional criterion to assign the preferred classification. If the source is red in optical, or has no confirmed optical counterpart, they can be either LMXBs or AGNs, with the same X-ray color criterion to assign the preferred classification.
We do not expect to match low-mass main sequence stars in M\,31, but OB-type or giant stars are possible. Additionally, the possibility that the detected XRBs are foreground sources is not ruled out. Therefore, if no counterpart is found, the XRB is considered a LMXB. However, if a possible counterpart is found, the XRB is then classified as a HMXB or LMXB based on the color of the counterpart. Despite the potential for a false positive match due to the limited apparent magnitude of the actual donor star, a match can still offer insights into the environment of the XRB, such as the presence of a star-forming region.

$\bullet$ The above classification of some types of sources, e.g., the X-ray soft fg star candidates, could be largely uncertain. We then use some auxiliary criteria to add some caveats to their classification, such as the X-ray-to-IR flux ratio.

\subsection{Noted sources of interest} \label{sec:NotedSrcs}  

In this section, we summarize a few noted X-ray sources which may be especially interesting in some aspects. This is only a very incomplete summary. We include some additional brief notes on many other sources in the final catalogue as will be described in \S\ref{sec:Catalog}.

\begin{figure}
\begin{center}
\includegraphics[width=1.0 \columnwidth]{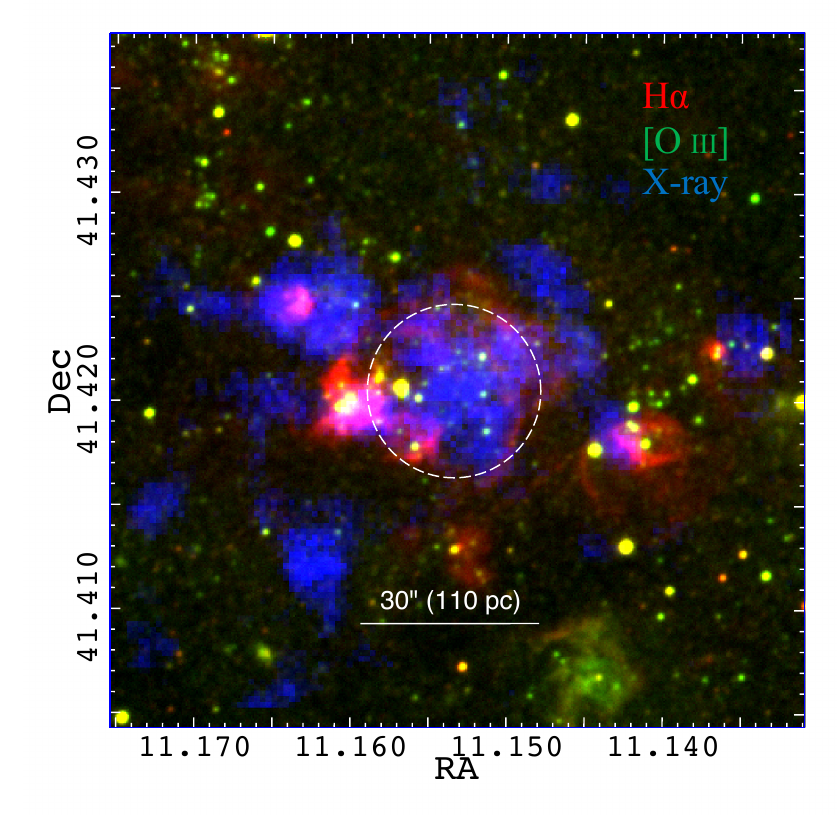}
\caption{False-color image of the newly discovered superbubble candidate J004436.8+412514. It is composed of $\mathrm{H{\alpha}}$ (red), [\ion{O}{3}] $\lambda5007\rm\AA$ (green), and the X-ray (blue, $0.2-1.0$ keV image which is smoothed with a 20 arcsec Gaussian kernel. 
A dashed circle with a diameter of $30^{\prime\prime}$ ($\sim110$ pc at a distance of 761 kpc) has been superimposed on the location of the super bubble candidate.} 
\label{fig:bubble}
\end{center}
\end{figure}

We discovered a $\sim110\rm~pc$ diameter superbubble candidate J004436.8+412514 in the eastern disk of M\,31 (Fig.~\ref{fig:bubble}).
The extended soft X-ray emission is apparently enclosed by a few shells and filaments bright in optical emission lines such as H$\alpha$ (optical images obtained from the LGGS project; Local Group Galaxies Survey, \citealt{Massey2006}).
The overlapping location of the extended X-ray source and the $\rm H\alpha$ shell structure suggests the possible existence of a hot super bubble. The \emph{XMM-Newton} spectra extracted from the entire bubble could be well fitted with an absorbed thermal plasma model (APEC, with $N_\mathrm{H}=2.4\times10^{21}~\mathrm{cm^{-2}}$, $kT=0.2~\mathrm{keV}$), the absorption corrected luminosity between 0.2-10 keV is $1.1\times10^{37}~\mathrm{erg~s^{-1}}$.

\begin{figure}
\begin{center}
\includegraphics[width=1.0 \columnwidth]{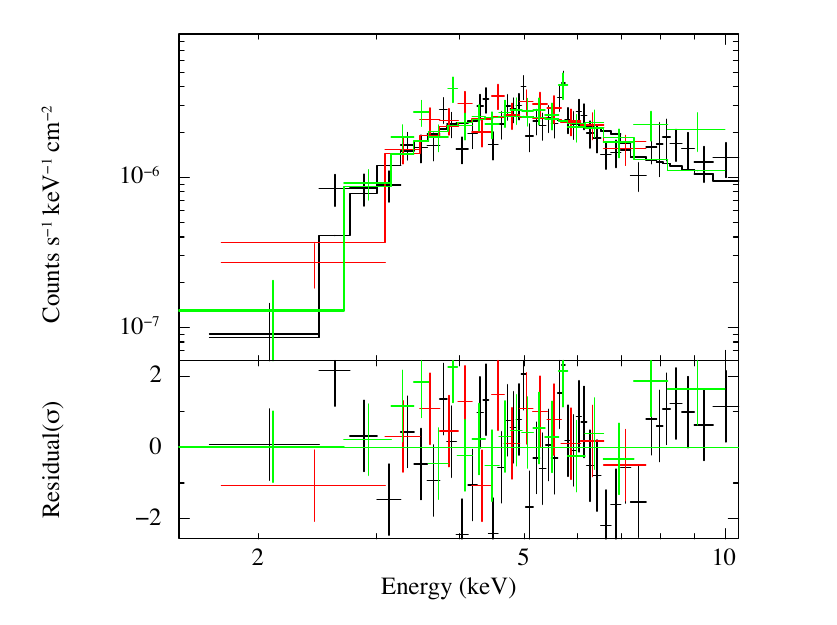}
\end{center}
\caption{The stacked pn, mos1 and mos2 spectrum of AGN candidate J004442.8+415340 are shown in black, red and green respectively. Also shown are the corresponding best-fit absorbed power-law model and the residuals. }\label{pic:spectrum_AGN}
\end{figure}

We detect a very hard point source J004442.8+415340 in the eastern disk direction of M\,31, with a $HR3=0.96\pm0.02$. The flux below 2 keV is negligible. The spectrum, as shown in Fig.~\ref{pic:spectrum_AGN}, can be well modeled using an absorbed power-law with $N_\mathrm{H}={1.7\pm0.2}\times 10^{23}~\mathrm{cm^{-2}}$ and $\Gamma=1.9\pm0.3$. The uncertainty is defined in the 90\% confidence region. The measured flux (and unabsorbed flux) between 2-8 keV is $\rm 1.2\times10^{-13}~ergs~s^{-1}~cm^{-2}$ ($\rm 3.0\times10^{-13}~ergs~s^{-1}~cm^{-2}$). Our analysis suggests that the source is likely a moderately obscured AGN.

\begin{figure*}
\includegraphics[width=1.0 \textwidth]{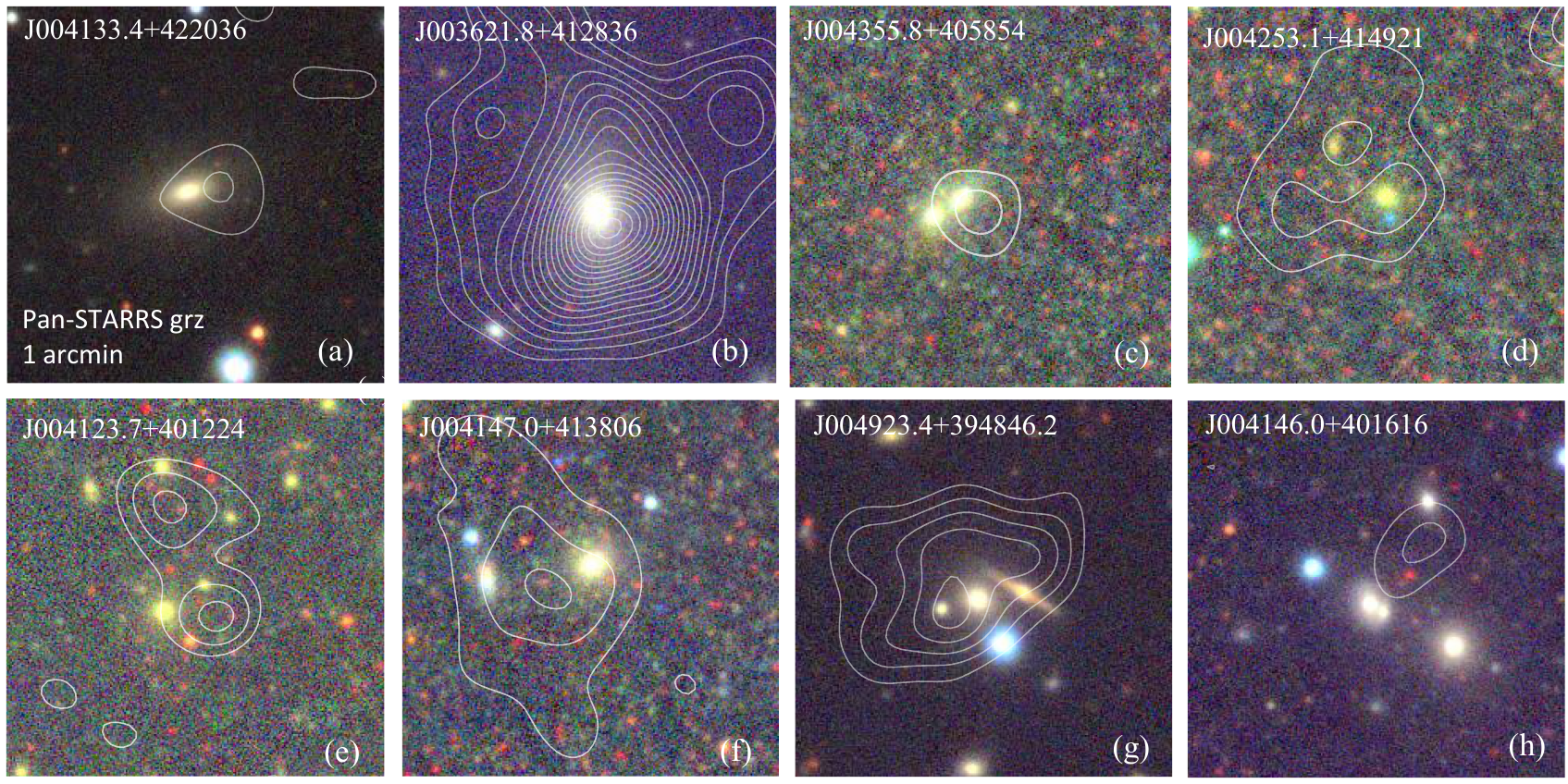}
\caption{The X-ray contour and optical image of 8 extended galaxies. The contours are the 2 sigma (background uncertainty) and above of the 0.5-4.5 keV X-ray count rate images smoothed with 5 arcsec Gaussian kernel. The $\mathrm{1\times1~arcmin^2}$ optical images are from Pan-STARRS g, r and z band. }
\label{fig:galaxy}
\end{figure*}

In this work, we detect eight extended X-ray sources associated with galaxies (see Fig.~\ref{fig:galaxy}). Two of them (a,b) have unique single galaxy counterparts. The remaining six (c-h) have multiple galaxy counterparts and have potential to be galaxy groups. Four sources' counterparts (a-d) have measured redshift and are therefore confirmed to be background sources. The extended X-ray source J004133.4+422036 (a) has an apparent soft X-ray luminosity of $2.06\times10^{41}~\mathrm{erg~s^{-1}}$ at the distance of its counterpart 2MASXJ00413355+4220359 ($z=0.093$, \citealt{Caldwell2009}). J003621.8+412836, associated with 2MASXJ00362166+4128333 at $z=0.064$ \citep{Bilicki2014}, has an apparent luminosity of ${3.7\times10^{40} ~\rm erg~s^{-1}}$. For J004355.8+405854, the apparent luminosity of the detected extended source is $\mathrm{3.49\times10^{40}~erg~s^{-1}}$ at a distance of $z=0.121$ \citep{Caldwell2009}, based on the redshift record of its counterpart NAME~[G2012b]C039. J004253.1+414921 has an apparent luminosity of $4.9\times10^{40}~\rm erg~s^{-1}$ at a distance of the counterpart 2MASXJ00414697+4138094 with $z=0.106$ \citep{Galleti2007}. This source also shows potential AGN activity based on its radio counterpart WSTB~37W145. In spite of the measured redshift, the limited luminosity of these sources excludes their possibilities as galaxy clusters. 
For the rest four extended X-ray source (e-h), the lack of measured redshift of the counterparts prevents estimation of their luminosity. These extended X-ray sources are still classified as galaxy in the catalogue, while we will note that they are extended and have one or more galaxy counterparts.

\section{X-ray Source Catalog}\label{sec:Catalog}

In this paper, we present the most complete catalogue of 4506 X-ray sources detected from an area of $\sim7.2\rm~deg^2$ around M\,31 (Fig.~\ref{fig:source map}). All reported sources have a limiting detection likelihood \texttt{EP\_DET\_ML}$>$6, 
which leads to a false detection rate of $<0.5\%$ (e.g., \citealt{NiQ2021}).

\cite{stieleDeepXMMNewtonSurvey2011a} detected 1948 X-ray sources in a comprehensive analysis of the archival XMM-Newton data covering the entire disk of M\,31. By incorporating 31 additional observations pointing slightly offset the disk of M\,31, we detect 2776 sources in the same region, including 1014 newly discovered X-ray sources. About 200 sources in \citet{stieleDeepXMMNewtonSurvey2011a}'s catalogue are rejected due to their low detection likelihood (e.g., $<30$). The archival Chandra data of M\,31 has a better angular resolution, but covers a small portion of the disk. We also include the information of the cross-matched Chandra sources (from the Chandra Source Catalog Release 2.0; \citealt{Evans2018}) in our catalogue for comparison.

For each source, the reported parameters can be classified into three categories. We first summarize the basic detection information, including the position, the detection likelihood, and the involved observations where the source is detected.
The second is the physical parameters of the source, count rate, flux, hardness ratio, extent etc. 
These parameters are not only reported for combined result from all involved exposures, but also for individual observation, instruments (all-EPIC, pn, MOS1, and MOS2), and energy bands (0.2-12.0 keV, 0.2-0.5 keV, 0.5-1.0 keV, 1.0-2.0 keV, 2.0-4.5 keV, and 4.5-12.0 keV).
The last is the classification, we list our preferred identification or classification of the source based on either the multi-wavelength cross-identification (\S\ref{sec:correlation}) or the brief multi-color classification (\S\ref{sec:diagram}). 
We describe the catalogue of the source parameters in Appendix~\ref{appendsec:CatalogColumn}. A machine-readable format of the table is provided online only.

Results of the classification in our X-ray source catalogue are also summarized in Table~\ref{Tab:classification}.
Based on the multi-wavelength cross-identifications, in our X-ray source catalogue, we identify 352 single stars in the foreground (fg stars; all regarded as firmly identified), 35 firmly identified GlCs and 31 SNRs (27 firmly identified, 4 as candidates) associated with M\,31, as well as 567 AGNs (62 firmly identified, 505 as candidates), 62 galaxies (59 firmly identified, 3 as candidates), and 6 GCls (1 firmly identified, 5 as candidates) in the background. 
We uniquely classify 236 fg star and 17 SSS candidates based on their X-ray colors. Since the LMXBs, HMXBs, and AGNs are not explicitly separated just based on their X-ray and optical colors, we often give them a preferred but not confirmed classification based on multiple criteria (based on the X-ray colors, X-ray-to-optical colors, and X-ray variability). 
The exceptions are some highly variable X-ray sources which are regarded as firmly identified XRBs. 83 of them are identified as LMXBs, and 1 of them is taken as HMXBs. 
The remaining 3116 X-ray sources either have too low S/N to calculate their X-ray colors or have not unique classification, so are regarded as unclassified.

We do not perform standard astrometry correction to individual observations before conducting source detection. This is partially because the astrometry correction, typically calculated based on cross-matching the X-ray and optical sources, could be largely uncertain, especially in the halo of M\,31 where only a few sources can be identified in optical. Instead, we directly apply the astrometry correction result for each observations after detecting the sources, using the suggested offset values from the 4XMM-DR12s catalogue \citep{Traulsen20}. For sources detected in overlapping observations, we weight the offset of different observations by the number of counts of the detected sources in the corresponding observations. We also list the updated source position (\texttt{ast\_RA} and \texttt{ast\_DEC}) considering the astrometry correction in the catalogue. Besides the original identification and classification result (\texttt{type}), the sources with an updated astrometry correction has a slightly different multi-wavelength cross-matching result. We also list this updated identification and classification result (\texttt{ast\_type}) in the catalogue for reference, but caution that this is based on a simple geometric correction, without updating the source detection parameters.

We publish the full source catalogue with the X-ray/multi-wavelength properties of the sources and their classifications as an online only table (\S\ref{appendsec:CatalogColumn},\S\ref{appendsec:catalog}, and Table~\ref{catatlog_tab:1}). This source catalogue is unique for the study of the X-ray source populations, as well as quantitatively estimating the contribution of X-ray sources in the unresolved X-ray emissions of an external galaxy. These topics will be discussed in the follow-up papers. We also summarize our X-ray source classification criteria in Table~\ref{Tab:classification}, which are potentially applicable to other galaxies.

\section*{Acknowledgements}
The authors would like to thank Prof. Junjie Mao (Tsinghua University) and Prof. B. Luo (Nanjing University) for helpful discussions.
This work was supported in part by the Ministry of Science and Technology of China through Grant 2018YFA0404502 and the National Natural Science Foundation of China through Grants E3GJ251110 and 11821303. RH acknowledges support from the China Scholarship Council. 
GP acknowledges funding from the European Research Council (ERC) under the European Union’s Horizon 2020 research and innovation programme (grant agreement No 865637) and from "Bando per il Finanziamento della Ricerca Fondamentale 2022 dell’Istituto Nazionale di Astrofisica (INAF): Canale: GO Large program"

This publication makes use of data products from the Wide-field Infrared Survey Explorer, which is a joint project of the University of California, Los Angeles, and the Jet Propulsion Laboratory/California Institute of Technology, funded by the National Aeronautics and Space Administration.
This work has made use of data from the European Space Agency (ESA) mission
{\it Gaia} (\url{www.cosmos.esa.int/gaia}), processed by the {\it Gaia}
Data Processing and Analysis Consortium (DPAC,
\url{www.cosmos.esa.int/web/gaia/dpac/consortium}). Funding for the DPAC
has been provided by national institutions, in particular the institutions
participating in the {\it Gaia} Multilateral Agreement.
This research has made use of the VizieR catalog access tool, CDS,  Strasbourg, France (DOI : 10.26093/cds/vizier). The original description of the VizieR service was published in 2000, A\&AS 143, 23

Data Availability:
The X-ray source catalogue is published online in machine-readable format. The original \emph{XMM-Newton} data used in the presented paper is available on the \emph{XMM-Newton} data archive. The reduced data could be shared by reasonable request to the corresponding author.

\bibliographystyle{aasjournal}
\bibliography{M31}

\begin{thebibliography}{}
\expandafter\ifx\csname natexlab\endcsname\relax\def\natexlab#1{#1}\fi
\providecommand{\url}[1]{\href{#1}{#1}}
\providecommand{\dodoi}[1]{doi:~\href{http://doi.org/#1}{\nolinkurl{#1}}}
\providecommand{\doeprint}[1]{\href{http://ascl.net/#1}{\nolinkurl{http://ascl.net/#1}}}
\providecommand{\doarXiv}[1]{\href{https://arxiv.org/abs/#1}{\nolinkurl{https://arxiv.org/abs/#1}}}

\bibitem[{{Bahramian} {et~al.}(2020){Bahramian}, {Strader}, {Miller-Jones},
  {Chomiuk}, {Heinke}, {Maccarone}, {Pooley}, {Shishkovsky}, {Tudor}, {Zhao},
  {Li}, {Sivakoff}, {Tremou}, \& {Buchner}}]{Bahramian20}
{Bahramian}, A., {Strader}, J., {Miller-Jones}, J. C.~A., {et~al.} 2020, \apj,
  901, 57, \dodoi{10.3847/1538-4357/aba51d}

\bibitem[{{Barmby} {et~al.}(2006){Barmby}, {Ashby}, {Bianchi}, {Engelbracht},
  {Gehrz}, {Gordon}, {Hinz}, {Huchra}, {Humphreys}, {Pahre},
  {P{\'e}rez-Gonz{\'a}lez}, {Polomski}, {Rieke}, {Thilker}, {Willner}, \&
  {Woodward}}]{Barmby06}
{Barmby}, P., {Ashby}, M.~L.~N., {Bianchi}, L., {et~al.} 2006, \apjl, 650, L45,
  \dodoi{10.1086/508626}

\bibitem[{{Bilicki} {et~al.}(2014){Bilicki}, {Jarrett}, {Peacock}, {Cluver}, \&
  {Steward}}]{Bilicki2014}
{Bilicki}, M., {Jarrett}, T.~H., {Peacock}, J.~A., {Cluver}, M.~E., \&
  {Steward}, L. 2014, \apjs, 210, 9, \dodoi{10.1088/0067-0049/210/1/9}

\bibitem[{{Binder} {et~al.}(2015){Binder}, {Williams}, {Eracleous},
  {Plucinsky}, {Gaetz}, {Anderson}, {Skillman}, {Dalcanton}, {Kong}, \&
  {Weisz}}]{Binder2015AJ}
{Binder}, B., {Williams}, B.~F., {Eracleous}, M., {et~al.} 2015, \aj, 150, 94,
  \dodoi{10.1088/0004-6256/150/3/94}

\bibitem[{{Blair} {et~al.}(2012){Blair}, {Winkler}, \& {Long}}]{Blair12}
{Blair}, W.~P., {Winkler}, P.~F., \& {Long}, K.~S. 2012, \apjs, 203, 8,
  \dodoi{10.1088/0067-0049/203/1/8}

\bibitem[{{Blair} {et~al.}(2015){Blair}, {Winkler}, {Long}, {Whitmore}, {Kim},
  {Soria}, {Kuntz}, {Plucinsky}, {Dopita}, \& {Stockdale}}]{Blair15}
{Blair}, W.~P., {Winkler}, P.~F., {Long}, K.~S., {et~al.} 2015, \apj, 800, 118,
  \dodoi{10.1088/0004-637X/800/2/118}

\bibitem[{{Bogd{\'a}n} \&
  {Gilfanov}(2008)}]{bogdanUnresolvedEmissionIonized2008}
{Bogd{\'a}n}, {\'A}., \& {Gilfanov}, M. 2008, \mnras, 388, 56,
  \dodoi{10.1111/j.1365-2966.2008.13391.x}

\bibitem[{{Bogd{\'a}n} \& {Gilfanov}(2010)}]{bogdanUnresolvedXrayEmission2010}
---. 2010, \mnras, 405, 209, \dodoi{10.1111/j.1365-2966.2010.16476.x}

\bibitem[{{Bozzetto} {et~al.}(2017){Bozzetto}, {Filipovi{\'c}}, {Vukoti{\'c}},
  {Pavlovi{\'c}}, {Uro{\v{s}}evi{\'c}}, {Kavanagh}, {Arbutina}, {Maggi},
  {Sasaki}, {Haberl}, {Crawford}, {Roper}, {Grieve}, \& {Points}}]{Bozzetto17}
{Bozzetto}, L.~M., {Filipovi{\'c}}, M.~D., {Vukoti{\'c}}, B., {et~al.} 2017,
  \apjs, 230, 2, \dodoi{10.3847/1538-4365/aa653c}

\bibitem[{{Brightman} {et~al.}(2013){Brightman}, {Silverman}, {Mainieri},
  {Ueda}, {Schramm}, {Matsuoka}, {Nagao}, {Steinhardt}, {Kartaltepe},
  {Sanders}, {Treister}, {Shemmer}, {Brandt}, {Brusa}, {Comastri}, {Ho},
  {Lanzuisi}, {Lusso}, {Nandra}, {Salvato}, {Zamorani}, {Akiyama}, {Alexander},
  {Bongiorno}, {Capak}, {Civano}, {Del Moro}, {Doi}, {Elvis}, {Hasinger},
  {Laird}, {Masters}, {Mignoli}, {Ohta}, {Schawinski}, \&
  {Taniguchi}}]{Brightman13}
{Brightman}, M., {Silverman}, J.~D., {Mainieri}, V., {et~al.} 2013, \mnras,
  433, 2485, \dodoi{10.1093/mnras/stt920}

\bibitem[{{Buchner} {et~al.}(2021){Buchner}, {Salvato}, {Budava{\`I}ri}, \&
  {Fotopoulou}}]{nway2021}
{Buchner}, J., {Salvato}, M., {Budava{\`I}ri}, T., \& {Fotopoulou}, S. 2021,
  {nway: Bayesian cross-matching of astronomical catalogs}, Astrophysics Source
  Code Library, record ascl:2102.014.
\newblock \doeprint{2102.014}

\bibitem[{{Caldwell} {et~al.}(2009){Caldwell}, {Harding}, {Morrison}, {Rose},
  {Schiavon}, \& {Kriessler}}]{Caldwell2009}
{Caldwell}, N., {Harding}, P., {Morrison}, H., {et~al.} 2009, \aj, 137, 94,
  \dodoi{10.1088/0004-6256/137/1/94}

\bibitem[{{Cash}(1979)}]{1979ApJ...228..939C}
{Cash}, W. 1979, \apj, 228, 939, \dodoi{10.1086/156922}

\bibitem[{{Chambers} {et~al.}(2016){Chambers}, {Magnier}, {Metcalfe},
  {Flewelling}, {Huber}, {Waters}, {Denneau}, {Draper}, {Farrow}, {Finkbeiner},
  {Holmberg}, {Koppenhoefer}, {Price}, {Rest}, {Saglia}, {Schlafly}, {Smartt},
  {Sweeney}, {Wainscoat}, {Burgett}, {Chastel}, {Grav}, {Heasley}, {Hodapp},
  {Jedicke}, {Kaiser}, {Kudritzki}, {Luppino}, {Lupton}, {Monet}, {Morgan},
  {Onaka}, {Shiao}, {Stubbs}, {Tonry}, {White}, {Ba{\~n}ados}, {Bell},
  {Bender}, {Bernard}, {Boegner}, {Boffi}, {Botticella}, {Calamida},
  {Casertano}, {Chen}, {Chen}, {Cole}, {Deacon}, {Frenk}, {Fitzsimmons},
  {Gezari}, {Gibbs}, {Goessl}, {Goggia}, {Gourgue}, {Goldman}, {Grant},
  {Grebel}, {Hambly}, {Hasinger}, {Heavens}, {Heckman}, {Henderson}, {Henning},
  {Holman}, {Hopp}, {Ip}, {Isani}, {Jackson}, {Keyes}, {Koekemoer}, {Kotak},
  {Le}, {Liska}, {Long}, {Lucey}, {Liu}, {Martin}, {Masci}, {McLean}, {Mindel},
  {Misra}, {Morganson}, {Murphy}, {Obaika}, {Narayan}, {Nieto-Santisteban},
  {Norberg}, {Peacock}, {Pier}, {Postman}, {Primak}, {Rae}, {Rai}, {Riess},
  {Riffeser}, {Rix}, {R{\"o}ser}, {Russel}, {Rutz}, {Schilbach}, {Schultz},
  {Scolnic}, {Strolger}, {Szalay}, {Seitz}, {Small}, {Smith}, {Soderblom},
  {Taylor}, {Thomson}, {Taylor}, {Thakar}, {Thiel}, {Thilker}, {Unger},
  {Urata}, {Valenti}, {Wagner}, {Walder}, {Walter}, {Watters}, {Werner},
  {Wood-Vasey}, \& {Wyse}}]{Chambers2016}
{Chambers}, K.~C., {Magnier}, E.~A., {Metcalfe}, N., {et~al.} 2016, arXiv
  e-prints, arXiv:1612.05560, \dodoi{10.48550/arXiv.1612.05560}

\bibitem[{{Chen} {et~al.}(2018){Chen}, {Brandt}, {Luo}, {Ranalli}, {Yang},
  {Alexander}, {Bauer}, {Kelson}, {Lacy}, {Nyland}, {Tozzi}, {Vito},
  {Cirasuolo}, {Gilli}, {Jarvis}, {Lehmer}, {Paolillo}, {Schneider}, {Shemmer},
  {Smail}, {Sun}, {Tanaka}, {Vaccari}, {Vignali}, {Xue}, {Banerji}, {Chow},
  {H{\"a}u{\ss}ler}, {Norris}, {Silverman}, \& {Trump}}]{Chen18}
{Chen}, C. T.~J., {Brandt}, W.~N., {Luo}, B., {et~al.} 2018, \mnras, 478, 2132,
  \dodoi{10.1093/mnras/sty1036}

\bibitem[{{Cutri} {et~al.}(2021){Cutri}, {Wright}, {Conrow}, {Fowler},
  {Eisenhardt}, {Grillmair}, {Kirkpatrick}, {Masci}, {McCallon}, {Wheelock},
  {Fajardo-Acosta}, {Yan}, {Benford}, {Harbut}, {Jarrett}, {Lake}, {Leisawitz},
  {Ressler}, {Stanford}, {Tsai}, {Liu}, {Helou}, {Mainzer}, {Gettngs},
  {Gonzalez}, {Hoffman}, {Marsh}, {Padgett}, {Skrutskie}, {Beck}, {Papin}, \&
  {Wittman}}]{2014yCat.2328....0C}
{Cutri}, R.~M., {Wright}, E.~L., {Conrow}, T., {et~al.} 2021, VizieR Online
  Data Catalog, II/328

\bibitem[{{Dalcanton} {et~al.}(2012){Dalcanton}, {Williams}, {Lang}, {Lauer},
  {Kalirai}, {Seth}, {Dolphin}, {Rosenfield}, {Weisz}, {Bell}, {Bianchi},
  {Boyer}, {Caldwell}, {Dong}, {Dorman}, {Gilbert}, {Girardi}, {Gogarten},
  {Gordon}, {Guhathakurta}, {Hodge}, {Holtzman}, {Johnson}, {Larsen}, {Lewis},
  {Melbourne}, {Olsen}, {Rix}, {Rosema}, {Saha}, {Sarajedini}, {Skillman}, \&
  {Stanek}}]{dalcantonPANCHROMATICHUBBLEANDROMEDA2012}
{Dalcanton}, J.~J., {Williams}, B.~F., {Lang}, D., {et~al.} 2012, \apjs, 200,
  18, \dodoi{10.1088/0067-0049/200/2/18}

\bibitem[{{Delchambre} {et~al.}(2022){Delchambre}, {Bailer-Jones},
  {Bellas-Velidis}, {Drimmel}, {Garabato}, {Carballo}, {Hatzidimitriou},
  {Marshall}, {Andrae}, {Dafonte}, {Livanou}, {Fouesneau}, {Licata},
  {Lindstrom}, {Manteiga}, {Robin}, {Silvelo}, {Abreu Aramburu}, {Alvarez},
  {Bakker}, {Bijaoui}, {Brouillet}, {Brugaletta}, {Burlacu}, {Casamiquela},
  {Chaoul}, {Chiavassa}, {Contursi}, {Cooper}, {Creevey}, {Dapergolas}, {de
  Laverny}, {Demouchy}, {Dharmawardena}, {Edvardsson}, {Fremat},
  {Garcia-Lario}, {Garcia-Torres}, {Gavel}, {Gomez}, {Gonzalez-Santamaria},
  {Heiter}, {Jean-Antoine Piccolo}, {Kontizas}, {Kordopatis}, {Korn},
  {Lanzafame}, {Lebreton}, {Lobel}, {Lorca}, {Magdaleno Romeo}, {Marocco},
  {Mary}, {Nicolas}, {Ordenovic}, {Pailler}, {Palicio}, {Pallas-Quintela},
  {Panem}, {Pichon}, {Poggio}, {Recio-Blanco}, {Riclet}, {Rybizki},
  {Santovena}, {Sarro}, {Schultheis}, {Segol}, {Slezak}, {Smart}, {Sordo},
  {Soubiran}, {Suveges}, {Thevenin}, {Torralba Elipe}, {Ulla}, {Utrilla},
  {Vallenari}, {van Dillen}, {Zhao}, \& {Zorec}}]{Delchambre22}
{Delchambre}, L., {Bailer-Jones}, C.~A.~L., {Bellas-Velidis}, I., {et~al.}
  2022, arXiv e-prints, arXiv:2206.06710, \dodoi{10.48550/arXiv.2206.06710}

\bibitem[{{Di Stefano} {et~al.}(2004){Di Stefano}, {Kong}, {Greiner},
  {Primini}, {Garcia}, {Barmby}, {Massey}, {Hodge}, {Williams}, {Murray},
  {Curry}, \& {Russo}}]{DiStefano04}
{Di Stefano}, R., {Kong}, A.~K.~H., {Greiner}, J., {et~al.} 2004, \apj, 610,
  247, \dodoi{10.1086/421696}

\bibitem[{{Esposito} {et~al.}(2013){Esposito}, {Motta}, {Pintore}, {Zampieri},
  \& {Tomasella}}]{Esposito2013}
{Esposito}, P., {Motta}, S.~E., {Pintore}, F., {Zampieri}, L., \& {Tomasella},
  L. 2013, \mnras, 428, 2480, \dodoi{10.1093/mnras/sts248}

\bibitem[{{Evans} \& {Civano}(2018)}]{Evans2018}
{Evans}, I.~N., \& {Civano}, F. 2018, Astronomy and Geophysics, 59, 2.17,
  \dodoi{10.1093/astrogeo/aty079}

\bibitem[{{Fabbiano}(2006)}]{Fabbiano06}
{Fabbiano}, G. 2006, \araa, 44, 323,
  \dodoi{10.1146/annurev.astro.44.051905.092519}

\bibitem[{{Flesch}(2015)}]{fleschHalfMillionQuasars2015}
{Flesch}, E.~W. 2015, \pasa, 32, e010, \dodoi{10.1017/pasa.2015.10}

\bibitem[{{Gaia Collaboration}(2022)}]{Gaia2022yCat}
{Gaia Collaboration}. 2022, VizieR Online Data Catalog, I/355

\bibitem[{{Gaia Collaboration} {et~al.}(2016){Gaia Collaboration}, {Prusti},
  {de Bruijne}, {Brown}, {Vallenari}, {Babusiaux}, {Bailer-Jones}, {Bastian},
  {Biermann}, {Evans}, \& et~al.}]{Gaia2016}
{Gaia Collaboration}, {Prusti}, T., {de Bruijne}, J.~H.~J., {et~al.} 2016,
  \aap, 595, A1, \dodoi{10.1051/0004-6361/201629272}

\bibitem[{{Gaia Collaboration} {et~al.}(2021){Gaia Collaboration}, {Brown},
  {Vallenari}, {Prusti}, {de Bruijne}, {Babusiaux}, {Biermann}, {Creevey},
  {Evans}, {Eyer}, \& et~al.}]{Gaia2021}
{Gaia Collaboration}, {Brown}, A.~G.~A., {Vallenari}, A., {et~al.} 2021, \aap,
  649, A1, \dodoi{10.1051/0004-6361/202039657}

\bibitem[{{Galleti} {et~al.}(2007){Galleti}, {Bellazzini}, {Federici},
  {Buzzoni}, \& {Fusi Pecci}}]{Galleti2007}
{Galleti}, S., {Bellazzini}, M., {Federici}, L., {Buzzoni}, A., \& {Fusi
  Pecci}, F. 2007, \aap, 471, 127, \dodoi{10.1051/0004-6361:20077788}

\bibitem[{{Galleti} {et~al.}(2004){Galleti}, {Federici}, {Bellazzini}, {Fusi
  Pecci}, \& {Macrina}}]{Galleti2004}
{Galleti}, S., {Federici}, L., {Bellazzini}, M., {Fusi Pecci}, F., \&
  {Macrina}, S. 2004, \aap, 416, 917, \dodoi{10.1051/0004-6361:20035632}

\bibitem[{{Galleti} {et~al.}(2014){Galleti}, {Federici}, {Bellazzini}, {Fusi
  Pecci}, {Macrina}, \& {Buzzoni}}]{Galleti2014yCat}
{Galleti}, S., {Federici}, L., {Bellazzini}, M., {et~al.} 2014, VizieR Online
  Data Catalog, V/143

\bibitem[{{Gilfanov}(2004)}]{Gilfanov04}
{Gilfanov}, M. 2004, \mnras, 349, 146, \dodoi{10.1111/j.1365-2966.2004.07473.x}

\bibitem[{{Green}(2019)}]{Green19}
{Green}, D.~A. 2019, Journal of Astrophysics and Astronomy, 40, 36,
  \dodoi{10.1007/s12036-019-9601-6}

\bibitem[{{Greiner}(2000)}]{Greiner00}
{Greiner}, J. 2000, \na, 5, 137, \dodoi{10.1016/S1384-1076(00)00018-X}

\bibitem[{{G{\"u}del}(2004)}]{Gudel04}
{G{\"u}del}, M. 2004, \aapr, 12, 71, \dodoi{10.1007/s00159-004-0023-2}

\bibitem[{{Heinke} {et~al.}(2005){Heinke}, {Grindlay}, {Edmonds}, {Cohn},
  {Lugger}, {Camilo}, {Bogdanov}, \& {Freire}}]{Heinke05}
{Heinke}, C.~O., {Grindlay}, J.~E., {Edmonds}, P.~D., {et~al.} 2005, \apj, 625,
  796, \dodoi{10.1086/429899}

\bibitem[{{Henze} {et~al.}(2012){Henze}, {Pietsch}, \&
  {Haberl}}]{Henze2012ATel}
{Henze}, M., {Pietsch}, W., \& {Haberl}, F. 2012, The Astronomer's Telegram,
  3921, 1

\bibitem[{{Henze} {et~al.}(2009){Henze}, {Pietsch}, {Haberl}, \&
  {Greiner}}]{Henze2009ATel}
{Henze}, M., {Pietsch}, W., {Haberl}, F., \& {Greiner}, J. 2009, The
  Astronomer's Telegram, 2356, 1

\bibitem[{{HI4PI Collaboration} {et~al.}(2016){HI4PI Collaboration}, {Ben
  Bekhti}, {Fl{\"o}er}, {Keller}, {Kerp}, {Lenz}, {Winkel}, {Bailin},
  {Calabretta}, {Dedes}, {Ford}, {Gibson}, {Haud}, {Janowiecki}, {Kalberla},
  {Lockman}, {McClure-Griffiths}, {Murphy}, {Nakanishi}, {Pisano}, \&
  {Staveley-Smith}}]{HI4PI2016}
{HI4PI Collaboration}, {Ben Bekhti}, N., {Fl{\"o}er}, L., {et~al.} 2016, \aap,
  594, A116, \dodoi{10.1051/0004-6361/201629178}

\bibitem[{{Irwin} {et~al.}(2003){Irwin}, {Athey}, \& {Bregman}}]{Irwin03}
{Irwin}, J.~A., {Athey}, A.~E., \& {Bregman}, J.~N. 2003, \apj, 587, 356,
  \dodoi{10.1086/368179}

\bibitem[{{Irwin} {et~al.}(2002){Irwin}, {Sarazin}, \& {Bregman}}]{Irwin02}
{Irwin}, J.~A., {Sarazin}, C.~L., \& {Bregman}, J.~N. 2002, \apj, 570, 152,
  \dodoi{10.1086/339734}

\bibitem[{{Jennings} {et~al.}(2014){Jennings}, {Williams}, {Murphy},
  {Dalcanton}, {Gilbert}, {Dolphin}, {Weisz}, \& {Fouesneau}}]{Jennings2014}
{Jennings}, Z.~G., {Williams}, B.~F., {Murphy}, J.~W., {et~al.} 2014, \apj,
  795, 170, \dodoi{10.1088/0004-637X/795/2/170}

\bibitem[{{Jonker} \& {Nelemans}(2004)}]{Jonker04}
{Jonker}, P.~G., \& {Nelemans}, G. 2004, \mnras, 354, 355,
  \dodoi{10.1111/j.1365-2966.2004.08193.x}

\bibitem[{{Kaur} {et~al.}(2012){Kaur}, {Henze}, {Haberl}, {Pietsch}, {Greiner},
  {Rau}, {Hartmann}, {Sala}, \& {Hernanz}}]{Kaur2012}
{Kaur}, A., {Henze}, M., {Haberl}, F., {et~al.} 2012, \aap, 538, A49,
  \dodoi{10.1051/0004-6361/201118025}

\bibitem[{{Kavanagh} {et~al.}(2020){Kavanagh}, {Sasaki}, {Breitschwerdt}, {de
  Avillez}, {Filipovi{\'c}}, {Galvin}, {Haberl}, {Hatzidimitriou}, {Henze},
  {Plucinsky}, {Saeedi}, {Sokolovsky}, \& {Williams}}]{2020A&A...637A..12K}
{Kavanagh}, P.~J., {Sasaki}, M., {Breitschwerdt}, D., {et~al.} 2020, \aap, 637,
  A12, \dodoi{10.1051/0004-6361/201937008}

\bibitem[{{Kim} \& {Fabbiano}(2004)}]{Kim04}
{Kim}, D.-W., \& {Fabbiano}, G. 2004, \apj, 611, 846, \dodoi{10.1086/422210}

\bibitem[{{Kim} {et~al.}(2009){Kim}, {Fabbiano}, {Brassington}, {Fragos},
  {Kalogera}, {Zezas}, {Jord{\'a}n}, {Sivakoff}, {Kundu}, {Zepf}, {Angelini},
  {Davies}, {Gallagher}, {Juett}, {King}, {Pellegrini}, {Sarazin}, \&
  {Trinchieri}}]{Kim09}
{Kim}, D.~W., {Fabbiano}, G., {Brassington}, N.~J., {et~al.} 2009, \apj, 703,
  829, \dodoi{10.1088/0004-637X/703/1/829}

\bibitem[{{Kong} {et~al.}(2003){Kong}, {DiStefano}, {Garcia}, \&
  {Greiner}}]{kongChandraStudiesRay2003}
{Kong}, A. K.~H., {DiStefano}, R., {Garcia}, M.~R., \& {Greiner}, J. 2003,
  \apj, 585, 298, \dodoi{10.1086/345947}

\bibitem[{{Kotov} {et~al.}(2006){Kotov}, {Trudolyubov}, \&
  {Vestrand}}]{Kotov2006}
{Kotov}, O., {Trudolyubov}, S., \& {Vestrand}, W.~T. 2006, \apj, 641, 756,
  \dodoi{10.1086/500630}

\bibitem[{{Lazzarini} {et~al.}(2021){Lazzarini}, {Williams}, {Durbin},
  {Dalcanton}, {Antoniou}, {Binder}, {Eracleous}, {Plucinsky}, {Sasaki}, \&
  {Vulic}}]{Lazzarini2021}
{Lazzarini}, M., {Williams}, B.~F., {Durbin}, M., {et~al.} 2021, \apj, 906,
  120, \dodoi{10.3847/1538-4357/abccca}

\bibitem[{{Lee} \& {Lee}(2014)}]{Lee2014}
{Lee}, J.~H., \& {Lee}, M.~G. 2014, \apj, 786, 130,
  \dodoi{10.1088/0004-637X/786/2/130}

\bibitem[{{Lehmer} {et~al.}(2010){Lehmer}, {Alexander}, {Bauer}, {Brandt},
  {Goulding}, {Jenkins}, {Ptak}, \& {Roberts}}]{Lehmer10}
{Lehmer}, B.~D., {Alexander}, D.~M., {Bauer}, F.~E., {et~al.} 2010, \apj, 724,
  559, \dodoi{10.1088/0004-637X/724/1/559}

\bibitem[{{Li} {et~al.}(2016){Li}, {Bregman}, {Wang}, {Crain}, \&
  {Anderson}}]{LiJ16}
{Li}, J.-T., {Bregman}, J.~N., {Wang}, Q.~D., {Crain}, R.~A., \& {Anderson},
  M.~E. 2016, \apj, 830, 134, \dodoi{10.3847/0004-637X/830/2/134}

\bibitem[{{Li} {et~al.}(2018){Li}, {Bregman}, {Wang}, {Crain}, \&
  {Anderson}}]{LiJ18}
---. 2018, \apjl, 855, L24, \dodoi{10.3847/2041-8213/aab2af}

\bibitem[{{Li} {et~al.}(2017){Li}, {Bregman}, {Wang}, {Crain}, {Anderson}, \&
  {Zhang}}]{LiJ17}
{Li}, J.-T., {Bregman}, J.~N., {Wang}, Q.~D., {et~al.} 2017, \apjs, 233, 20,
  \dodoi{10.3847/1538-4365/aa96fc}

\bibitem[{{Li} {et~al.}(2008){Li}, {Li}, {Wang}, {Irwin}, \& {Rossa}}]{LiJ08}
{Li}, J.-T., {Li}, Z., {Wang}, Q.~D., {Irwin}, J.~A., \& {Rossa}, J. 2008,
  \mnras, 390, 59, \dodoi{10.1111/j.1365-2966.2008.13749.x}

\bibitem[{{Li} {et~al.}(2021{\natexlab{a}}){Li}, {Wang}, {Yang}, {Bregman},
  {Fan}, \& {Zhang}}]{LiJ21}
{Li}, J.-T., {Wang}, F., {Yang}, J., {et~al.} 2021{\natexlab{a}}, \mnras, 504,
  2767, \dodoi{10.1093/mnras/stab1042}

\bibitem[{{Li} \& {Wang}(2013{\natexlab{a}})}]{LiJ13a}
{Li}, J.-T., \& {Wang}, Q.~D. 2013{\natexlab{a}}, \mnras, 428, 2085,
  \dodoi{10.1093/mnras/sts183}

\bibitem[{{Li} \& {Wang}(2013{\natexlab{b}})}]{LiJ13b}
---. 2013{\natexlab{b}}, \mnras, 435, 3071, \dodoi{10.1093/mnras/stt1501}

\bibitem[{{Li} {et~al.}(2021{\natexlab{b}}){Li}, {Riess}, {Busch}, {Casertano},
  {Macri}, \& {Yuan}}]{liSub2DistanceM312021}
{Li}, S., {Riess}, A.~G., {Busch}, M.~P., {et~al.} 2021{\natexlab{b}}, \apj,
  920, 84, \dodoi{10.3847/1538-4357/ac1597}

\bibitem[{{Li} \& {Wang}(2007)}]{liChandraDetectionDiffuse2007}
{Li}, Z., \& {Wang}, Q.~D. 2007, \apjl, 668, L39, \dodoi{10.1086/522674}

\bibitem[{{Liu} {et~al.}(2007){Liu}, {van Paradijs}, \& {van den
  Heuvel}}]{Liu07}
{Liu}, Q.~Z., {van Paradijs}, J., \& {van den Heuvel}, E.~P.~J. 2007, \aap,
  469, 807, \dodoi{10.1051/0004-6361:20077303}

\bibitem[{{Long} {et~al.}(2010){Long}, {Blair}, {Winkler}, {Becker}, {Gaetz},
  {Ghavamian}, {Helfand}, {Hughes}, {Kirshner}, {Kuntz}, {McNeil}, {Pannuti},
  {Plucinsky}, {Saul}, {T{\"u}llmann}, \& {Williams}}]{Long10}
{Long}, K.~S., {Blair}, W.~P., {Winkler}, P.~F., {et~al.} 2010, \apjs, 187,
  495, \dodoi{10.1088/0067-0049/187/2/495}

\bibitem[{{Lyke} {et~al.}(2020){Lyke}, {Higley}, {McLane}, {Schurhammer},
  {Myers}, {Ross}, {Dawson}, {Chabanier}, {Martini}, {Busca}, {Mas des
  Bourboux}, {Salvato}, {Streblyanska}, {Zarrouk}, {Burtin}, {Anderson},
  {Bautista}, {Bizyaev}, {Brandt}, {Brinkmann}, {Brownstein}, {Comparat},
  {Green}, {de la Macorra}, {Mu{\~n}oz Guti{\'e}rrez}, {Hou}, {Newman},
  {Palanque-Delabrouille}, {P{\^a}ris}, {Percival}, {Petitjean}, {Rich},
  {Rossi}, {Schneider}, {Smith}, {Vivek}, \& {Weaver}}]{SDSSQSOdr16}
{Lyke}, B.~W., {Higley}, A.~N., {McLane}, J.~N., {et~al.} 2020, \apjs, 250, 8,
  \dodoi{10.3847/1538-4365/aba623}

\bibitem[{{Maggi} {et~al.}(2016){Maggi}, {Haberl}, {Kavanagh}, {Sasaki},
  {Bozzetto}, {Filipovi{\'c}}, {Vasilopoulos}, {Pietsch}, {Points}, {Chu},
  {Dickel}, {Ehle}, {Williams}, \& {Greiner}}]{Maggi16}
{Maggi}, P., {Haberl}, F., {Kavanagh}, P.~J., {et~al.} 2016, \aap, 585, A162,
  \dodoi{10.1051/0004-6361/201526932}

\bibitem[{{Maggi} {et~al.}(2019){Maggi}, {Filipovi{\'c}}, {Vukoti{\'c}},
  {Ballet}, {Haberl}, {Maitra}, {Kavanagh}, {Sasaki}, \& {Stupar}}]{Maggi19}
{Maggi}, P., {Filipovi{\'c}}, M.~D., {Vukoti{\'c}}, B., {et~al.} 2019, \aap,
  631, A127, \dodoi{10.1051/0004-6361/201936583}

\bibitem[{{Massey} {et~al.}(2006){Massey}, {Olsen}, {Hodge}, {Strong},
  {Jacoby}, {Schlingman}, \& {Smith}}]{Massey2006}
{Massey}, P., {Olsen}, K.~A.~G., {Hodge}, P.~W., {et~al.} 2006, \aj, 131, 2478,
  \dodoi{10.1086/503256}

\bibitem[{{Miller} {et~al.}(2004){Miller}, {Fabian}, \& {Miller}}]{Miller04}
{Miller}, J.~M., {Fabian}, A.~C., \& {Miller}, M.~C. 2004, \apjl, 614, L117,
  \dodoi{10.1086/425316}

\bibitem[{{Mineo} {et~al.}(2012){Mineo}, {Gilfanov}, \& {Sunyaev}}]{Mineo12}
{Mineo}, S., {Gilfanov}, M., \& {Sunyaev}, R. 2012, \mnras, 419, 2095,
  \dodoi{10.1111/j.1365-2966.2011.19862.x}

\bibitem[{{Mukai}(2017)}]{Mukai17}
{Mukai}, K. 2017, \pasp, 129, 062001, \dodoi{10.1088/1538-3873/aa6736}

\bibitem[{{Ni} {et~al.}(2021){Ni}, {Brandt}, {Chen}, {Luo}, {Nyland}, {Yang},
  {Zou}, {Aird}, {Alexander}, {Bauer}, {Lacy}, {Lehmer}, {Mallick}, {Salvato},
  {Schneider}, {Tozzi}, {Traulsen}, {Vaccari}, {Vignali}, {Vito}, {Xue},
  {Banerji}, {Chow}, {Comastri}, {Del Moro}, {Gilli}, {Mullaney}, {Paolillo},
  {Schwope}, {Shemmer}, {Sun}, {Timlin}, \& {Trump}}]{NiQ2021}
{Ni}, Q., {Brandt}, W.~N., {Chen}, C.-T., {et~al.} 2021, \apjs, 256, 21,
  \dodoi{10.3847/1538-4365/ac0dc6}

\bibitem[{{Pietsch} {et~al.}(2005{\natexlab{a}}){Pietsch}, {Fliri}, {Freyberg},
  {Greiner}, {Haberl}, {Riffeser}, \& {Sala}}]{PietschOpticalnovae2005}
{Pietsch}, W., {Fliri}, J., {Freyberg}, M.~J., {et~al.} 2005{\natexlab{a}},
  \aap, 442, 879, \dodoi{10.1051/0004-6361:20053127}

\bibitem[{{Pietsch} {et~al.}(2005{\natexlab{b}}){Pietsch}, {Freyberg}, \&
  {Haberl}}]{Pietsch05}
{Pietsch}, W., {Freyberg}, M., \& {Haberl}, F. 2005{\natexlab{b}}, \aap, 434,
  483, \dodoi{10.1051/0004-6361:20041990}

\bibitem[{{Remillard} \& {McClintock}(2006)}]{Remillard06}
{Remillard}, R.~A., \& {McClintock}, J.~E. 2006, \araa, 44, 49,
  \dodoi{10.1146/annurev.astro.44.051905.092532}

\bibitem[{{Revnivtsev} {et~al.}(2008){Revnivtsev}, {Churazov}, {Sazonov},
  {Forman}, \& {Jones}}]{Revnivtsev08}
{Revnivtsev}, M., {Churazov}, E., {Sazonov}, S., {Forman}, W., \& {Jones}, C.
  2008, \aap, 490, 37, \dodoi{10.1051/0004-6361:200809889}

\bibitem[{{Revnivtsev} {et~al.}(2009){Revnivtsev}, {Sazonov}, {Churazov},
  {Forman}, {Vikhlinin}, \& {Sunyaev}}]{Revnivtsev09}
{Revnivtsev}, M., {Sazonov}, S., {Churazov}, E., {et~al.} 2009, \nat, 458,
  1142, \dodoi{10.1038/nature07946}

\bibitem[{{Revnivtsev} {et~al.}(2011){Revnivtsev}, {Sazonov}, {Forman},
  {Churazov}, \& {Sunyaev}}]{Revnivtsev11}
{Revnivtsev}, M., {Sazonov}, S., {Forman}, W., {Churazov}, E., \& {Sunyaev}, R.
  2011, \mnras, 414, 495, \dodoi{10.1111/j.1365-2966.2011.18411.x}

\bibitem[{{Revnivtsev} {et~al.}(2007){Revnivtsev}, {Vikhlinin}, \&
  {Sazonov}}]{Revnivtsev07}
{Revnivtsev}, M., {Vikhlinin}, A., \& {Sazonov}, S. 2007, \aap, 473, 857,
  \dodoi{10.1051/0004-6361:20066850}

\bibitem[{{Rosner} {et~al.}(1985){Rosner}, {Golub}, \& {Vaiana}}]{Rosner85}
{Rosner}, R., {Golub}, L., \& {Vaiana}, G.~S. 1985, \araa, 23, 413,
  \dodoi{10.1146/annurev.aa.23.090185.002213}

\bibitem[{{Salvato} {et~al.}(2018){Salvato}, {Buchner}, {Budav{\'a}ri},
  {Dwelly}, {Merloni}, {Brusa}, {Rau}, {Fotopoulou}, \&
  {Nandra}}]{salvatoFindingCounterpartsAllsky2018}
{Salvato}, M., {Buchner}, J., {Budav{\'a}ri}, T., {et~al.} 2018, \mnras, 473,
  4937, \dodoi{10.1093/mnras/stx2651}

\bibitem[{{Sana} {et~al.}(2006){Sana}, {Rauw}, {Naz{\'e}}, {Gosset}, \&
  {Vreux}}]{Sana06}
{Sana}, H., {Rauw}, G., {Naz{\'e}}, Y., {Gosset}, E., \& {Vreux}, J.~M. 2006,
  \mnras, 372, 661, \dodoi{10.1111/j.1365-2966.2006.10847.x}

\bibitem[{{Sasaki} {et~al.}(2012){Sasaki}, {Pietsch}, {Haberl},
  {Hatzidimitriou}, {Stiele}, {Williams}, {Kong}, \& {Kolb}}]{Sasaki2012}
{Sasaki}, M., {Pietsch}, W., {Haberl}, F., {et~al.} 2012, \aap, 544, A144,
  \dodoi{10.1051/0004-6361/201219025}

\bibitem[{{Sasaki} {et~al.}(2018){Sasaki}, {Haberl}, {Henze}, {Saeedi},
  {Williams}, {Plucinsky}, {Hatzidimitriou}, {Karampelas}, {Sokolovsky},
  {Breitschwerdt}, {de Avillez}, {Filipovi{\'c}}, {Galvin}, {Kavanagh}, \&
  {Long}}]{sasakiDeepXMMNewtonObservations2018}
{Sasaki}, M., {Haberl}, F., {Henze}, M., {et~al.} 2018, \aap, 620, A28,
  \dodoi{10.1051/0004-6361/201833588}

\bibitem[{{Sazonov} {et~al.}(2006){Sazonov}, {Revnivtsev}, {Gilfanov},
  {Churazov}, \& {Sunyaev}}]{Sazonov06}
{Sazonov}, S., {Revnivtsev}, M., {Gilfanov}, M., {Churazov}, E., \& {Sunyaev},
  R. 2006, \aap, 450, 117, \dodoi{10.1051/0004-6361:20054297}

\bibitem[{{Schmitt} {et~al.}(1995){Schmitt}, {Fleming}, \&
  {Giampapa}}]{Schmitt95}
{Schmitt}, J. H.~M.~M., {Fleming}, T.~A., \& {Giampapa}, M.~S. 1995, \apj, 450,
  392, \dodoi{10.1086/176149}

\bibitem[{{Shaw Greening} {et~al.}(2009){Shaw Greening}, {Barnard}, {Kolb},
  {Tonkin}, \& {Osborne}}]{greeningXraySpectralSurvey2009}
{Shaw Greening}, L., {Barnard}, R., {Kolb}, U., {Tonkin}, C., \& {Osborne},
  J.~P. 2009, \aap, 495, 733, \dodoi{10.1051/0004-6361/200809864}

\bibitem[{{Singh} \& {Pandey}(2022)}]{Singh22}
{Singh}, G., \& {Pandey}, J.~C. 2022, \apj, 934, 20,
  \dodoi{10.3847/1538-4357/ac7716}

\bibitem[{{Stiele} {et~al.}(2008){Stiele}, {Pietsch}, {Haberl}, \&
  {Freyberg}}]{stieleTimeVariabilityXray2008}
{Stiele}, H., {Pietsch}, W., {Haberl}, F., \& {Freyberg}, M. 2008, \aap, 480,
  599, \dodoi{10.1051/0004-6361:20078858}

\bibitem[{{Stiele} {et~al.}(2011){Stiele}, {Pietsch}, {Haberl},
  {Hatzidimitriou}, {Barnard}, {Williams}, {Kong}, \&
  {Kolb}}]{stieleDeepXMMNewtonSurvey2011a}
{Stiele}, H., {Pietsch}, W., {Haberl}, F., {et~al.} 2011, \aap, 534, A55,
  \dodoi{10.1051/0004-6361/201015270}

\bibitem[{{Sun} {et~al.}(2009){Sun}, {Voit}, {Donahue}, {Jones}, {Forman}, \&
  {Vikhlinin}}]{Sun09}
{Sun}, M., {Voit}, G.~M., {Donahue}, M., {et~al.} 2009, \apj, 693, 1142,
  \dodoi{10.1088/0004-637X/693/2/1142}

\bibitem[{{Supper} {et~al.}(2001){Supper}, {Hasinger}, {Lewin}, {Magnier}, {van
  Paradijs}, {Pietsch}, {Read}, \& {Tr{\"u}mper}}]{2001A&A...373...63S}
{Supper}, R., {Hasinger}, G., {Lewin}, W.~H.~G., {et~al.} 2001, \aap, 373, 63,
  \dodoi{10.1051/0004-6361:20010495}

\bibitem[{{Supper} {et~al.}(1997){Supper}, {Hasinger}, {Pietsch}, {Truemper},
  {Jain}, {Magnier}, {Lewin}, \& {van Paradijs}}]{1997A&A...317..328S}
{Supper}, R., {Hasinger}, G., {Pietsch}, W., {et~al.} 1997, \aap, 317, 328

\bibitem[{{Tamm} {et~al.}(2012){Tamm}, {Tempel}, {Tenjes}, {Tihhonova}, \&
  {Tuvikene}}]{Tamm12}
{Tamm}, A., {Tempel}, E., {Tenjes}, P., {Tihhonova}, O., \& {Tuvikene}, T.
  2012, \aap, 546, A4, \dodoi{10.1051/0004-6361/201220065}

\bibitem[{{Traulsen} {et~al.}(2019){Traulsen}, {Schwope}, {Lamer}, {Ballet},
  {Carrera}, {Coriat}, {Freyberg}, {Michel}, {Motch}, {Rosen}, {Webb},
  {Ceballos}, {Koliopanos}, {Kurpas}, {Page}, \&
  {Watson}}]{traulsenXMMNewtonSerendipitousSurvey2019}
{Traulsen}, I., {Schwope}, A.~D., {Lamer}, G., {et~al.} 2019, \aap, 624, A77,
  \dodoi{10.1051/0004-6361/201833938}

\bibitem[{{Traulsen} {et~al.}(2020){Traulsen}, {Schwope}, {Lamer}, {Ballet},
  {Carrera}, {Ceballos}, {Coriat}, {Freyberg}, {Koliopanos}, {Kurpas},
  {Michel}, {Motch}, {Page}, {Watson}, \& {Webb}}]{Traulsen20}
---. 2020, \aap, 641, A137, \dodoi{10.1051/0004-6361/202037706}

\bibitem[{{Traulsen} {et~al.}(2022){Traulsen}, {Schwope}, {Lamer}, {Ballet},
  {Carrera}, {Ceballos}, {Coriat}, {Freyberg}, {Koliopanos}, {Kurpas},
  {Michel}, {Motch}, {Page}, {Watson}, \& {Webb}}]{Traulsen22}
---. 2022, VizieR Online Data Catalog, IX/66

\bibitem[{{Trinchieri} \& {Fabbiano}(1991)}]{1991ApJ...382...82T}
{Trinchieri}, G., \& {Fabbiano}, G. 1991, \apj, 382, 82, \dodoi{10.1086/170696}

\bibitem[{{van Speybroeck} {et~al.}(1979){van Speybroeck}, {Epstein}, {Forman},
  {Giacconi}, {Jones}, {Liller}, \&
  {Smarr}}]{vanspeybroeckObservationsXraySources1979}
{van Speybroeck}, L., {Epstein}, A., {Forman}, W., {et~al.} 1979, \apjl, 234,
  L45, \dodoi{10.1086/183106}

\bibitem[{{Voss} \& {Gilfanov}(2007)}]{vossStudyPopulationLMXBs2007}
{Voss}, R., \& {Gilfanov}, M. 2007, \aap, 468, 49,
  \dodoi{10.1051/0004-6361:20066614}

\bibitem[{Watson {et~al.}(2009)Watson, Schr{\"o}der, Fyfe, Page, Lamer, Mateos,
  Pye, Sakano, Rosen, Ballet, Barcons, Barret, Boller, Brunner, Brusa,
  Caccianiga, Carrera, Ceballos, Della~Ceca, Denby, Denkinson, Dupuy, Farrell,
  Fraschetti, Freyberg, Guillout, Hambaryan, Maccacaro, Mathiesen, McMahon,
  Michel, Motch, Osborne, Page, Pakull, Pietsch, Saxton, Schwope, Severgnini,
  Simpson, Sironi, Stewart, Stewart, Stobbart, Tedds, Warwick, Webb, West,
  Worrall, \& Yuan}]{watsonXMMNewtonSerendipitousSurvey2009}
Watson, M.~G., Schr{\"o}der, A.~C., Fyfe, D., {et~al.} 2009, Astronomy \&
  Astrophysics, 493, 339

\bibitem[{{Webb} {et~al.}(2020){Webb}, {Coriat}, {Traulsen}, {Ballet}, {Motch},
  {Carrera}, {Koliopanos}, {Authier}, {de la Calle}, {Ceballos}, {Colomo},
  {Chuard}, {Freyberg}, {Garcia}, {Kolehmainen}, {Lamer}, {Lin}, {Maggi},
  {Michel}, {Page}, {Page}, {Perea-Calderon}, {Pineau}, {Rodriguez}, {Rosen},
  {Santos Lleo}, {Saxton}, {Schwope}, {Tom{\'a}s}, {Watson}, \&
  {Zakardjian}}]{Webb20}
{Webb}, N.~A., {Coriat}, M., {Traulsen}, I., {et~al.} 2020, \aap, 641, A136,
  \dodoi{10.1051/0004-6361/201937353}

\bibitem[{{Wenger} {et~al.}(2000){Wenger}, {Ochsenbein}, {Egret}, {Dubois},
  {Bonnarel}, {Borde}, {Genova}, {Jasniewicz}, {Lalo{\"e}}, {Lesteven}, \&
  {Monier}}]{Wenger2000}
{Wenger}, M., {Ochsenbein}, F., {Egret}, D., {et~al.} 2000, \aaps, 143, 9,
  \dodoi{10.1051/aas:2000332}

\bibitem[{{White} {et~al.}(2019){White}, {Long}, {Becker}, {Blair}, {Helfand},
  \& {Winkler}}]{White19}
{White}, R.~L., {Long}, K.~S., {Becker}, R.~H., {et~al.} 2019, \apjs, 241, 37,
  \dodoi{10.3847/1538-4365/ab0e89}

\bibitem[{{Williams} {et~al.}(2014){Williams}, {Lang}, {Dalcanton}, {Dolphin},
  {Weisz}, {Bell}, {Bianchi}, {Byler}, {Gilbert}, {Girardi}, {Gordon},
  {Gregersen}, {Johnson}, {Kalirai}, {Lauer}, {Monachesi}, {Rosenfield},
  {Seth}, \& {Skillman}}]{williamsPANCHROMATICHUBBLEANDROMEDA2014}
{Williams}, B.~F., {Lang}, D., {Dalcanton}, J.~J., {et~al.} 2014, \apjs, 215,
  9, \dodoi{10.1088/0067-0049/215/1/9}

\bibitem[{{Williams} {et~al.}(2018){Williams}, {Lazzarini}, {Plucinsky},
  {Sasaki}, {Antoniou}, {Vulic}, {Eracleous}, {Long}, {Binder}, {Dalcanton},
  {Lewis}, \& {Weisz}}]{williamsComparingChandraHubble2018}
{Williams}, B.~F., {Lazzarini}, M., {Plucinsky}, P.~P., {et~al.} 2018, \apjs,
  239, 13, \dodoi{10.3847/1538-4365/aae37d}

\bibitem[{{Zhang} {et~al.}(2011){Zhang}, {Gilfanov}, {Voss}, {Sivakoff},
  {Kraft}, {Brassington}, {Kundu}, {Jord{\'a}n}, \& {Sarazin}}]{Zhang2011}
{Zhang}, Z., {Gilfanov}, M., {Voss}, R., {et~al.} 2011, \aap, 533, A33,
  \dodoi{10.1051/0004-6361/201116936}

\bibitem[{{Zhao} \& {Heinke}(2022)}]{Zhao22}
{Zhao}, J., \& {Heinke}, C.~O. 2022, \mnras, 511, 5964,
  \dodoi{10.1093/mnras/stac442}

\bibitem[{{Zhu} {et~al.}(2018){Zhu}, {Li}, \& {Morris}}]{Zhu18}
{Zhu}, Z., {Li}, Z., \& {Morris}, M.~R. 2018, \apjs, 235, 26,
  \dodoi{10.3847/1538-4365/aab14f}

\bibitem[{{Zuo} {et~al.}(2008){Zuo}, {Li}, \& {Liu}}]{Zuo08}
{Zuo}, Z.-Y., {Li}, X.-D., \& {Liu}, X.-W. 2008, \mnras, 387, 121,
  \dodoi{10.1111/j.1365-2966.2008.12974.x}

\end{thebibliography}

\appendix
\renewcommand\thetable{A.\arabic{table}}
\setcounter{table}{0}

\section{XMM-Newton Data Used in the NEW-ANGELS Program}
\label{appendsec:XMMDatalist}

\setlength{\LTcapwidth}{1.0 \linewidth}

\begin{longtable*}[c]{l l l l p{1.0cm} p{1.5cm} p{1.5cm} p{1.5cm} p{0.8cm} p{0.8cm} p{0.8cm} }

 \caption{ The \emph{XMM-Newton} observations used in the New-ANGELS program. The total good time interval (GTI) of the mos1, mos2 and PN cameras are in unit of ks. M and T1 in the filter columns refer to the median and thin1 filter.} \\

 \hline
 \multicolumn{11}{ c }{Begin of Table} \\
 \hline
 ObsID & RA & DEC & Start Date & PI & $t_{\mathrm{M1}}$ (GTI)  & $t_{\mathrm{M2}}$ (GTI)  & $t_{\mathrm{PN}}$(GTI)   & filter$\rm _{M1}$ & filter$\rm _{M2}$ & filter$\rm _{PN}$ \\
 \hline
 \endfirsthead

 \multicolumn{11}{c}{Continuation of Table \ref{table:XMMDatalist}} \\
 \hline
 ObsID & RA & DEC & Start Date & PI & $t_{\mathrm{M1}}^a$ (GTI) & $t_{\mathrm{M2}}$ (GTI) & $t_{\mathrm{PN}}$(GTI) & filter$\rm _{M1}$ & filter$\rm _{M2}$ & filter$\rm _{PN}$ \\
 \hline
 \endhead
 \hline
 \endfoot

 \hline
 \multicolumn{11}{ c }{End of Table} \\
 \hline\hline
 \endlastfoot

0109270101 & 00 42 42.9 & +41 15 46.0 & 2001-06-29 &  Mason & 50.8 (28.4) & 50.8 (30.8) & 45.5 (25.1) & M & M & M  \\
0109270301 & 00 45 20.0 & +41 56 09.0 & 2002-01-26 &  Mason & 55.6 (26.6) & 55.9 (26.9) & 27.5 (25.4) & M & M & M \\
0109270401 & 00 46 38.0 & +42 16 20.0 & 2002-06-29 &  Mason & 53.3 (48.9) & 53.9 (50.1) & 91.6 (64.2) & M & M & M \\
0109270701 & 00 44 00.9 & +41 35 57.0 & 2002-01-05 &  Mason & 57.4 (56.8) & 57.4 (57.0) & 54.9 (54.5) & M & M & M \\
0112570101 & 00 42 42.9 & +41 15 46.0 & 2002-01-06 &  Watson & 62.5 (59.6) & 63.7 (59.9) & 53.2 (50.1) & T1 & T1 & T1 \\
0112570201 & 00 41 24.9 & +40 55 35.0 & 2002-01-12 &  Watson & 66.5 (55.5) & 66.5 (55.3) & 59.6 (51.3) & T1 & T1 & T1 \\
0112570301 & 00 40 06.0 & +40 35 24.0 & 2002-01-24 &  Watson & 53.1 (28.5) & 53.3 (30.3) & 45.4 (21.5) & T1 & T1 & T1 \\
0112570401 & 00 42 42.9 & +41 15 46.0 & 2000-06-25 &  Watson & 30.0 (29.9) & 30.0 (29.5) & 26.3 (26.1) & M & M & M  \\
0151580201 & 00 43 11.9 & +39 47 60.0 & 2003-02-03 &  Di Stefano &  & flared &    &  &  &  \\
0151580301 & 00 45 59.0 & +40 42 36.0 & 2003-02-03 &  Di Stefano &  & flared &    &  &  &  \\
0151580401 & 00 46 06.9 & +41 20 58.0 & 2003-02-06 &  Di Stefano & 13.2 (12.8) & 13.2 (12.8) & 11.5 (11.1) & M & M & M \\
0151581201 & 00 43 11.9 & +39 47 60.0 & 2003-07-01 &  Di Stefano &  & flared &    &  &  &  \\
0151581301 & 00 45 59.0 & +40 42 36.0 & 2003-07-01 &  Di Stefano & 6.4(2.2) & 6.4(2.3) &  & M & M &    \\
0202230301 & 00 42 38.5 & +41 16 03.8 & 2004-07-17 &  Barnard &  & flared &    \\
0204790401 & 00 40 22.2 & +41 41 09.0 & 2004-01-02 &  Di Stefano &  & flared &    \\
0300910201 & 00 41 53.3 & +40 21 18.0 & 2005-08-01 &  Piconcelli &  & flared &    \\
0402560101 & 00 38 52.8 & +40 15 00.0 & 2006-06-28 &  Pietsch & 42.1 (6.1) & 9.1 (3.6) & 42.1 (6.7) & M & M & T1 \\
0402560201 & 00 43 28.8 & +40 55 12.0 & 2006-06-30 &  Pietsch &  & flared &    &  &  &  \\
0402560301 & 00 40 43.2 & +41 17 60.0 & 2006-07-01 &  Pietsch & 61.4 (48.0) & 61.4 (49.4) & 57.6 (39.0) & M & M & T1 \\
0402560401 & 00 42 16.8 & +40 37 12.0 & 2006-07-08 &  Pietsch &  & flared &    &  &  &  \\
0402560501 & 00 39 40.7 & +40 58 48.0 & 2006-07-20 &  Pietsch & 58.7 (54.8) & 58.7 (56.3) & 47.6 (28.6) & M & M & T1\\
0402560601 & 00 40 45.6 & +40 21 05.1 & 2006-07-28 &  Pietsch & 49.4 (32.4) & 49.2 (32.2) & 42.6 (25.1) & M & M & T1\\
0402560701 & 00 39 02.4 & +40 37 48.0 & 2006-07-23 &  Pietsch & 61.1 (26.0) & 61.7 (30.7) & 51.2 (18.3)& M & M & T1 \\
0402560801 & 00 40 06.0 & +40 35 24.0 & 2006-12-25 &  Pietsch & 52.1 (49.8) & 52.1 (49.0) & 50.8 (44.1) & M & M & T1\\
0402560901 & 00 41 52.7 & +41 36 36.0 & 2006-12-26 &  Pietsch & 49.0 (45.6) & 49.0 (45.3) & 44.6 (40.8) & M & M & T1\\
0402561001 & 00 44 38.4 & +41 12 01.0 & 2006-12-30 &  Pietsch & 54.8 (52.0) & 54.7 (53.3) & 45.8 (42.2) & M & M & T1\\
0402561101 & 00 43 09.6 & +41 55 12.0 & 2007-01-01 &  Pietsch & 57.8 (47.7) & 57.9 (47.1) & 57.6 (42.5) & M & M & T1\\
0402561201 & 00 45 43.1 & +41 31 48.0 & 2007-01-02 &  Pietsch & 54.4 (41.4) & 54.4 (40.7) & 53.0 (35.7) & M & M & T1\\
0402561301 & 00 44 45.5 & +42 09 36.0 & 2007-01-03 &  Pietsch & 54.3 (36.4) & 54.3 (36.0) & 40.3 (29.6) & M & M & T1\\
0402561401 & 00 46 38.4 & +41 53 60.0 & 2007-01-04 &  Pietsch & 52.4 (46.1) & 52.4 (46.5) & 51.0 (43.4) & M & M & T1\\
0402561501 & 00 45 20.0 & +41 56 09.0 & 2007-01-05 &  Pietsch & 45.4 (43.8) & 45.4 (43.8) & 43.0 (37.9) & M & M & T1\\
0404060201 & 00 45 19.5 & +40 57 47.0 & 2006-07-03 &  Bregman & 35.3 (21.1) & 35.3 (22.3) & 33.6 (16.1) & T1 & T1 & T1\\
0405320501 & 00 42 44.3 & +41 16 09.4 & 2006-07-02 &  Pietsch & 21.4 (13.4) & 21.7 (15.1) & 13.3 (9.5) & M & M & T1 \\
0405320601 & 00 42 44.3 & +41 16 09.4 & 2006-08-09 &  Pietsch &  & flared &    &  &  &  \\
0405320801 & 00 42 44.3 & +41 16 09.4 & 2007-01-16 &  Pietsch &  & flared &    &  &  &  \\
0405320901 & 00 42 44.3 & +41 16 09.4 & 2007-02-05 &  Pietsch & 16.4 (16.0) & 16.6 (16.3) & 15.0 (13.2)  \\
0410582001 & 00 40 59.1 & +41 15 51.2 & 2007-07-25 &  Schartel & 15.9 (14.9) & 15.9 (15.5) & 13.4 (10.5) & M & M & T1\\
0505720201 & 00 42 44.3 & +41 16 09.4 & 2007-12-29 &  Pietsch & 27.2 (26.9) & 27.2 (27.1) & 25.7 (25.6) & M & M & T1 \\
0505720301 & 00 42 44.3 & +41 16 09.4 & 2008-01-08 &  Pietsch & 26.9 (26.4) & 26.9 (26.3) & 24.4 (22.8) & M & M & T1\\
0505720401 & 00 42 44.3 & +41 16 09.4 & 2008-01-18 &  Pietsch & 22.4 (21.6) & 22.4 (21.5) & 20.8 (17.2) & M & M & T1\\
0505720501 & 00 42 44.3 & +41 16 09.4 & 2008-01-27 &  Pietsch & 21.5 (14.2) & 21.5 (15.0) & 15.0 (7.8) & M & M & T1\\
0505720601 & 00 42 44.3 & +41 16 09.4 & 2008-02-07 &  Pietsch &  & flared &    &  &  &  \\
0505760101 & 00 38 52.8 & +40 15 00.0 & 2007-07-24 &  Pietsch & 35.5 (27.6) & 55.7 (29.7) & 34.8 (17.8) & M & M & T1\\
0505760201 & 00 43 28.8 & +40 55 12.0 & 2007-07-22 &  Pietsch & 57.5 (42.4) & 57.5 (47.1) & 34.6 (23.8) & M & M & T1\\
0505760301 & 00 42 16.8 & +40 37 12.0 & 2007-12-28 &  Pietsch & 44.2 (42.7) & 44.4 (44.0) & 40.1 (39.5) & M & M & T1\\
0505760401 & 00 40 45.6 & +40 21 05.1 & 2007-12-25 &  Pietsch & 28.6 (28.5) & 28.6 (28.4) & 27.1 (26.3) & M & M & T1\\
0505760501 & 00 39 02.4 & +40 37 48.0 & 2007-12-31 &  Pietsch & 38.6 (34.5) & 38.6 (35.1) & 37.0 (25.8) & M & M & T1\\
0505900101 & 00 38 19.5 & +41 47 15.0 & 2007-07-21 &  Tanvir &  & flared &    &  &  &  \\
0505900301 & 00 38 04.5 & +40 44 39.0 & 2007-12-27 &  Tanvir &  & flared &    &  &  &  \\
0505900801 & 00 38 04.5 & +40 44 39.0 & 2008-02-09 &  Tanvir &  & flared &    &  &  &  \\
0511380101 & 00 38 52.8 & +40 15 00.0 & 2008-01-02 &  Pietsch & 45.6 (45.4) & 45.6 (45.4) & 44.0 (41.4) & M & M & T1\\
0511380201 & 00 43 28.8 & +40 55 12.0 & 2008-01-05 &  Pietsch &  & flared &    &  &  &  \\
0511380301 & 00 39 40.7 & +40 58 48.0 & 2008-01-06 &  Pietsch & 35.4 (33.1) & 35.4 (33.0) & 33.9 (27.4) & M & M & T1\\
0511380601 & 00 43 28.8 & +40 55 12.0 & 2008-02-09 &  Pietsch &  & flared &    &  &  &  \\
0551690201 & 00 42 44.3 & +41 16 09.4 & 2008-12-30 &  Pietsch &  & flared &    &  &  &  \\
0551690301 & 00 42 44.3 & +41 16 09.4 & 2009-01-09 &  Pietsch & 21.6 (20.8) & 21.6 (20.6) & 20.0 (18.3) & M & M & T1\\
0551690401 & 00 42 44.3 & +41 16 09.4 & 2009-01-15 &  Pietsch & 17.1 (5.3) & 11.4 (5.1) & 9.4 (3.5) & M & M & T1\\
0551690501 & 00 42 44.3 & +41 16 09.4 & 2009-01-27 &  Pietsch &  & flared &    &  &  &  \\
0551690601 & 00 42 44.3 & +41 16 09.4 & 2009-02-04 &  Pietsch &  & flared &    &  &  &  \\
0560180101 & 00 43 19.9 & +41 13 46.6 & 2008-07-18 &  Schartel & 21.1 (9.2) & 19.9 (8.8) & 16.5 (4.4)  & M & M & T1 \\
0600660201 & 00 42 44.3 & +41 16 09.4 & 2009-12-28 &  Pietsch & 18.5 (18.1) & 18.5 (17.8) & 16.9 (16.1)  \\
0600660301 & 00 42 44.3 & +41 16 09.4 & 2010-01-07 &  Pietsch & 17.0 (16.9) & 17.0 (16.9) & 15.4 (14.8)  \\
0600660401 & 00 42 44.3 & +41 16 09.4 & 2010-01-15 &  Pietsch & 16.8 (16.7) & 16.5 (16.4) & 12.7 (7.7)  &  &  &  \\
0600660501 & 00 42 44.3 & +41 16 09.4 & 2010-01-25 &  Pietsch & 19.3 (13.7) & 19.4 (16.1) & 17.8 (17.8)  &  &  &  \\
0600660601 & 00 42 44.3 & +41 16 09.4 & 2010-02-02 &  Pietsch & 17.0 (16.0) & 17.0 (16.2) & 15.4 (10.2)  &  &  &  \\
0650560201 & 00 42 44.3 & +41 16 09.4 & 2010-12-26 &  Pietsch &  & flared &    &  &  &  \\
0650560301 & 00 42 44.3 & +41 16 09.4 & 2011-01-04 &  Pietsch & 33.1 (28.6) & 33.1 (30.9) & 30.6 (17.9)  & M & M & T1\\
0650560401 & 00 42 44.3 & +41 16 09.4 & 2011-01-14 &  Pietsch & 21.8 (13.1) & 22.1 (15.4) & 21.5 (10.1)  & M & M & T1\\
0650560501 & 00 42 44.3 & +41 16 09.4 & 2011-01-25 &  Pietsch & 23.6 (22.8) & 23.6 (22.8) & 9.2 (5.5)  & M & M & T1\\
0650560601 & 00 42 44.3 & +41 16 09.4 & 2011-02-03 &  Pietsch & 23.3 (22.0) & 23.3 (22.5) & 18.7 (11.0)  & M & M & T1\\
0652500101 & 00 46 55.0 & +40 31 00.0 & 2011-02-01 &  Di Stefano &  & flared &    &  &  &  \\
0652500201 & 00 49 05.0 & +40 17 32.0 & 2011-02-01 &  Di Stefano & 21.6 (12.1) & 21.6 (12.2) & 20.0 (8.3)  & T1 & T1 & T1\\
0652500301 & 00 53 27.9 & +39 49 46.0 & 2011-02-03 &  Di Stefano &  & flared &    &  &  &  \\
0655620301 & 00 43 45.5 & +41 07 54.7 & 2011-08-01 &  Henze &  & flared &    &  &  &  \\
0655620401 & 00 44 01.7 & +41 04 23.8 & 2012-01-21 &  Henze & 20.5 (12.5) & 20.6 (12.8) & 9.1 (8.7)  & M & M & T1\\
0672130101 & 00 42 41.8 & +40 51 54.6 & 2011-06-27 &  Wang & 100.5 (85.9) & 100.6 (90.8) & 97.8 (53.3)  & T1 & T1 & T1\\
0672130501 & 00 42 41.8 & +40 51 54.6 & 2011-07-13 &  Wang & 49.4 (27.1) & 49.4 (29.6) & 47.6 (12.8)  & T1 & T1 & T1\\
0672130601 & 00 42 41.8 & +40 51 54.6 & 2011-07-05 &  Wang & 95.9 (74.5) & 92.4 (78.9) & 95.4 (58.8)  & T1 & T1 & T1\\
0672130701 & 00 42 41.8 & +40 51 54.6 & 2011-07-07 &  Wang & 85.7 (78.4) & 90.8 (80.1) & 82.0 (62.4)  & T1 & T1 & T1\\
0674210201 & 00 42 44.3 & +41 16 09.4 & 2011-12-28 &  Pietsch & 20.6 (20.6) & 20.6 (20.6) & 18.9 (18.5)  & M & M & T1\\
0674210601 & 00 42 44.3 & +41 16 09.4 & 2012-01-31 &  Pietsch &  & flared &    &  &  &  \\
0690600401 & 00 42 52.0 & +41 31 07.8 & 2012-06-26 &  Barnard & 114.3 (89.8) & 114.4 (96.6) & 103.5 (57.0)  & T1 & T1 & T1\\
0700380501 & 00 42 43.7 & +41 25 18.5 & 2012-07-28 &  Schartel & 11.6 (11.6) & 11.6 (11.6) & 10.0 (9.9)  & M & M & T1\\
0700380601 & 00 42 43.7 & +41 25 18.5 & 2012-08-08 &  Schartel & 20.8 (20.1) & 20.2 (19.2) & 19.6 (18.3)  & M & M & T1\\
0701981201 & 00 44 02.0 & +41 25 44.4 & 2013-02-08 &  Schartel & 23.6 (19.8) & 23.6 (19.8) & 21.9 (17.0)  & T1 & T1 & T1\\
0727960401 & 00 42 52.4 & +41 16 31.2 & 2013-07-06 &  Schartel & 10.6 (10.3) & 10.6 (10.4) & 9.0 (9.0)  &  &  &  \\
0744350301 & 00 43 26.7 & +41 23 33.3 & 2014-08-09 &  Henze & 23.4 (23.2) & 23.4 (23.0) & 21.8 (21.0)  & M & M & T1\\
0744350901 & 00 43 26.7 & +41 23 33.3 & 2015-02-01 &  Henze & 11.3 (7.9) & 11.3 (8.2) & 9.9 (9.8)  & M & M & T1\\
0761970101 & 00 43 37.9 & +41 19 19.0 & 2015-06-27 &  Hornschemeier &  & flared &    &  &  &  \\
0763120101 & 00 44 21.9 & +41 31 16.6 & 2015-06-28 &  Sasaki & 96.7 (95.1) & 96.6 (95.2) & 94.7 (85.0)  & M & M & T1\\
0763120201 & 00 44 21.9 & +41 31 16.6 & 2016-01-21 &  Sasaki & 21.2 (8.5) & 80.9 (52.5) & 21.4 (10.3)  & M & M & T1\\
0763120301 & 00 44 49.3 & +41 49 31.1 & 2015-08-11 &  Sasaki & 101.3 (100.7) & 101.6 (100.9) & 99.9 (96.9)  & M & M & T1\\
0763120401 & 00 44 54.3 & +41 49 12.7 & 2016-01-01 &  Sasaki & 98.9 (69.6) & 99.0 (73.2) & 97.4 (50.8)  & M & M & T1\\
0764030301 & 00 42 57.7 & +41 08 12.3 & 2016-01-16 &  Henze &  & flared &    &  &  &  \\
0764030401 & 00 42 57.7 & +41 08 12.3 & 2016-02-09 &  Henze & 11.1 (10.6) & 11.6 (11.4) & 7.0 (5.6)  & T1 & T1 & T1\\
0784000101 & 00 45 28.8 & +41 54 09.9 & 2016-12-26 &  Henze & 40.5 (13.6) & 40.5 (14.5) & 36.1 (6.4)  & T1 & T1 & T1\\
0784000201 & 00 45 28.8 & +41 54 09.9 & 2016-12-28 &  Henze & 57.5 (40.0) & 58.3 (40.9) & 56.0 (35.4)  & T1 & T1 & T1\\
0790830101 & 00 43 31.9 & +41 13 14.3 & 2017-01-21 &  Yukita & 58.7 (39.2) & 58.6 (39.5) & 57.0 (36.4)  & M & M & M \\
0800730101 & 00 37 02.3 & +43 13 20.9 & 2017-06-28 &  Li & 18.1 (15.6) & 22.8 (18.5) & 14.3 (6.6)  & T1 & T1 & T1\\
0800730201 & 00 35 39.4 & +42 52 34.0 & 2017-07-25 &  Li & 15.6 (15.6) & 15.6 (15.5) & 11.7 (11.4)  & T1 & T1 & T1\\
0800730301 & 00 34 06.2 & +42 34 03.8 & 2017-07-14 &  Li & 23.7 (20.4) & 24.6 (23.7) & 13.8 (11.5)  & T1 & T1 & T1\\
0800730401 & 00 35 05.4 & +42 11 46.9 & 2017-07-27 &  Li & 15.6 (15.3) & 15.6 (15.6) & 11.7 (11.7)  & T1 & T1 & T1\\
0800730501 & 00 36 27.1 & +42 28 34.6 & 2017-07-28 &  Li & 24.5 (17.7) & 24.5 (18.6) & 14.1 (11.5)  & T1 & T1 & T1\\
0800730601 & 00 37 41.8 & +42 48 47.7 & 2017-08-12 &  Li & 15.6 (13.5) & 15.6 (13.3) & 11.1 (7.7)  & T1 & T1 & T1\\
0800730701 & 00 38 31.0 & +42 25 33.8 & 2017-08-12 &  Li & 15.5 (15.4) & 15.6 (15.4) & 10.8 (10.8)  & T1 & T1 & T1\\
0800730801 & 00 37 14.0 & +42 09 03.0 & 2017-08-10 &  Li & 15.7 (15.6) & 15.7 (15.7) & 11.8 (11.8)  & T1 & T1 & T1\\
0800730901 & 00 35 55.5 & +41 53 21.6 & 2018-01-07 &  Li & 19.7 (19.6) & 19.7 (19.6) & 15.8 (14.8)  & T1 & T1 & T1\\
0800731001 & 00 36 43.7 & +41 33 24.1 & 2017-08-10 &  Li & 15.6 (15.4) & 15.6 (15.5) & 11.7 (11.7)  & T1 & T1 & T1\\
0800731101 & 00 39 10.6 & +42 01 39.4 & 2018-01-07 &  Li & 20.9 (20.7) & 20.9 (20.8) & 17.3 (16.5)  & T1 & T1 & T1\\
0800731201 & 00 40 13.9 & +42 22 05.6 & 2018-01-10 &  Li & 17.6 (6.3) & 17.6 (8.0) & 4.4 (3.7)  & T1 & T1 & T1\\
0800731301 & 00 42 27.3 & +42 13 16.8 & 2018-01-11 &  Li & 11.4 (10.9) & 11.4 (11.1) & 7.5 (6.6)  & T1 & T1 & T1\\
0800731401 & 00 41 08.2 & +41 58 34.9 & 2018-01-11 &  Li & 15.2 (7.0) & 15.1 (7.3) & 11.7 (4.5)  & T1 & T1 & T1\\
0800731501 & 00 38 58.6 & +41 29 21.5 & 2017-08-12 &  Li & 15.6 (15.5) & 15.6 (15.4) & 11.7 (11.4)  & T1 & T1 & T1\\
0800731601 & 00 37 48.6 & +41 12 46.3 & 2017-08-02 &  Li & 21.5 (21.4) & 21.5 (21.5) & 17.6 (16.2)  & T1 & T1 & T1\\
0800731701 & 00 43 24.4 & +40 18 50.8 & 2018-01-11 &  Li &  & flared &    &  &  &  \\
0800731801 & 00 45 00.6 & +40 36 11.2 & 2018-01-12 &  Li & 17.6 (3.8) & 17.6 (5.4) & 8.9 (0.7)  & T1 & T1 & T1\\
0800731901 & 00 46 32.3 & +40 56 27.6 & 2018-01-13 &  Li & 17.6 (17.5) & 17.6 (17.6) & 13.7 (13.7)  & T1 & T1 & T1\\
0800732001 & 00 47 44.2 & +41 19 12.1 & 2018-01-13 &  Li & 17.6 (17.6) & 17.6 (17.5) & 13.6 (13.5)  & T1 & T1 & T1\\
0800732101 & 00 48 44.9 & +41 00 00.0 & 2018-01-13 &  Li & 19.2 (15.7) & 21.1 (16.9) & 16.8 (11.2)  & T1 & T1 & T1\\
0800732201 & 00 48 06.0 & +40 44 34.9 & 2018-01-14 &  Li & 15.6 (10.0) & 15.6 (10.2) & 11.7 (5.6)  & T1 & T1 & T1\\
0800732301 & 00 45 23.1 & +40 10 48.9 & 2018-01-15 &  Li & 15.6 (11.7) & 15.6 (12.7) & 11.7 (7.4)  & T1 & T1 & T1\\
0800732401 & 00 47 03.6 & +40 04 29.5 & 2018-01-15 &  Li & 15.3 (6.0) & 15.3 (6.4) & 11.2 (1.8)  & T1 & T1 & T1\\
0800732501 & 00 49 41.6 & +40 42 43.4 & 2018-01-15 &  Li & 16.9 (11.9) & 16.9 (12.3) & 12.9 (7.6)  & T1 & T1 & T1\\
0800732601 & 00 50 32.4 & +40 24 05.7 & 2018-01-16 &  Li & 8.7 (6.4) & 9.1 (7.9) & 5.6 (3.5)  & T1 & T1 & T1\\
0800732701 & 00 49 03.6 & +40 00 34.6 & 2018-01-17 &  Li & 15.6 (10.1) & 15.6 (11.0) & 11.6 (4.8)  & T1 & T1 & T1\\
0800732801 & 00 48 17.8 & +39 42 55.5 & 2018-01-17 &  Li & 15.6 (12.3) & 15.6 (12.2) & 11.6 (6.7)  & T1 & T1 & T1\\
0800732901 & 00 50 11.8 & +39 49 41.7 & 2018-01-19 &  Li & 15.3 (12.5) & 15.3 (13.2) & 11.7 (7.6)  & T1 & T1 & T1\\
0800733001 & 00 51 23.6 & +40 05 56.9 & 2018-01-19 &  Li & 11.8 (11.5) & 11.8 (11.8) & 11.4 (11.2)  & T1 & T1 & T1\\
 0800733101 & 00 38 31.0 & +42 25 33.8 & 2018-01-03 &  Li &  & flared &   \\ 
 total      &            &             &            &     & 3082.9 (2491.6) & 3133.9 (2602.0) & 2710.2 (1927.4) 
\label{table:XMMDatalist}
\end{longtable*} %

\clearpage

\section{Astrometric Accuracy}\label{appendsec:astrometry}

\begin{figure}
\begin{center}
\includegraphics[width=1.0 \columnwidth]{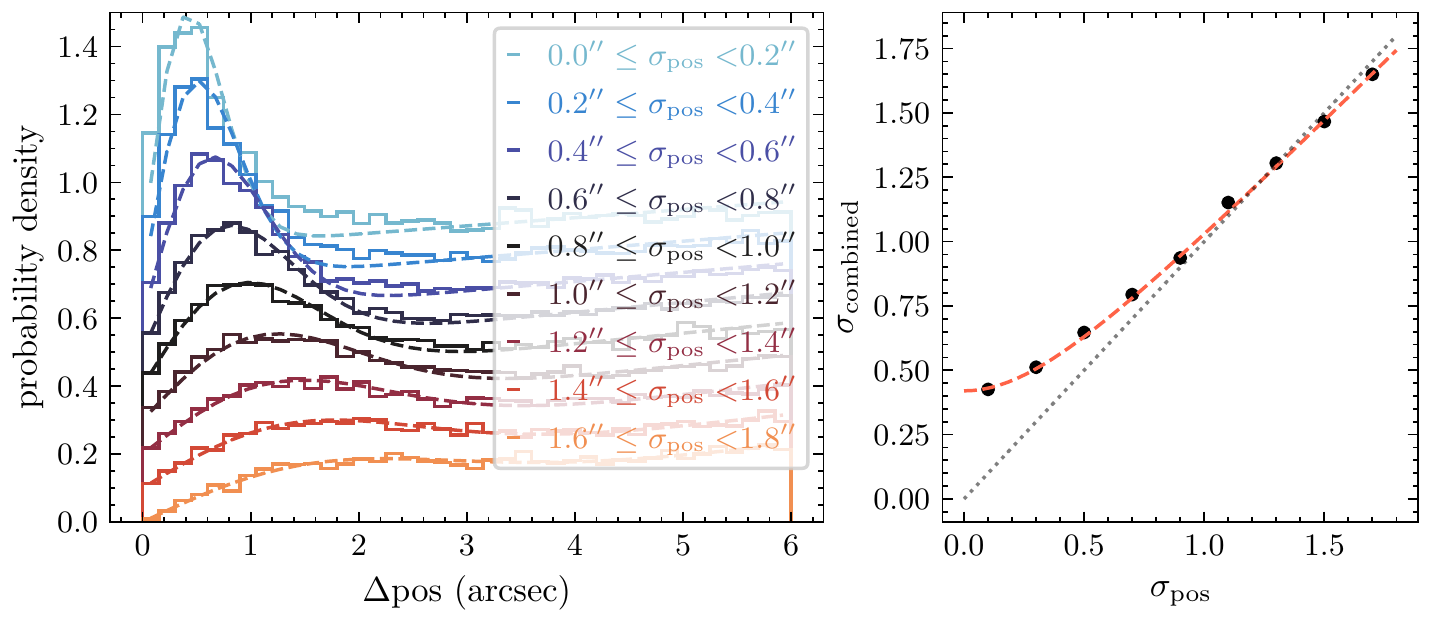}
\end{center}
\caption{Left: The probability density function (PDF) of the offset $\Delta \mathrm{pos}$ between \emph{XMM-Newton} X-ray sources and GAIA optical sources at the different ranges of the position statistical uncertainty $\sigma_{\mathrm{pos}}$ of the X-ray sources. The corresponding best-fit models are shown in dashed lines. The distribution have been shifted vertically for the ease of comparison. 
Right: Total position uncertainty $\sigma_{\mathrm{combined}}$ as the function of $\sigma_{\mathrm{pos}}$.
The $\sigma_{\mathrm{combined}}$ is the $\sigma$ of Rayleigh distribution. The measured $\sigma_{\mathrm{combined}}$
is in black dot, while the red dashed line is the best-fit model and the black dotted line represent the uncorrected $\sigma_{\mathrm{pos}}$.}
\label{fig:astrometry}
\end{figure}

We initially use a $\Delta \mathrm{pos}<3.44~\sigma_{\mathrm{pos}}$ criterion to match X-ray sources, but this approach proved inadequate for exceptionally bright sources. This is due to the fact that the statistical position uncertainty, $\sigma_{\mathrm{pos}}$, underestimated the actual uncertainty for these sources, resulting in failed matches. 
To address this issue, we match the 4XMM-DR11s serendipitous source catalogue from stacks \citep{Traulsen20,Traulsen22} with GAIA DR3 catalogue \citep{Gaia2022yCat} to apply a correction to the position uncertainty.
We filter the GAIA DR3 catalogue with Gmag$<$18 and PSS$>$0.9, which excludes optical faint sources and limits our selection to fg stars, as some fg stars are particularly bright in X-ray observations.
In Fig.~\ref{fig:astrometry}, we clearly see the evolution of the distribution of the offset $\Delta \mathrm{pos}$ between the X-ray sources and GAIA sources with the position uncertainty $\sigma_{\mathrm{pos}}$. The distribution can be modeled using a Rayleigh function and a proportional function. The Rayleigh function represents true counterparts, while the proportional function accounts for the false positive matches. We take the fitted $\sigma$ of Rayleigh function as the $\sigma_{\mathrm{combined}}$. The relation between the $\sigma_{\mathrm{combined}}$ and the statistical position uncertainty $\sigma_{\mathrm{pos}}$ is shown in the right panel of Fig.~\ref{fig:astrometry}. At low $\sigma_{\mathrm{pos}}$, which is often the case for bright X-ray sources, the deviation of $\sigma_{\mathrm{combined}}$ from $\sigma_{\mathrm{pos}}$ is evident. However, at the large $\sigma_{\mathrm{pos}}$ end, $\sigma_{\mathrm{combined}}$ is almost unchanged.

We fit the measured $\sigma_{\mathrm{combined}}$, and get the best fit model (Equation.~\ref{eq:position_uncerainty}) $$\sigma_{\mathrm{combined}}=\sqrt{0.89\times \sigma_{\mathrm{pos}}^2 +0.42^2}$$
Then we modify the matching criterion based on distance as $\Delta \mathrm{pos}<3.44~\sigma_{\mathrm{combined}}$.
Since we use the same source detection tool \texttt{edetect\_stack} in this work as in 4XMM-DR11s,
the empirical relation of the $\sigma_{\mathrm{combined}}$ can be adopted in this work.

\section{Cross-matching with NWAY}\label{appendsec:CrossMatchGAIA}

\begin{figure}
\begin{center}
\includegraphics[width=0.9 \columnwidth]{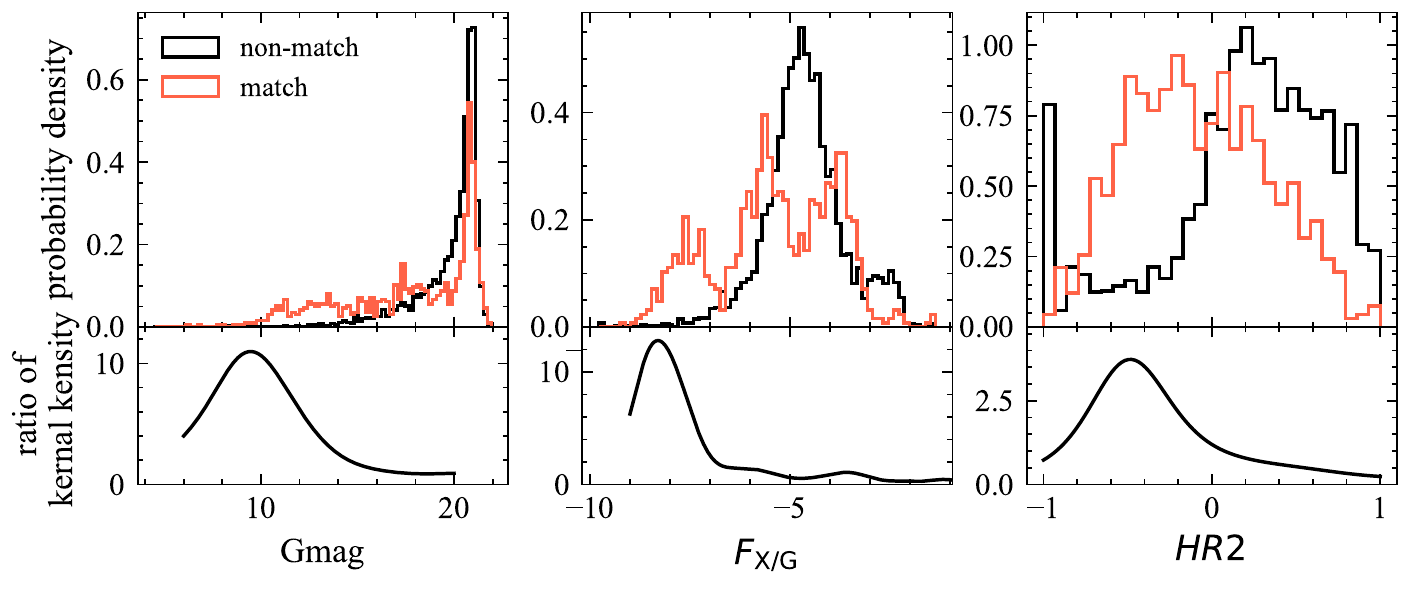}
\end{center}
\caption{The probability density of GAIA Gmag, $F_{\mathrm{X/G}}$(flux ratio between 0.5-2.0 keV to GAIA G band), and $HR2$ for secure matches and non-matches based on distance. Also shown are the corresponding ratio of the kernel density of these parameters.}
\label{fig:nway_kde}
\end{figure}

\begin{figure}
\begin{center}
\includegraphics[width=0.9 \columnwidth]{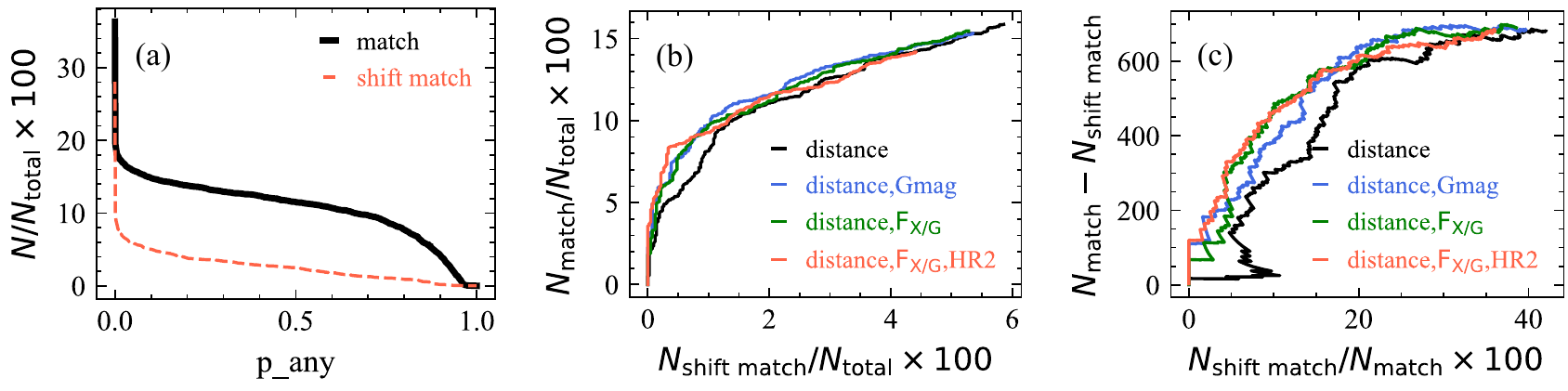}
\end{center}
\caption{a. The fraction of matched sources and matches of shifted X-ray sources as the function of p\_any (the probability of matching defined in NWAY) based on the distance; b. The fraction of the matched sources as the function of the fraction of the matches of shifted sources; c. The estimated number of secure matches as the function of the fraction of false positive matches. In b and c, we demonstrate the comparison of matching performance between different priors: no prior (distance only), Gmag, $F_{\mathrm{X/G}}$, and $F_{\mathrm{X/G}}$ with $HR2$.}
\label{fig:nway_performance}
\end{figure}

In our study, we utilize NWAY \citep{salvatoFindingCounterpartsAllsky2018}, a Bayesian cross-matching algorithm, to match fg stars with our X-ray source list. Since it is possible for one X-ray source to match with multiple stars within the distance criterion due to the numerous stars in GAIA catalogue. NWAY takes into account positional uncertainties and assigns probabilities to potential matches based on source properties and positional agreement. The priors of these properties can be automatically assigned based on the sources of secure matches and non-matches in the field. Compared to distance-based matching algorithms, NWAY can handle complex fields and provides more reliable and informative matches.

To select reference star catalogues, we filter GAIA sources with PSS $>$ 0.9. Initially, we match GAIA sources based solely on distance and compare the GAIA parameter distributions of the matched sources with those of non-matched sources. Fig.~\ref{fig:nway_kde} shows the parameter distributions and highlights a significant trend where counterpart stars are optically bright, have a relatively small flux ratio between 0.5-2.0 keV and GAIA G band ($F_{\mathrm{X/G}}$), and the matched X-ray source are soft.

Next, we consider $Gmag$, $F_{\mathrm{X/G}}$, and $HR2$ as prior parameters and re-run the NWAY algorithm. To test the reliability of our method, we randomly shift our X-ray source list and perform matching with the same prior parameters. In Fig.~\ref{fig:nway_performance}a, we present the fraction of matched sources ($N_{\mathrm{match}}/N_{\mathrm{total}}$) and matches to shifted X-ray sources ($N_{\mathrm{shift~match}}/N_{\mathrm{total}}$) as a function of p\_any. $N_{\mathrm{total}}$ is the total number of the X-ray sources to be matched. p\_any is the posterior probability of the X-ray source having any correct counterparts defined by NWAY. In Fig.~\ref{fig:nway_performance}b, We present the function of the fraction of matched sources as the fraction of matches to shifted sources by varying p\_any. The different priors, including no prior (distance only), $Gmag$, $F_{\mathrm{X/G}}$, and $F_{\mathrm{X/G}}+HR2$, are compared. 
We estimate the true counterpart as the function of the false match fraction in Fig.~\ref{fig:nway_performance}c. Increasing the range of p\_any can include more true counterparts but comes at the cost of a higher fraction of false matches. However, there is still an upper limit to the number of true sources. The introduction of other parameters has resulted in a noticeable improvement in matches.
We can clearly observe the improvement in the matches after introducing other parameters. 
As Gmag and $F_{\mathrm{X/G}}$ are correlated, we ultimately choose to use $F_{\mathrm{X/G}}$ for its better performance. Additionally, despite HR2 not resulting in a significant improvement in the final matching performance, we still took it into consideration since it represents the X-ray color preference of stars. Ultimately, we selected p\_any$>$0.9 to identify foreground stars in X-ray sources, which corresponded to only 5\% of false matches.


\section{Description of the New-ANGELS catalogue columns}\label{appendsec:CatalogColumn}

\begin{enumerate}

\item Column 1, \texttt{SRCID}: Source identifier. 
\item Column 2, \texttt{OBS\_ID}: Identification pf XMM-Newton observation.
\item Column 3, \texttt{N\_OBS}: Number of observations involved in the stack.  
\item Column 4, \texttt{N\_CONTRIB}: Number of observations in which the source was fitted. 
\item Column 5, \texttt{N\_EXP}: Number of instruments in which the source was fitted.
\item Column 6-8, \texttt{RA}, \texttt{DEC} and \texttt{RADEC\_ERR}: Right ascension and Declination (J2000, in degree) and square root of squared sum of their 1-sigma uncertaintys (in arcsec).
\item Column 9-10, \texttt{ast\_RA} and \texttt{ast\_DEC}: Astrometry corrected Right ascension and Declination (J2000, in degree).
\item Column 11-12, \texttt{RAOFFSET} and \texttt{DECOFFSET}: RA and DEC offset of the observation preferred in 4XMM-DR12s.

\item Column 13-14, \texttt{LII} and \texttt{BII}: Galactic longitude and latitude (in degree).
\item Column 15, \texttt{DIST\_NN}: Distance (in arcsec) to the nearest neighbouring detected source.
\item Column 16, \texttt{N\_BLEND}: Number of neighbouring sources being fit simultaneously. 
\item Column 17-18, 29-30, 41-42, 53-54, \texttt{EP\_FLUX}, \texttt{PN\_FLUX}, \texttt{M1\_FLUX}, \texttt{M2\_FLUX}, \texttt{EP\_FLUX\_ERR}, \texttt{PN\_FLUX\_ERR}, \texttt{M1\_FLUX\_ERR} and  \texttt{M2\_FLUX\_ERR}: Total flux ($\mathrm{
erg~cm^{-2}~s^{-2}}$) of all-EPIC, pn, MOS1, and MOS2 and their 1-sigma uncertainty in the energy range of 0.2-12.0 keV. 
The flux and flux uncertainty are derived from counts rates and errors according to energy conversion factors. ($\rm{ ECF = \frac{count~rate}{flux}}$, in units of $10^{11} \rm~cts~cm^{-2}/erg$).
The factor for each detector and energy band is different depend on the detector, pattern selection and filter used during the observation. So the ECFs (fluxes and energy conversion factors) were derived with \texttt{XSPEC} using on-axis response matrices assuming a power law model with photon index $\Gamma=1.7$ and the Galactic foreground absorption of $N_H=7 \times 10^{20}\mathrm{cm^{-2}}$ (Stark et al. 1992, see also PFH2005) to be the universal source spectrum for the ECF calculation.

\item Column 19-28, 31-40, 43-52, 55-64, \texttt{EP\_n\_FLUX}, \texttt{PN\_n\_FLUX}, \texttt{M1\_n\_FLUX}, \texttt{M2\_n\_FLUX}, \texttt{EP\_n\_FLUX\_ERR}, \texttt{PN\_n\_FLUX\_ERR}, \texttt{M1\_n\_FLUX\_ERR} and \texttt{M2\_n\_FLUX\_ERR}: The flux ($\mathrm{erg~cm^{-2}s^-1)}$) of all-EPIC, pn, MOS1, and MOS2 in energy band n and their 1-sigma uncertainty. Energy band 1, 2, 3, 4, and 5 cover 0.2-0.5 keV, 0.5-1.0 keV, 1.0-2.0 keV, 2.0-4.5 keV, and 4.5-12.0 keV respectively. The flux and the flux uncertainty of all-EPIC are weighted from that of pn, MOS1, and MOS2.

\begin{equation}
\begin{aligned}
\rm EP\_n\_FLUX = & \rm \frac{PN\_n\_FLUX \times (PN\_n\_FLUX\_ERR)^{-2}}{(PN\_n\_FLUX\_ERR)^{-2}+(M1\_n\_FLUX\_ERR)^{-2}+(M2\_n\_FLUX\_ERR)^{-2}} \\
& \rm + \frac{M1\_n\_FLUX \times (M1\_n\_FLUX\_ERR)^{-2}}{(PN\_n\_FLUX\_ERR)^{-2}+(M1\_n\_FLUX\_ERR)^{-2}+(M2\_n\_FLUX\_ERR)^{-2}} \\
& \rm + \frac{M2\_n\_FLUX \times (M2\_n\_FLUX\_ERR)^{-2}}{(PN\_n\_FLUX\_ERR)^{-2}+(M1\_n\_FLUX\_ERR)^{-2}+(M2\_n\_FLUX\_ERR)^{-2}}
\end{aligned}
\end{equation}

\begin{equation}
\begin{aligned}
\rm EP\_n\_FLUX\_ERR = & \rm ((\frac{PN\_n\_FLUX\_ERR \times (PN\_n\_FLUX\_ERR)^{-2}}{(PN\_n\_FLUX\_ERR)^{-2}+(M1\_n\_FLUX\_ERR)^{-2}+(M2\_n\_FLUX\_ERR)^{-2}})^2 \\
& \rm +(\frac{M1\_n\_FLUX\_ERR \times (M1\_n\_FLUX\_ERR)^{-2}}{(PN\_n\_FLUX\_ERR)^{-2}+(M1\_n\_FLUX\_ERR)^{-2}+(M2\_n\_FLUX\_ERR)^{-2}})^2 \\
& \rm +(\frac{M2\_n\_FLUX\_ERR \times (M2\_n\_FLUX\_ERR)^{-2}}{(PN\_n\_FLUX\_ERR)^{-2}+(M1\_n\_FLUX\_ERR)^{-2}+(M2\_n\_FLUX\_ERR)^{-2}})^2)^{-2}
\end{aligned}
\end{equation}

\item Column 65-68, 79-80, 91-92, \texttt{EP\_RATE}, \texttt{PN\_RATE}, \texttt{M1\_RATE}, \texttt{M2\_RATE}, \texttt{EP\_RATE\_ERR}, \texttt{PN\_RATE\_ERR}, \texttt{M1\_RATE\_ERR} and \texttt{M2\_RATE\_ERR}: Count rate ($\mathrm{counts~s^{-1}}$) of all-EPIC, pn, mos1 and mos2 detectors and their 1-sigma uncertainty.

\item Column 69-78, 81-90, 93-102 \texttt{PN\_n\_RATE}, \texttt{M1\_n\_RATE}, \texttt{M2\_n\_RATE}, \texttt{PN\_n\_RATE\_ERR}, \texttt{M1\_n\_RATE\_ERR} and \texttt{M2\_n\_RATE\_ERR}: Count rate ($\mathrm{counts~s^{-1}}$) of pn, mos1 and mos2 detectors in energy band n and their 1-sigma uncertainty.

\item Column 103-110, \texttt{EP\_CTS}, \texttt{PN\_CTS}, \texttt{M1\_CTS}, \texttt{M2\_CTS}, \texttt{EP\_CTS\_ERR}, \texttt{PN\_CTS\_ERR}, \texttt{M1\_CTS\_ERR} and  \texttt{M2\_CTS\_ERR}: Number of counts ($\mathrm{counts}$) of all-EPIC, pn, mos1 and mos2 detectors and their 1-sigma uncertainty.

\item Column 111, 117, 123, 129 \texttt{EP\_DET\_ML}, \texttt{PN\_DET\_ML}, \texttt{M1\_DET\_ML} and \texttt{M2\_DET\_ML}:  Equivalent maximum detection likelihood of all-EPIC, pn, mos1 and mos2 detectors. The likelihood is calculated as
\begin{equation}
\mathrm{DET\_ML}=-\ln (1-\Gamma(\frac{\nu}{2}, \sum_{i=1}^{n} \frac{\Delta C_{i}}{2})) \\
\label{eqn:ML}
\end{equation}
where $C$ is C-statistic (\citealt{1979ApJ...228..939C}). $n$ is  the number of images used for the detection of the same source. $\nu$ is the degree of freedom, $\nu=n+2$ for point source, $\nu=n+3$ for the extend source. 
While combining the detection from different instrument or observations, $\Delta C$ are summed up in equation \ref{eqn:ML}.
In principal, the detection likelihood also follows $L=-\ln(p)$, where the p is the probability of resulting in at least the observed counts by Poissonian random fluctuation. The probability $p$ is calculated from incomplete Gamma function $P(a,b)$, where a is the raw source counts and b is the raw background counts in the detection cell. 
 (For more detail of detection likelihood, see \citealt{watsonXMMNewtonSerendipitousSurvey2009}, \citealt{traulsenXMMNewtonSerendipitousSurvey2019}) 

\item Column 112-116, 118-122, 124-128, 130-134 \texttt{EP\_n\_DET\_ML}, \texttt{PN\_n\_DET\_ML}, \texttt{M1\_n\_DET\_ML} and \texttt{M2\_n\_DET\_ML}: Equivalent maximum detection likelihood ($n=1$ in equation \ref{eqn:ML}) of all-EPIC, pn, MOS1 and MOS2 detectors in energy band n, n=1, 2, 3, 4 and 5. 

\item Column 135-136, \texttt{EXTENT} and \texttt{EXTENT\_ERR}: Extent radius ($\mathrm{arcsec}$) ans its 1-sigma uncertainty. The parameters are obtained when fitting the extended source with a convolution of the instrumental PSF and an extent model ($\beta$ model with fixed $\beta=2/3$). ($\nu=n+3$ in equation \ref{eqn:ML}). 
\item Column 137, \texttt{EXTENT\_ML}: Likelihood of the detection being extended when fitting with a
convolution of the instrumental PSF and an extent model ($\beta$ model with fixed $\beta=2/3$). ($\nu=n+3$ in equation \ref{eqn:ML}). 
\item Column 138-169, \texttt{EP\_HRn}, \texttt{PN\_HRn}, \texttt{M1\_HRn}, \texttt{M2\_HRn}, \texttt{EP\_HRn\_ERR}, \texttt{PN\_HRn\_ERR}, \texttt{M1\_HRn\_ERR} and \texttt{M2\_HRn\_ERR}: Hardness ratio and corresponding 1-sigma uncertainty of all-EPIC, pn, mos1, and mos2 detectors between energy band n and n+1, n=1, 2, 3 and 4. The hardness ratios are defined as:${HR}_n=\frac{B_{n+1}-B_{n}}{B_{n+1}+B_{n}}$, $B_{n}$ is the count rate in energy band n. After computing the hardness ratios, we can plot the hardness ratio diagram to classify the spectra.

\item Column 170, 176, 182, \texttt{PN\_EXP}, \texttt{M1\_EXP} and \texttt{M2\_EXP}: PSF-weighted exposure time ($\mathrm{second}$) of the detection in pn, mos1 and mos2.
\item Column 171-175, 177-181, 183-187, \texttt{PN\_n\_EXP}, \texttt{M1\_n\_EXP} and \texttt{M2\_n\_EXP}: PSF-weighted exposure ($\mathrm{second}$) of the detection in energy band n of pn, mos1, and mos2.

\item Column 188, 194, 100, \texttt{PN\_BG}, \texttt{M1\_BG} and \texttt{M2\_BG}: Background map ($\mathrm{counts~pixel^{-1}}$) at detection position on pn, mos1 and mos2.
\item Column 189-193, 195-199, 201-205, \texttt{PN\_n\_BG}, \texttt{M1\_n\_BG} and \texttt{M2\_n\_BG}: Background map ($\mathrm{counts~pixel^{-1}}$) in energy band n of pn, mos1 and mos2 at the detection position.
\item Column 206-209, \texttt{EP\_ONTIME}, \texttt{PN\_ONTIME}, \texttt{M1\_ONTIME} and \texttt{M2\_ONTIME}: Total good exposure time ($\mathrm{second}$) of all-EPIC, pn, mos1 and mos2 detectors.
\item Column 210-212, \texttt{PN\_PILEUP}, \texttt{M1\_PILEUP} and \texttt{M2\_PILEUP}: Estimate of the pile-up level in pn, mos1 and mos2. It is calculated in terms of source count rate and instrument readout. For point-like sources, it is defined as the source count rate times the frame time, divided by an instrument-specific pile-up threshold. For extended sources, it is derived from the source counts per instrument pixel times the frame time, divided by the number of instrument pixels covered by an image pixel and by a pile-up threshold.
\item Column 213-215, \texttt{PN\_MASKFRAC}, \texttt{M1\_MASKFRAC} and \texttt{M2\_MASKFRAC}: PSF-weighted detector coverage in pn, mos1 and mos2.
\item Column 216, \texttt{DIST\_REF}: Distance (arcmin) to the reference coordinates of the field. 
\item Column 217-220, \texttt{EP\_OFFAX}, \texttt{PN\_OFFAX}, \texttt{M1\_OFFAX} and \texttt{M2\_OFFAX}: Offaxis angle (arcmin) between source position and aim point all-EPIC, pn, mos1, and mos2 detectors.
\item Column 221-235, \texttt{PN\_n\_VIG}, \texttt{M1\_n\_VIG} and \texttt{M2\_n\_VIG}: Vignetting factor in energy band n  of pn, mos1 and mos2 detectors for the detection.
\item Column 236, \texttt{STACK\_FLAG}: Integer representation of the stack detection flags.
\item Column 237, \texttt{OVERLAP}: True, if \texttt{N\_CONTRIB}$\geq2$ at the source position.
\item Column 238, \texttt{VAR\_CHI2}: Reduced chi square of all-EPIC inter-observation variability. \texttt{VAR\_CHI2}=$\frac{1}{n-1} \sum_{k=1}^{n}\left(\frac{F_{k}-F_{\mathrm{EPIC}}}{\sigma_{k}}\right)^{2}$, $F_{k}$ is \texttt{EP\_FLUX} for each observation, $\sigma_{k}$ is \texttt{EP\_FLUX\_ERR}. $F_{EPIC}$ is the mean \texttt{EP\_FLUX} for all detections. 
\item Column 239-243, \texttt{VAR\_CHI2\_n}: Reduced chi square of all-EPIC band n inter-observation variability.

\item Column 244, \texttt{VAR\_PROB}: Probability that the all-EPIC flux variability is consistent with zero. \texttt{VAR\_PROB} $=\int_{\chi^{2}}^{\infty} \frac{x^{v / 2-1} e^{-x / 2}}{2^{v / 2} \Gamma(v / 2)} d x$, $\chi^2$ is \texttt{VAR\_CHI2} at $\nu=n-1$ degrees of freedom. The high \texttt{VAR\_PROB} means that the flux of the source is almost constant throughout the observations.

\item Column 245-249, \texttt{VAR\_PROB\_n}: Probability that the all-EPIC band n flux variability is consistent with zero.
\item Column 250, \texttt{FRATIO} and \texttt{FRATIO\_ERR}: The ratio and corresponding 1-sigma uncertainty between the highest and the lowest (non-zero, non-null) mean EPIC flux. \texttt{FRATIO} $=F_{\max } / F_{\min }$, $F_{\max }$ and $F_{\min }$ are the maximum and minimum of \texttt{EP\_FLUX} in detections. 
\item Column 252-261, \texttt{FRATIO\_n} and \texttt{FRATIO\_n\_ERR}: The ratio and corresponding 1-sigma uncertainty between the highest and the lowest (non-zero, non-null) mean EPIC flux in energy band n.
\item Column 262, \texttt{FLUXVAR}: The Largest all-EPIC flux difference in terms of sigma.  \texttt{FLUXVAR}$=\max _{k, l \in[1, n]} \frac{\left|F_{k}-F_{l}\right|}{\sqrt{\sigma_{k}^{2}+\sigma_{l}^{2}}}$, $F_{k}$ is \texttt{EP\_RATE}, $\sigma_{k}$ is \texttt{EP\_RATE\_ERR}.

\item Column 263-267, \texttt{FLUXVAR\_n}: The Largest all-EPIC band n flux difference in energy band n in terms of sigma.  
\item Column 269-270, \texttt{MJD\_FIRST} and \texttt{MJD\_LAST}: Modified Julian Date JD-2400000.5 (in days) when starting and ending the observation.

\item Column 270, \texttt{REVOLUT}: XMM-Newton revolution number (in orbit) of the observation.
\item Column 271, \texttt{PA\_PNT}: Mean position angle (degree) of the spacecraft.
\item Column 272-274, \texttt{PN\_SUBMODE}, \texttt{M1\_SUBMODE} and \texttt{M2\_SUBMODE}: instrument submode of pn, mos1 and mos2 used for the observation.
\item Column 275-277, \texttt{PN\_FILTER}, \texttt{M1\_FILTER} and \texttt{M2\_FILTER}: Filter used for the pn, mos1 and mos2.
\item Column 278, \texttt{type}: The identification or classification type assigned to each source. The classification types include fg star, AGN, galaxy, GlC, GCl, SNR, HMXB, LMXB, and SSS. The classification of each candidate source is enclosed within angle brackets ($<>$) for easy identification.
\item Column 279, \texttt{ast\_type}: The same as \texttt{type}, but the source is matched with new source position (\texttt{ast\_RA} and \texttt{ast\_DEC}).
\item Column 280, \texttt{note}: additional comments.
\item Column 281, \texttt{XMMLPt}: the source number of matched sources in \citet{stieleDeepXMMNewtonSurvey2011a}.
\item Column 282, \texttt{CSC\_name}: The name of the matched source in the Chandra Source Catalog Release 2.0 \citep{Evans2018}.
\end{enumerate}


\section{An example of the source catalogue table}\label{appendsec:catalog}
\begin{sidewaystable}
\caption{An example of the source catalogue table that contains only one source covered by two observations. Each column has three rows for this source. The first row is the summary row which contains the parameters obtained based on all of the related observations. The second and third rows contain the results based on two individual observations.}

    \tiny
    \begin{tabular}{|*{10}{c|}} 
    \hline\hline
    \multicolumn{10}{|c|}{Begin of Table}       \\
    \hline\hline
SRCID & OBS\_ID & N\_OBS & N\_CONTRIB & N\_EXP & RA & DEC & ast\_RA & ast\_DEC & RAOFFSET  \\
\hline
23010927070100304 &   & 31 & 8 & 20 & 11.15348030 & 41.42053952 & 11.09348034 & 42.28709875 & 0.05999996  \\
\hline
23010927070100304 & 0109270701 & -- & -- & 3 & 11.15348030 & 41.42053952 & -- & -- & 0.0  \\
\hline
23010927070100304 & 0402561001 & -- & -- & 3 & 11.15348030 & 41.42053952 & -- & -- & 0.0  \\
\hline\hline
DECOFFSET & RADEC\_ERR & LII & BII & DIST\_NN & N\_BLEND & EP\_FLUX & EP\_FLUX\_ERR & EP\_1\_FLUX & EP\_1\_FLUX\_ERR  \\
\hline
-0.86655923 & 1.1616716 & 121.5576631 & -21.43299982 & 28.409655 & 1 & 9.7772e-14 & 3.1135e-14 & 4.7373e-15 & 5.0262e-16  \\
\hline
0.0 & 1.1616716 & 121.5576631 & -21.43299982 & 28.409655 & 1 & 7.9128e-14 & 4.4409e-14 & 7.7186e-15 & 2.1086e-15  \\
\hline
0.0 & 1.1616716 & 121.5576631 & -21.43299982 & 28.409655 & 1 & 1.8289e-13 & 1.0293e-13 & 5.4740e-15 & 1.4772e-15  \\
\hline\hline
PN\_FLUX & PN\_FLUX\_ERR & PN\_1\_FLUX & PN\_1\_FLUX\_ERR & M1\_FLUX & M1\_FLUX\_ERR & M1\_1\_FLUX & M1\_1\_FLUX\_ERR & M2\_FLUX & M2\_FLUX\_ERR  \\
\hline
9.1766e-14 & 4.2648e-14 & 4.3386e-15 & 5.5695e-16 & 1.1074e-13 & 6.4199e-14 & 6.5578e-15 & 1.9507e-15 & 1.1598e-13 & 6.8233e-14  \\
\hline
5.7552e-14 & 6.3138e-14 & 6.1542e-15 & 3.1553e-15 & 9.7584e-14 & 8.6648e-14 & 1.0393e-14 & 4.2934e-15 & 1.4657e-13 & 9.3396e-14  \\
\hline
2.1863e-13 & 1.5237e-13 & 5.9267e-15 & 1.7166e-15 & 1.5368e-13 & 1.3875e-13 & 3.8231e-15 & 3.9790e-15 & 1.6361e-14 & 7.6997e-15  \\
\hline\hline
M2\_1\_FLUX & M2\_1\_FLUX\_ERR & EP\_RATE & EP\_RATE\_ERR & PN\_RATE & PN\_RATE\_ERR & PN\_1\_RATE & PN\_1\_RATE\_ERR & M1\_RATE & M1\_RATE\_ERR  \\
\hline
6.4477e-15 & 1.4557e-15 & 0.03645468 & 0.00219558 & 0.02337551 & 0.00186468 & 0.00409002 & 0.00052355 & 0.00643091 & 0.00096972  \\
\hline
7.8904e-15 & 3.7736e-15 & 0.04422306 & 0.00716647 & 0.02952519 & 0.00663382 & 0.00515083 & 0.00264084 & 0.00602423 & 0.00186651  \\
\hline
4.5875e-15 & 4.2368e-15 & 0.04965767 & 0.01016562 & 0.03871836 & 0.00946021 & 0.00564509 & 0.00163499 & 0.00815423 & 0.00262398  \\
\hline\hline
M1\_1\_RATE & M1\_1\_RATE\_ERR & M2\_RATE & M2\_RATE\_ERR & M2\_1\_RATE & M2\_1\_RATE\_ERR & EP\_CTS & EP\_CTS\_ERR & PN\_CTS & PN\_CTS\_ERR  \\
\hline
0.00100714 & 0.00030009 & 0.00664825 & 0.00063500 & 0.00099870 & 0.00022413 & 3891.196 & 205.65431 & 2536.822 & 179.87935  \\
\hline
0.00158575 & 0.00065509 & 0.00867364 & 0.00196641 & 0.00120115 & 0.00057445 & 501.36707 & 64.01653 & 193.49692 & 39.751995  \\
\hline
0.00058333 & 0.00060712 & 0.00278509 & 0.00263800 & 0.00069836 & 0.00064496 & 618.26685 & 91.57865 & 439.2531 & 78.04734  \\
\hline\hline
M1\_CTS & M1\_CTS\_ERR & M2\_CTS & M2\_CTS\_ERR & EP\_DET\_ML & EP\_1\_DET\_ML & PN\_DET\_ML & PN\_1\_DET\_ML & M1\_DET\_ML & M1\_1\_DET\_ML  \\
\hline
467.70544 & 62.865623 & 886.6685 & 77.36295 & 519.283 & 40.153515 & 378.32312 & 32.29504 & 28.289425 & 2.6725852  \\
\hline
119.68079 & 33.35646 & 188.18935 & 37.486553 & 38.77565 & -- & 15.839467 & 2.4071922 & 4.0498304 & 4.0510645  \\
\hline
131.51355 & 35.5845 & 47.50021 & 32.078136 & 51.64108 & -- & 42.96344 & 7.913733 & 12.368129 & 0.53715014  \\
\hline\hline
M2\_DET\_ML & M2\_1\_DET\_ML & EXTENT & EXTENT\_ERR & EXTENT\_ML & EP\_HR1 & EP\_HR1\_ERR & PN\_HR1 & PN\_HR1\_ERR & M1\_HR1  \\
\hline
126.46535 & 6.2184834 & 27.585667 & 1.0059265 & 335.00296 & 0.59011996 & 0.03774169 & 0.57942075 & 0.04802216 & 0.48118436  \\
\hline
19.935226 & 2.939534 & 27.585667 & 1.0059265 & 335.00296 & 0.43357438 & 0.15567033 & 0.5555008 & 0.24856497 & -0.15709119  \\
\hline
0.41799778 & 0.6974814 & 27.585667 & 1.0059265 & 335.00296 & 0.5311424 & 0.14330031 & 0.488368 & 0.16499938 & 0.76985604  \\
\hline\hline
M1\_HR1\_ERR & M2\_HR1 & PN\_EXP & PN\_1\_EXP & M1\_EXP & M1\_1\_EXP & M2\_EXP & M2\_1\_EXP & PN\_BG & PN\_1\_BG  \\
\hline
0.12708639 & 0.6452527 & 519375.5 & 109760.27 & 349907.03 & 76494.08 & 632365.75 & 133364.56 & 12.555466 & 1.3695921  \\
\hline
0.35245 & 0.5969048 & 30627.36 & 6542.403 & 98723.5 & 21758.725 & 106946.56 & 23466.432 & 5e-05 & 1e-05  \\
\hline
0.34081677 & 0.38698247 & 58203.832 & 12877.205 & 80653.89 & 17832.574 & 77865.1 & 17235.486 & 1.4581821 & 0.13973232  \\
\hline\hline
M1\_BG & M1\_1\_BG & M2\_BG & M2\_1\_BG & EP\_ONTIME & PN\_ONTIME & M1\_ONTIME & M2\_ONTIME & PN\_PILEUP & M1\_PILEUP  \\
\hline
2.2322547 & 0.23470314 & 3.3413908 & 0.37062544 & 341184.5 & 271198.94 & 200632.45 & 275560.44 & 0.00924482 & 0.00125283  \\
\hline
0.5171118 & 0.06214709 & 0.5573624 & 0.06325618 & 56904.203 & 54506.547 & 56576.934 & 56904.203 & 1.0634e-07 & 0.00057253  \\
\hline
0.6422717 & 0.05786567 & 0.69377303 & 0.05424231 & 53185.09 & 42057.21 & 51987.848 & 53185.09 & 0.00096860 & 0.00075341  \\
\hline\hline
M2\_PILEUP & PN\_MASKFRAC & M1\_MASKFRAC & M2\_MASKFRAC & DIST\_REF & EP\_OFFAX & PN\_OFFAX & M1\_OFFAX & M2\_OFFAX & PN\_1\_VIG  \\
\hline
0.00110123 & 0.915966 & 0.91859525 & 0.91738135 & 22.996138 & -- & -- & -- & -- & --  \\
\hline
0.00074969 & 0.2550554 & 0.91859525 & 0.91738135 & 22.996138 & 11.488442 & 12.455518 & 12.512937 & 11.488442 & 0.41582617  \\
\hline
0.00061821 & 0.8852963 & 0.8975847 & 0.874431 & 22.996138 & 13.433759 & 14.05184 & 13.433759 & 14.252835 & 0.35662016  \\
\hline\hline
M1\_1\_VIG & M2\_1\_VIG & OVERLAP & STACK\_FLAG & VAR\_CHI2 & VAR\_PROB & FRATIO & FRATIO\_ERR & FLUXVAR & MJD\_FIRST  \\
\hline
-- & -- & True & -1 & 1.2860699 & 0.25247739 & 8.347179 & 25.2418 & 1.2803135 & 52279.30103  \\
\hline
0.39645877 & 0.44362473 & True & -1 & -- & -- & -- & -- & -- & 52279.30103  \\
\hline
0.35751617 & 0.33471233 & True & -1 & -- & -- & -- & -- & -- & 54099.81458  \\
\hline\hline
MJD\_LAST & REVOLUT & PA\_PNT & PN\_SUBMODE & M1\_SUBMODE & M2\_SUBMODE & PN\_FILTER & M1\_FILTER & M2\_FILTER & type  \\
\hline
57409.95817 & -- & -- &   &   &   &   &   &   & superbubble  \\
\hline
52279.93645 & 380 & 251.26717 & PrimeFullWindow & PrimeFullWindow & PrimeFullWindow & Medium & Medium & Medium & --  \\
\hline
54100.55166 & 1293 & 252.63655 & PrimeFullWindow & PrimeFullWindow & PrimeFullWindow & Thin1 & Medium & Medium & --  \\
\hline\hline
ast\_type & note & XMMLPt & CSC\_name & & & & & &  \\
\hline
superbubble & superbubble & -- & -- & & & & & &  \\
\hline
-- & -- & -- & -- & & & & & &  \\
\hline
-- & -- & -- & -- & & & & & &  \\
\hline\hline
  \end{tabular}
\label{catatlog_tab:1}
\end{sidewaystable}

\end{document}